\documentclass[usenatbib]{aa}  
\usepackage{float}
\usepackage{tikz}
\usepackage{lipsum,adjustbox}
\usetikzlibrary{arrows,positioning}
\usepackage{graphicx}
\usepackage[switch]{lineno}
\newcommand{\lensfit}{\textit{lens}fit\xspace}
\newcommand{\metacal}{{\sc MetaCalibration}\xspace}
\newcommand{\metadetect}{{\sc MetaDetection}\xspace}
\newcommand{\se}{{\sc SourceExtractor}\xspace}
\newcommand{\diff}{\mathrm{d}}

\usepackage{txfonts}

\usepackage{amsmath}	
\usepackage{amssymb}	
\usepackage{hyperref}
\usepackage[nameinlink,capitalise]{cleveref}
\usepackage{soul}

\hypersetup{
    colorlinks   = true,
    urlcolor     = blue, 
    linkcolor    = blue, 
    citecolor   = blue 
}
\usepackage{multirow}
\usepackage{tabularx}
\usepackage[group-separator={,}]{siunitx}

\usepackage{bm}
\usepackage{xcolor}

\newcommand{\bb}[1]{\left[ #1 \right]}

\begin{document} 

\title{KiDS-1000 cosmic shear reanalysis using \metacal}
\titlerunning{KiDS-1000 \metacal reanalysis} 
\authorrunning{M. Yoon et al.}
\author{Mijin~Yoon,
          \inst{1,2}\thanks{E-mail: myoon@strw.leidenuniv.nl}
          Henk~Hoekstra,\inst{1}
        Shun-Sheng~Li,\inst{1,2}
        Konrad~Kuijken,\inst{1}  
        Lance~Miller\inst{3},
        Hendrik~Hildebrandt\inst{2},
        Catherine~Heymans\inst{2,5},
        Benjamin~Joachimi\inst{4},
        Angus~H.~Wright\inst{2},
        Marika~Asgari\inst{6},
        Jan~Luca~van~den~Busch\inst{2}, 
        Robert~Reischke\inst{2,7}, and
        Benjamin~St\"{o}lzner\inst{2}}
          
\institute{
Leiden Observatory, Leiden University, PO Box 9513, 2300 RA Leiden, the Netherlands
\and Ruhr University Bochum, Faculty of Physics and Astronomy, Astronomical Institute (AIRUB), German Centre for Cosmological Lensing, 44780 Bochum, Germany 
\and Department of Physics, University of Oxford, Denys Wilkinson Building, Keble Road, Oxford OX1 3RH, UK
\and Department of Physics and Astronomy, University College London, Gower Street, London WC1E 6BT, UK
\and Institute for Astronomy, University of Edinburgh, Blackford Hill, Edinburgh, EH9 3HJ, UK
\and School of Mathematics, Statistics and Physics, Newcastle University, Herschel Building, NE1 7RU, Newcastle-upon-Tyne, UK
\and Argelander-Institut für Astronomie, Universität Bonn, Auf dem Hügel 71, D-53121 Bonn, Germany}

\date\today

\abstract{A number of cosmic shear studies have reported results that are in mild tension with Planck cosmic microwave background measurements. To explore if this can be caused by biases in shear estimation, we revisit data analyses from the Kilo-Degree Survey (KiDS) using an alternative shape measurement pipeline that is more robust to uncertainties in the calibration. To this end, we present an implementation of \metacal and compare its performance to that of \lensfit, which has been used in previous data analyses. We find that the multiplicative bias is reduced, in particular for the most distant redshifts, as derived from multi-band image simulations designed to match the KiDS data. For all tomographic bins, we obtain a multiplicative bias $|m|<0.017$, with negligible additive bias. Importantly, the calibration has a negligible sensitivity to key galaxy properties. The resulting robust shear estimates were used to obtain cosmological parameter constraints. We find that the parameter, $S_8\equiv \sigma_8 \sqrt{\Omega_\mathrm{m}/0.3} =0.789_{-0.024}^{+0.020}$, is consistent with the previous KiDS-1000 \lensfit constraint
of $S_8=0.776^{+0.029 +0.002}_{-0.027-0.003}$ (statistical and  systematic errors). Thanks to the higher effective source density of the \metacal catalogue, the constraining power is improved by about 28\%.
The difference in $S_8$ with the Planck value remains at a similar level (1.8$\sigma$), implying that it is not caused by the shear measurements.} 


\maketitle

\section{Introduction}

The growth of the large-scale structure depends on the expansion history of the Universe and the theory of gravity. Comparison to models of structure formation can help shed light on the composition of the Universe or constrain modified gravity theories. Although galaxies can be used to trace the distribution of matter, to fully exploit their constraining power, we must link their observable properties to the underlying dark matter distribution. Alternatively, it is possible to quantify the statistics of the total matter distribution directly because density inhomogeneities distort space-time, leading to coherent patterns in the observed shapes of distant galaxies, a phenomenon called weak gravitational lensing.
The study of the cosmological weak lensing signal -- or cosmic shear -- is a major driver for the next generation of large imaging surveys, such as {\it Euclid} \citep{Mellier2024}, 
the Legacy Survey of Space and Time \citep[LSST;][]{LSST2019}, the {\it Nancy Grace Roman} Space Telescope \citep{Akeson2019}, and the Chinese Space Station Telescope \citep[CSST;][]{Zhan2011, Gong2019}. 

Much remains to be learned from the analysis of the precursors to these large projects, such as the impact of 
analysis choices \citep[e.g.][]{KiDS_DES_2023}. A major motivation to scrutinise these data is that many lensing surveys prefer a less clumpy distribution of matter in the low-redshift Universe compared to that expected based on cosmic microwave background (CMB) measurements by {\it Planck} \citep{PlanckParams2018}.
To leading order, the correlations between the shapes of galaxy pairs are sensitive to the variance in the projected matter distribution. This can be quantified by the parameter $S_8 \equiv \sigma_8 \sqrt{\Omega_{\mathrm{m}}/0.3}$, which depends on $\sigma_8$, the amplitude of the linear density fluctuations within a sphere of radius $8\,h^{-1}\,{\rm Mpc}$ at the present day, and $\Omega_{\rm m}$, the mean matter density relative to the critical density.
Although the significance varies \citep[from $1\sigma$ to $3\sigma$;][]{Asgari2021, Amon2022, Secco2022, XiangchongLi2023,Dalal2023}, the tendency towards a lower amplitude of density fluctuations is intriguing. If confirmed, this may suggest the need to modify the current baseline $\Lambda$ cold dark matter ($\Lambda$CDM) model. Various alternatives have already been proposed to alleviate this discrepancy \citep[see][for a summary]{DiValentino2025}. 

Interestingly, the final analysis of the Kilo-Degree Survey \citep[KiDS;][]{Kuijken2015, Wright2024} agrees with the CMB measurements \citep{Wright2025b}. It may, however, be too soon to close the issue, because uncertainties in the modelling of the observed cosmic shear signal remain. First, intrinsic alignments of physically close galaxies also contribute to shape alignments \citep[see e.g.][for a review]{Joachimi2015}, while much of the constraining power comes from relatively small scales. 
A major challenge for the modelling on these scales is that the matter distribution is affected by feedback processes that suppress the predicted lensing signal \cite[e.g.][]{Semboloni2011}. Uncertainties in the modelling have thus been put forward to explain the lower $S_8$ values \citep{AmonEfstathiou2022}, although we note that other ways to probe the low-redshift matter distribution find similarly low values \citep{Dvonik2023,Miyatake2023}. 

Regardless of the myriad possibilities, it is worth recalling that accurate shapes and distances for the galaxies are a fundamental input. In this paper, we revisit the shape measurements using the 
4th KiDS data release \citep[DR4;][]{Kuijken2019}, which we refer to as KiDS-1000, while performance is determined using SKiLLS \citep[SURFS-based KiDS-Legacy-Like Simulations;][]{Li2023a}, a set of realistic multi-band image simulations developed to calibrate the shear estimates for the final KiDS cosmic shear analysis \citep[KiDS-Legacy;][]{Wright2025b}. Our work follows \cite{Li2023b} in that we do not change the calibration procedure of the photometric redshift (photo-z) distributions for the sources -- another area of major activity, which has been studied extensively in \cite{vdBusch2022} for KiDS-1000 and developed further in \cite{Wright2025a} for KiDS-Legacy. We used the result of \cite{vdBusch2022} to account for the use of a different algorithm to measure shapes.

To date, all KiDS lensing analyses are based on shape measurements from the model-fitting algorithm \lensfit \citep{Miller2013, FenechConti2017}. Performance evaluations using simulated data find that the average biases in the shear estimates are generally small and can be calibrated \citep{FenechConti2017, Kannawadi2019, Li2023a}.
However, the simulations do not reproduce the actual observations perfectly (see \cref{fig:com_skills_k1000}), which may lead to residual biases in the calibration of \lensfit. Although \cite{Li2023b} show that
the systematic uncertainty in $S_8$ is not limited by the fidelity of the simulations, the tests of the \textit{lens}fit pipeline suggest increased bias for the most distant redshift bins. 

In this paper, we revisit the problem of shape estimation by considering an alternative to \lensfit. We present a new pipeline for KiDS that uses \metacal \citep{Huff2017, Sheldon2017} to determine the lensing shear. This approach uses the actual data to determine the response to a shear. Studies using simulated images suggest it is a promising way forward \citep{Sheldon2017, Hoekstra2021a, Zhang2023, Sheldon2023}. 

The structure of the paper is as follows. We describe the KiDS-1000 data and simulated SKiLLS data in \cref{sec:data}. In \cref{sec:shear_calibration}, we describe our implementation of \metacal and discuss the motivations behind some of the pipeline choices we made. The performance of our algorithm was examined using SKiLLS in \cref{sec:comparison}, while we tested the robustness of the KiDS-1000 two-point statistics in \cref{sec:datavector}.  The inputs for the cosmological analysis are discussed in \cref{sec:datavector}, and the cosmological and astrophysical constraints from the \metacal shear catalogue are presented in \cref{sec:cosmology}.

\section{Data}
\label{sec:data}

Our ultimate aim is to use our \metacal pipeline as part  of KiDS-Legacy, the final analysis of KiDS based on the 5th data release \citep[DR5;][]{Wright2024}. To optimise the pipeline and to evaluate the performance, we followed \cite{Li2023b} and present results here that instead use DR4, which we summarise in \cref{sec:KiDS-1000}.  Moreover, the performance of the shear estimation is quantified using SKiLLS \citep{Li2023a}, a set of simulated multi-band images that was developed for the calibration of \lensfit \citep{Miller2013} for KiDS-Legacy. These simulated data are described in \cref{sec:SKiLLS}.

\subsection{KiDS-1000 observation data}
\label{sec:KiDS-1000}

The Kilo-Degree Survey (KiDS) is a European Southern Observatory (ESO) public survey that was completed in 2019. It surveyed 1350 deg$^2$ in the $ugri$ bands using the wide-field camera OmegaCAM mounted on the 
 VLT Survey Telescope (VST). The KiDS observations were 
complemented by near-infrared (NIR) imaging in the $ZYJHK_s$ bands from the VISTA Kilo-degree Infrared Galaxy (VIKING) survey. The resulting nine-band data enables high-fidelity photo-$z$ estimation, ideal for cosmic shear tomography. 

In this study, we used data from KiDS DR4 \citep{Kuijken2019}, which covers a total of 1006 deg$^2$ of sky. Similar to previous lensing studies with these data \citep{Asgari2021, vdBusch2022,Li2023b}, we used the \textit{r} band for the shape measurements, because they are the deepest (5$\sigma$ limiting AB magnitude of 24.9) and have the best image quality (mean seeing of 0.7 arcsecond).
We did not change the model for the point spread function (PSF), because the tests presented in \cite{Giblin2021} suggest it is sufficiently accurate.\footnote{We note that \lensfit internally excludes chip images with fewer than 30 stars to ensure the accuracy of the PSF model. This criterion excludes 756 individual chip images (out of a total of 159,136). As not all exposures in a pointing are affected, this mask only applies to about 30 pointings.}

As described in \cref{app:source_selection}, the selection of sources follows that of previous DR4 studies as much as possible.
However, because of differences in the shape measurement algorithms, the final weighted results differ somewhat. As discussed in \cref{app:psf_residual_check}, we find that some pointings show significant PSF residuals, which we removed from the analysis. The resulting effective area covers 771.9 deg$^2$ after applying the masks, slightly smaller than the previous KiDS analyses 777.4 deg$^2$ \citep{Giblin2021, Li2023b}.  

Robust colours for the sources were measured using the Gaussian Aperture and PSF (GAaP) pipeline \citep{Kuijken2015}. The resulting matched photometry in the nine bands was used to determine photometric redshifts, $z_{\rm B}$ with BPZ \citep{Benitez2000}. The sample of sources was split based on $z_{\rm B}$ into six redshift bins for a tomographic lensing analysis (see \cref{tab:mbias} for the bin definitions). 

As described in \cite{Wright2020, Hildebrandt2021}, a self-organising map (SOM) was used to identify the subsamples of galaxies for which the underlying redshift distributions could be robustly calibrated using deep spectroscopic reference catalogues. This resulting `gold-sample' selected based on `gold-flag', was used for the inference of cosmological parameters
(see \cref{sec:datavector} for more details). 
Previous cosmological analyses, presented in \cite{Heymans2021, Asgari2020, Li2023b}, were limited to $z_{\rm B}\le 1.2$ because of concerns about the fidelity of the $n(z)$ calibration of the 6th bin ($1.2\le z_{\rm B}\le 2.0$) due to the limited spectroscopic calibration catalogues. In this paper, we examine the possibility of including a sixth bin but do not consider it for our cosmological parameter estimation in \cref{sec:cosmology} to enable a more direct comparison to previous work.

\subsection{KiDS multi-band image simulation - SKiLLS}
\label{sec:SKiLLS}

Simulated imaging data are essential to quantify the biases of shear measurement algorithms. The results, however, are only meaningful if the simulations capture the observing conditions and galaxy morphologies sufficiently well \citep{Hoekstra2015,Hoekstra2017}. 
For the final analysis of KiDS data, \cite{Li2023a} developed SKiLLS, which features the first multi-band image simulations for KiDS that unify the shear and redshift calibration. We refer the interested reader to \cite{Li2023a} for details, in particular the improvements over previous setups \citep{FenechConti2017, Kannawadi2019}. Here, we summarise only its main features.

The input galaxy catalogue of SKiLLS is based on an N-body simulation, the Synthetic UniveRses For Surveys \cite[SURFS,][]{2018MNRAS.475.5338E}. 
We limited our sample to galaxies with $z<2.5$ to enable a robust assignment of galaxy properties, such as realistic spectral energy distributions (SEDs), a key ingredient for multi-band image simulations.
The galaxy properties were computed using the open-source semi-analytic model named \textsc{Shark} \citep{Lagos18}.
As described in \cite{Li2023a}, the stellar mass-to-light ratios in the simulation needed to be adjusted to match the magnitude distributions of the actual data in all nine bands. Importantly, the simulations include realistic clustering of galaxies, which is important, because the `blending' of sources along the line of sight is an important contributor to shear bias \citep[e.g.][]{Hoekstra2021a,Li2023a}.

Following \cite{Kannawadi2019}, SKiLLS assumes that the surface brightness profiles of the galaxies can be  described by a S\'{e}rsic profile \citep{Sersic1963} with three parameters: the effective radius (galaxy size), S\'{e}rsic index (galaxy brightness concentration), and axis ratio (galaxy ellipticity). The values of these structural parameters are based on the catalogue of \cite{Griffith2012}, which was derived from images taken by the Advanced Camera for Surveys (ACS) on the \textit{Hubble} Space Telescope (HST). Specifically, we used the results from the Cosmic Evolution Survey \citep[COSMOS;][]{Scoville2007} after selecting objects with a good shape fit and a reasonable size.
The corresponding redshift information for the sources was extracted from the catalogue compiled by \cite{vdBusch2022}, which is based on high-quality photometric and spectroscopic surveys in the COSMOS field.

However, where \cite{Kannawadi2019} aimed to emulate KiDS observations of the COSMOS field, the objective of SKiLLS is to simulate a larger sample of sources covering sufficient area in multiple bands. To this end, \cite{Li2023a} developed a vine-copula-based algorithm \citep[see e.g.][]{Joe2014, Czado2019} that allowed galaxies to be generated from the SURFS-\textsc{Shark} input, rather than repeatedly sampling from the measured values. The model learned the distributions of structural parameters and their mutual dependence and correlations with redshift and magnitude using the HST measurements as underlying truth. As the version of SURFS-\textsc{Shark} that was used covers about 108 $\deg^2$, SKiLLS simulated a total 108 KiDS pointings. To reduce shape noise, we also created versions with the sources rotated by 90 degrees. To capture the impact of varying star density, stellar catalogues were generated using the TRILEGAL  model \citep{Girardi2005} at six different Galactic coordinates evenly spaced across the KiDS-1000 footprint \citep[see][for details]{Li2023a}. 

Another important aspect of SKiLLS is that the multi-band images are generated with realistic seeing conditions drawn from the survey data in each band. Moreover, SKiLLS mimics realistic observational features such as 8\texttimes4 CCDs covering a single pointing (${\sim}1~{\rm deg}^2$) with corresponding CCD gaps and dither patterns. 
For the $r$-band images, which are used for the galaxy shape measurements, SKiLLS uses PSF models, represented as 2D polynomial functions derived from KiDS-1000 data \citep{Giblin2021}. These were constructed for each exposure and chip using the centres of the CCD images (for 32 images and 5 exposures). 
For the other optical bands, which were used for photometry, a Moffat profile for the PSF was adopted. The model parameters were determined from bright stars identified in the ASTRO-WISE\footnote{ASTRO-WISE is a KiDS pipeline built for photometry measurement. (\url{http://www.astro-wise.org/})} images. 

\begin{figure}
\includegraphics[width = 0.45\textwidth]{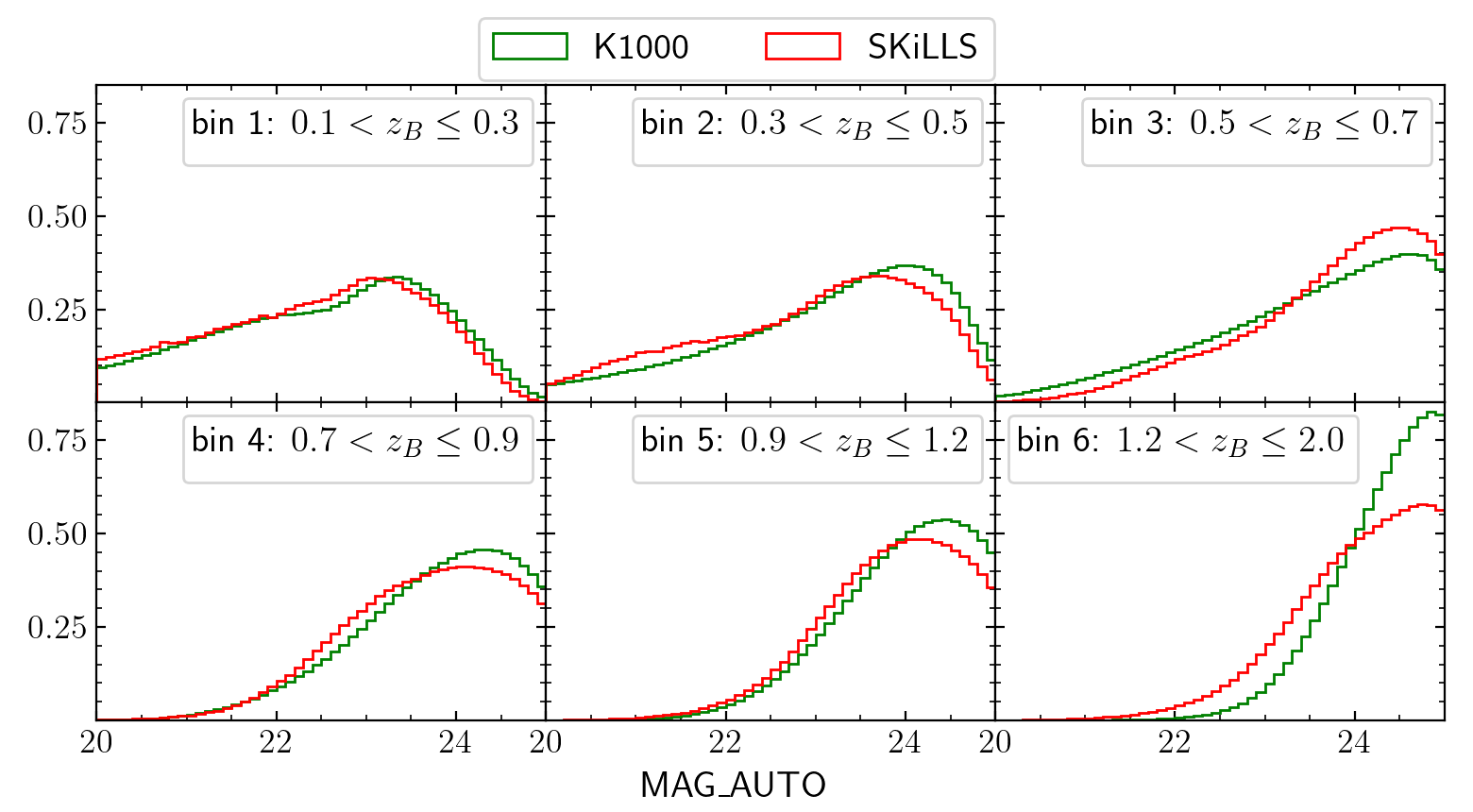}
\includegraphics[width = 0.45\textwidth]{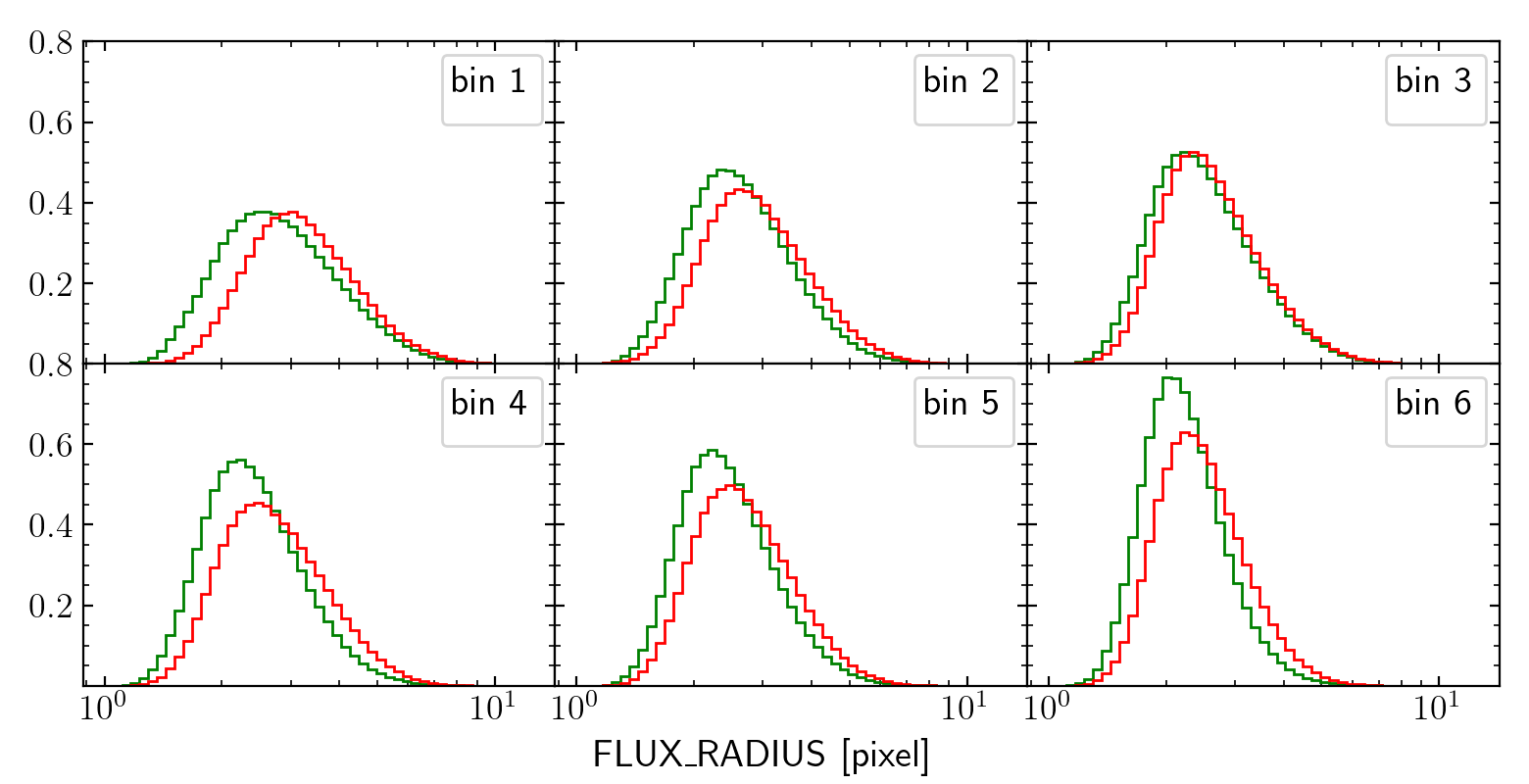}
\includegraphics[width = 0.45\textwidth]{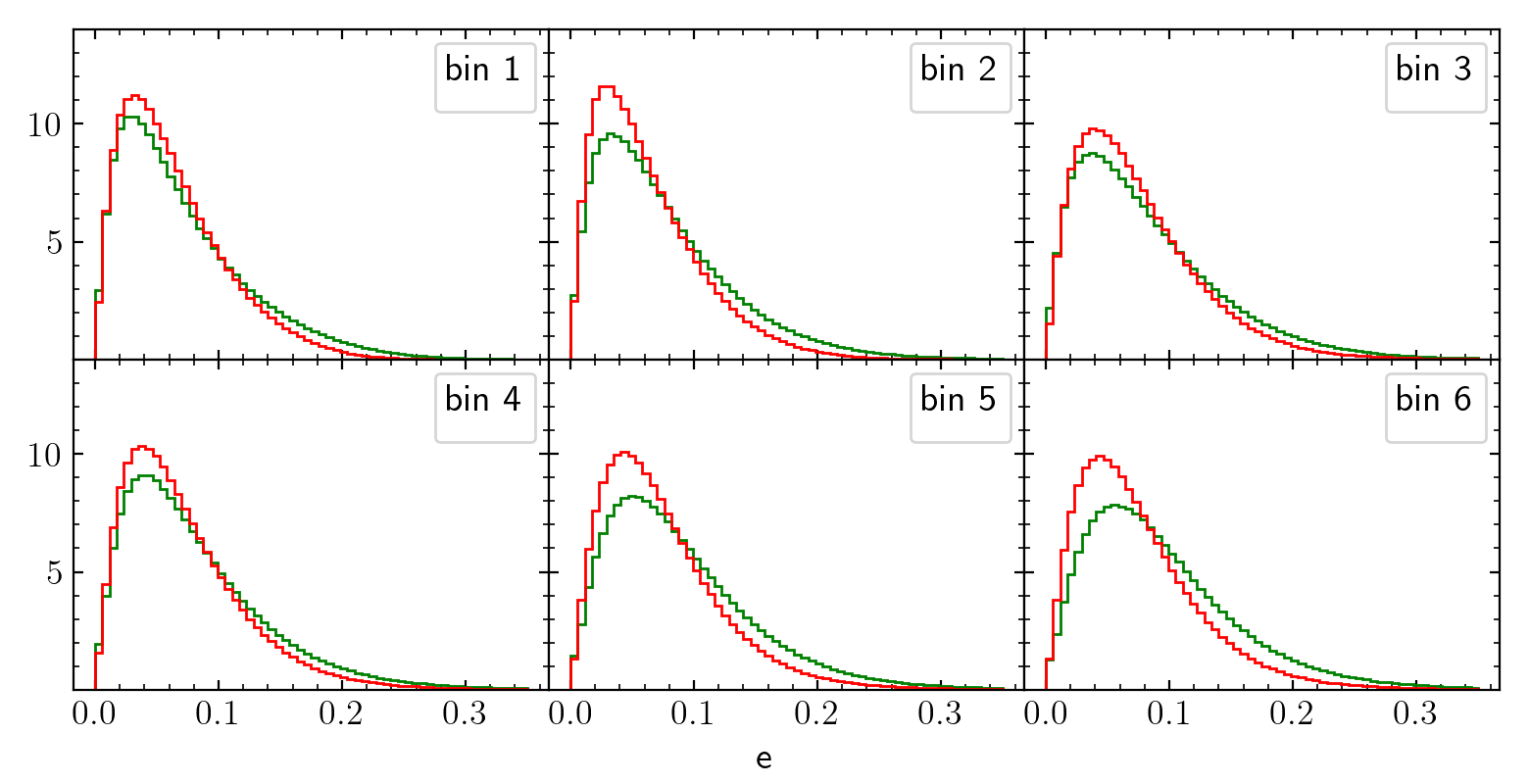}
\caption{Comparison of the source features between the KiDS-1000 and SKiLLS data selected for the metacal catalogue for each tomo-bin (weighting is not applied): r-band magnitude (top), size (middle), and ellipticity (bottom). The sixth bin shows the most deviation for the r-band magnitude and ellipticity. Regarding size, the deviation gradually increases for high redshift bins.}
\label{fig:com_skills_k1000}
\end{figure}

Thanks to the multi-band images, SKiLLS allows us to determine photometric redshifts in a fully realistic fashion, and subsequently to split the source sample into tomographic redshift bins. For this purpose, we used the GAaP (Gaussian Aperture and PSF) pipeline, which measures accurate multi-band colours by accounting for PSF variations across filters and improves the signal-to-noise ratio by down-weighting the noise-dominated outskirts of galaxies~\citep{Kuijken2015,Kuijken2019}. Photometric redshifts were then estimated with the public Bayesian photometric redshift (BPZ; \citealt{Benitez2000}) code, using the re-calibrated template set from \citet{Capak2004PhDT} together with the Bayesian redshift prior from \citet{Raichoor2014ApJ}. This enables a detailed comparison of key observables between the simulated and KiDS-1000 data. \Cref{fig:com_skills_k1000} shows the distributions for the six tomographic redshift bins as a function of apparent $r$-band magnitude (\texttt{MAG\_AUTO}; top panels), half-light radius (\texttt{FLUX\_RADIUS}; middle panels) as measured by \texttt{SourceExtractor}
\citep{Bertin1996}, as well as galaxy shapes (using \cref{eqn:e_definition}; see \cref{sec:shear_calibration} for details). Despite significant effort to match the simulated output to the data, some deviations in galaxy properties remain. These differences are more pronounced after redshift binning, implying that correlations between morphology and galaxy SEDs need to be improved. Nonetheless, the magnitude distributions match well, except for the sixth bin, which indicates that SKiLLS lacks faint high-redshift sources. Moreover, the simulated data tend to lack small sources, while the distributions in galaxy shapes also do not quite match.

Ideally, one might like to adjust the input properties such that the simulated data match the observations. However, the tests and sensitivity analyses by \cite{Li2023a} demonstrated that uncertainties in the input do not significantly change the calibration, and thus have a negligible impact on the cosmological inference, although this may change as more survey data are included. Therefore, we followed previous KiDS analyses and mitigated the impact of these differences using an empirical reweighting scheme based on measured galaxy properties.
As discussed below, reducing the sensitivity of the shear estimation algorithm to limitations in the simulated data is our main motivation to explore the prospects of \metacal in this paper.

\section{Shear measurement pipeline}
\label{sec:shear_calibration}

The observed galaxy images are modified by instrumental effects, such as detector imperfections or telescope optics, or by atmospheric turbulence in the case of ground-based telescopes \citep[see][for a review]{Mandelbaum2018}. The challenge is to recover the true underlying correlations of galaxy ellipticities. 
One approach is to fit galaxy models to the data and find the best match. This is the approach used by \lensfit \citep{Miller2013}. A benefit of fitting methods is that masked data or additional instrumental effects can be naturally accounted for. However, mismatches between the adopted model and the true morphologies can lead to model bias \citep[e.g.][]{Kacprzak2014}.

Alternatively, one can try to infer galaxy ellipticities from the moments of the observed surface brightness distribution. Such moment-based approaches \citep[e.g.][]{Kaiser1995, Melchior2011} are computationally efficient. For instance, the complex polarisation (or third eccentricity), $e=e_1+\mathrm{i}\,e_2$,  can be defined as
\begin{equation}
e = \frac{Q_{11} - Q_{22} + 2\,\mathrm{i} \,Q_{12}}{Q_{11} + Q_{22}},
\label{eqn:e_definition}
\end{equation}
where the (weighted) quadrupole moments, $Q_{ij}$, are defined as
\begin{equation}
Q_{ij} = \int \mathrm{d}^2\bm{x}\, x_i \, x_j \,W(\bm{x})\,I(\bm{x}),
\label{eqn:q_definition}
\end{equation}
where $I(\bm{x})$ is the observed surface brightness distribution of a source and $W(\bm{x})$ is a weight function. 

Arguably, \cref{eqn:e_definition} is the simplest way to quantify the shape of an object, but the observed values are biased because of the convolution with the PSF. In the case of unweighted moments, this can be easily corrected for
\citep{Valdes1983}. In practice, however, a compact weight function is needed to suppress the contribution of noise in the observed image, biasing the estimate for $e$ even more. The bias in $e$ depends on the higher-order moments of the surface brightness distribution. These can be used to obtain a corrected estimate for the shear \citep[e.g.][]{Kaiser1995, Luppino1997, Hoekstra1998}, although biases remain because of ellipticity gradients and blending \citep{Hoekstra2015}. Moreover, as shown in \cite{FenechConti2017}, and studied in more detail in \cite{Hoekstra2021a}, shear biases are already introduced when objects are detected. Hence, even a perfect shape measurement method results in a biased shear estimate for a sample of sources. 

To evaluate the performance of shape measurement algorithms, simulated data can be used \citep[e.g.][]{STEP1, STEP2, GREAT08}. If the image simulations match the data perfectly,\footnote{We implicitly assumed that the PSF was perfectly determined.} one could simply use the average observed $e_i^{\rm obs}$ for an ensemble of sources and calibrate any bias using the simulated data, for which the shear is known, using
\begin{equation}
\left\langle e_i^{\mathrm{obs}} \right\rangle = (1 + m_i)\,\gamma_i^{\mathrm{true}} + c_i ,
\label{Eqn:shear_calibration}
\end{equation}
where $c_i$ quantifies the so-called additive bias, which corresponds to a preferred orientation, arising from an anisotropic PSF or detector effects.
The multiplicative bias, $m_i$, quantifies the bias in the amplitude of the signal. Although both biases need to be accounted for, in the following we focus on the multiplicative bias, because it cannot be determined empirically. Its value is determined using image simulations. \cref{Eqn:shear_calibration} assumes a linear relation between the average observed shape $\langle e_i^{\rm{obs}}\rangle$, and the shear, $\gamma_i^{\rm{true}}$, which is a reasonable assumption for small shears. However, it is computationally efficient to use larger shears for the shear calibration simulations, while the shear can also be large around massive structures \citep[e.g.][]{Li2024}. We therefore studied the possible impact of non-linearity of the algorithm in \cref{app:test_high_amp_shear}.

As demonstrated by \cref{fig:com_skills_k1000}, it is difficult to obtain a perfect match between simulated and observed data.
However, the shear bias inferred from simulated data is only meaningful if they resemble the observations sufficiently well \citep{Hoekstra2015}. 
\cite{Kannawadi2019} and \cite{Li2023a} explored the implications of this in more detail, highlighting the trade-off between the realism of the simulated data and the sensitivity of the algorithm to deviations. 
As discussed in \cite{Kannawadi2019}, the residual multiplicative shear bias ($\Delta m$) for a sample of galaxies depends on observable quantities,
$\boldsymbol{D}$, such as the signal-to-noise ratio (S/N) or size, as well as hidden features, $\boldsymbol{h}$, such as undetected blends. A mismatch of properties ($q$) between the actual and simulated galaxy populations,
$\Delta q(\boldsymbol{D},\boldsymbol{h}) \coloneqq q_{\rm true}(\boldsymbol{D},\boldsymbol{h}) - q_{\rm sim}(\boldsymbol{D},\boldsymbol{h})$, implies a difference in multiplicative bias:
\begin{equation}
\Delta m \coloneqq m_{\mathrm{true}} - m_{\mathrm{sim}} = b(\vec{D}, \vec{h})\, \Delta q(\vec{D}, \vec{h}) ,
\label{eqn:m_inaccuracy}
\end{equation}
where $b(\boldsymbol{D},\boldsymbol{h})$ quantifies the sensitivity of the multiplicative bias to a small change in the sample:
\begin{equation}
b(\boldsymbol{D},\boldsymbol{h}) \coloneqq \frac{\delta m[q]}{\delta q(\boldsymbol{D},\boldsymbol{h})}.
\label{eqn:sensitivity}
\end{equation}
\Cref{eqn:m_inaccuracy} shows that if the samples match perfectly, the inferred shear bias is correct, irrespective of the shape measurement method. Alternatively, a method that is insensitive to the parameters that influence the shear bias also yields a meaningful result. 
The sensitivity of a method can be reduced by mapping the bias as a function of $\boldsymbol{D}$,
such as S/N and size \citep{FenechConti2017, Kannawadi2019}, but the sensitivity function cannot be determined perfectly as long as the hidden features, $\boldsymbol{h}$, their correlation with the explicit features, $\boldsymbol{D}$, and their impact on the bias are uncertain. 

An alternative, which aims to reduce the reliance on image simulations, is provided by \metacal \citep{Huff2017, Sheldon2017}. As discussed in more detail below, it uses the observed images themselves, so that $\Delta q\approx 0$ by construction. Given its promise, \metacal has been adopted as the main shear calibration process for the Dark Energy Survey Y1 \citep{Zuntz2018} and Y3 analyses \citep{Gatti2021}, and the LSST \citep{LSST2019}. The main advantage of this approach is that the calibration is less sensitive to the choice of galaxy population properties that are set to be inputs of the image simulation for shear calibration. This has motivated us to consider \metacal as an alternative for KiDS as well.

\metacal infers the response to shear from the data themselves by manipulation of postage stamps of the observed galaxies. We only need to assume that it is possible to construct a sheared version of the true image \citep{Huff2017}, which is the case for well-sampled data \citep[but also see][for the case of undersampled data]{Kannawadi2021}. We verified that the KiDS PSF models were sampled sufficiently well, so that the performance of \metacal would not degraded. 

\begin {figure}[t]
\centering
\includegraphics[width = 0.45\textwidth]{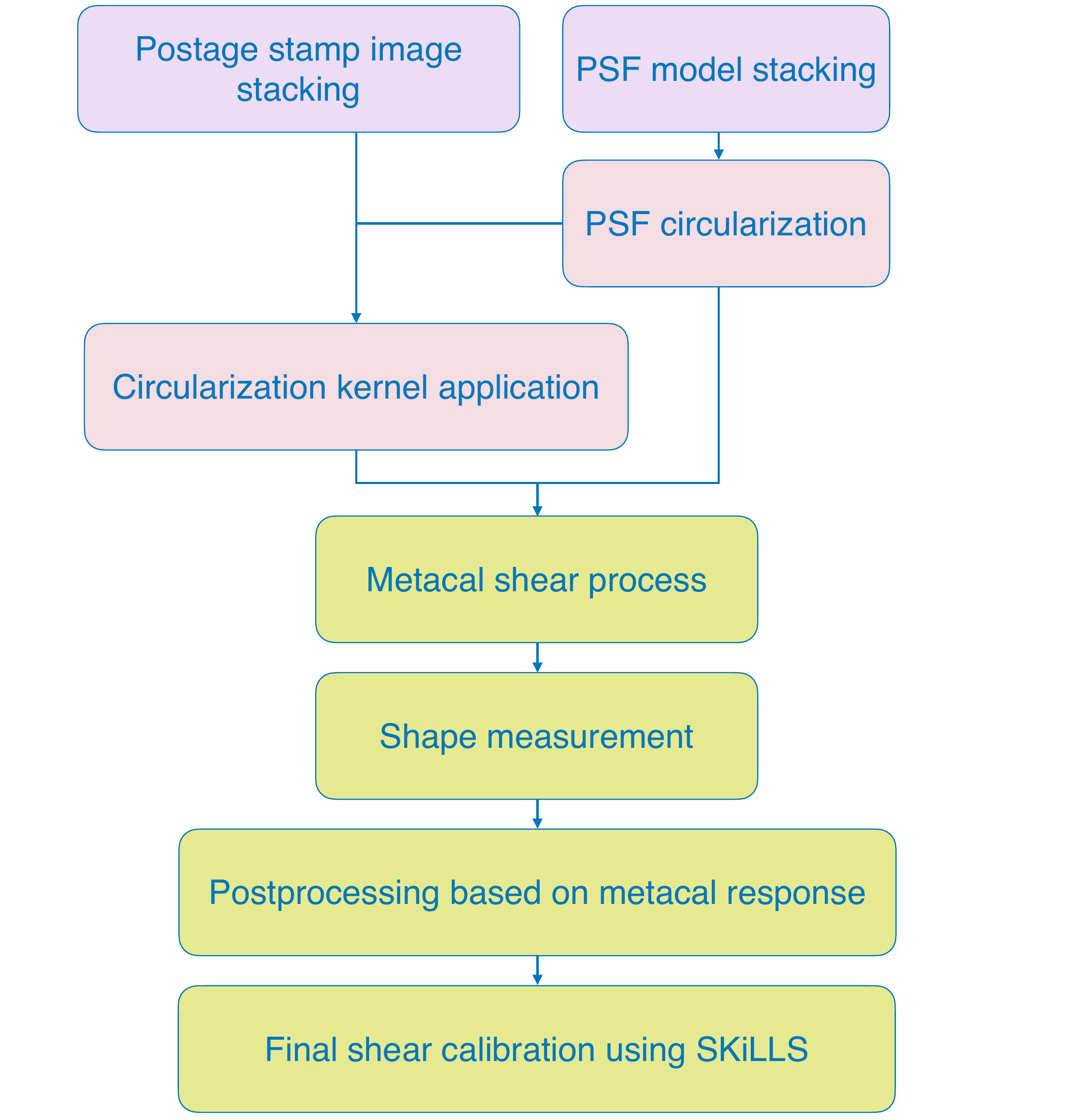}
\caption{Schematic work flow of the \metacal pipeline, which begins with the same input catalogue used in previous KiDS-1000 analyses.}
\label{fig:Workflow}
\end{figure}

We provide details of our specific implementation in \cref{sec:setup} but review the key steps in the following (also see \cref{fig:Workflow} for all the steps in the calibration pipeline). The postage stamp image was (i) deconvolved using a model for the PSF, (ii) sheared, and then (iii) reconvolved with a slightly dilated PSF to avoid the noise blowing up \citep[see][for details]{Sheldon2017}.
The shape estimate of the resulting sheared image is given by
\begin{equation}
\boldsymbol{e} \approx \left. \boldsymbol{e} \right|_{\gamma = 0} +
\left. \frac{\partial \boldsymbol{e}}{\partial \boldsymbol{\gamma}} \right|_{\gamma = 0} \boldsymbol{\gamma} + O(\boldsymbol{\gamma}^2)
\equiv \left. \boldsymbol{e} \right|_{\gamma = 0} + 
\bm{\mathsf{R}}^\gamma \boldsymbol{\gamma} + O(\boldsymbol{\gamma}^2) .
\label{eqn:metacal}
\end{equation}
where $\bm{\mathsf{R}}^\gamma$ is the $2\times 2$ shear response tensor. We can estimate its components by measuring the shapes of the galaxies in the sheared images and computing
\begin{equation}
    {\mathsf R}^\gamma_{ij} \equiv\frac{\partial e_i}{\partial \gamma_j} \approx  \frac{{e^{j+}_i - e^{j-}_i }}{\Delta \gamma_j}, \label{eq:response}   
\end{equation}
where the subscripts indicate the two shear components, and the superscript the sign of the applied shear, so that `$j+$' means that the image was sheared by $+\gamma_j$; hence, $\Delta\gamma_j=2\gamma_j$. In practice, the off-diagonal elements vanish on average, while the diagonal elements tend to be equal. We therefore can capture the response by using $R^\gamma\equiv \mathrm{Tr}(\bm{\mathsf{R}}^\gamma)/2$.

\metacal works well for isolated galaxies, but in the case of blended sources some residual bias remains \citep{Hoekstra2021a}. 
Moreover, the selection of objects already introduces shear bias \citep{Hirata2003, FenechConti2017}, and \metacal cannot undo this. The choice of weight function for the shape measurement can further exacerbate any selection bias \citep[e.g.][]{Sheldon2017}. 
We minimised such additional selection bias by adopting 
an axisymmetric Gaussian with a fixed width for 
$W(\bm{x})$. 

In principle, it is also possible to include the shear response to the detection process using an approach called \metadetect \citep{Sheldon2020, Zhang2023, Sheldon2023}. 
However, implementing this would require a major overhaul of the whole shear calibration pipeline, because it needs to be applied to all bands in order to avoid selection biases during the photometric redshift estimation. Moreover, \metadetect, cannot capture the residual bias caused by blended sources at different redshift \citep{MacCrann2022, Li2023a}. Given the
required level of accuracy, in combination with the realism of SKiLLS \citep{Li2023a,Li2023b},  the implementation of \metadetect is not warranted for our KiDS analysis.

Point spread function anisotropy is the main contributor to additive shear bias, and thus needs to be carefully accounted for. Different approaches can be used to do so. For instance, if the shape measurement itself does not remove the PSF anisotropy, the response can be determined with additional image manipulations in which the PSF is sheared \citep{Sheldon2017}. Alternatively, \cite{Hoekstra2021b} proposed to correct the observed polarisations (\cref{eqn:e_definition}) using the Kaiser–Squires–Broadhurst (KSB) formalism \citep{Kaiser1995, Hoekstra1998}. However, to fully remove the PSF anisotropy, an empirically determined boost factor is required. As detailed in \cref{sec:setup}, we adopted another approach here. We used the PSF model to derive a convolution kernel that renders the PSF isotropic. In principle, it removes additive bias arising from the PSF, and we only required \metacal to determine the shear response $R^\gamma$.

In \cref{fig:Workflow} we show the schematic workflow of our \metacal pipeline. It begins with the same input catalogue as used in previous KiDS-1000 analyses that used \lensfit, and we apply the same selections where possible (see \cref{app:source_selection} for details). Moreover, we use the same PSF model, which was tested extensively in \cite{Giblin2021}. The same work flow was used to analyse the SKiLLS data and to determine the residual multiplicative bias after \metacal. Below, we describe our implementation in more detail.

\begin{figure}
    \includegraphics[width = 0.495\textwidth]{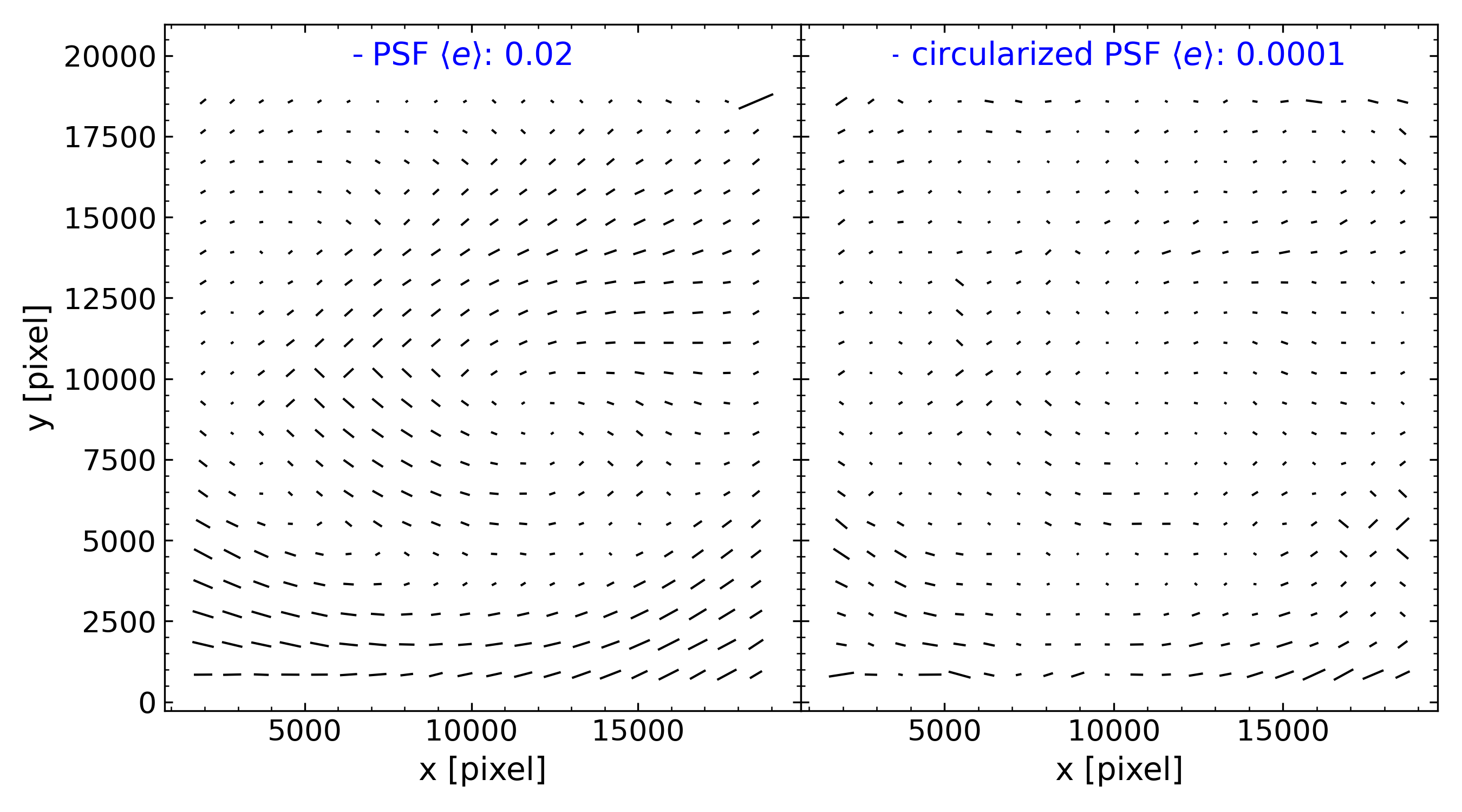}
    \caption{Performance of the circularising procedure applied to a representative KiDS-1000 pointing (at RA=9.4$^\circ$, Dec =-32.1$^\circ$) with the ellipticities indicated by the bar length (for reference, see the blue bar at the top) and the ratio of $e_1$/$e_2$ as their angle. The model PSFs, evaluated at the positions of the detected galaxies (with their average shown on the grid), exhibit some variation across the field of view, although the overall level of PSF anisotropy is small ($|e| \lesssim 0.02$: left panel). After circularisation, the PSF shapes are consistent with zero ($|e| \lesssim 10^{-4}$: right panel). Please note that the ellipticity scale in the right panel is 100 times greater than in the left panel.}

\label{fig:psf_circ_one_pointing}
\end{figure}

\subsection{Shape measurement setup}
\label{sec:setup}

We exploited the fact that \metacal can be used in combination with any shape measurement. Naively, one might prefer an estimator with an intrinsically small bias. However, more important is that the results are robust against limitation of the image simulations that are used to quantify the biases. Here, we adopted \cref{eqn:e_definition} using weighted moments. Although simple, several choices remain to be made, such as the choice of the weight function, how to account for anisotropic and spatially varying PSF, and how to combine the information from the (typically) five $r$-band exposures in KiDS observations.

As a first step, we created postage stamps with a fixed size of 48\texttimes48 pixels for each galaxy and exposure. We only retained complete stamps, thus removing any instances where a galaxy is located close to the chip boundaries in an exposure. To minimise the flux from neighbouring galaxies contaminating the shape measurement, we used the segmentation maps generated by \se to mask those objects (replacing the pixel values of adjacent objects with zero). We find that this is sufficient to address the impact of neighbouring objects, as further filtering of recognised blends has a negligible impact on the shear bias (see \cref{app:blending} for more details).

In principle, it is also possible to measure shapes on the individual images, but we decided to use stacked postage stamp images. A major advantage is that the stacking limits the \metacal step to only five image manipulations (as opposed to 25 in the case of processing all individual exposures). Moreover, the stacking procedure is fast, while the reduced noise in the images improves the shape measurement. We note that the galaxy images are well sampled, and hence the resampling does not introduce any bias or loss of information. 

\begin{figure*}
\sidecaption
\includegraphics[width=6cm]{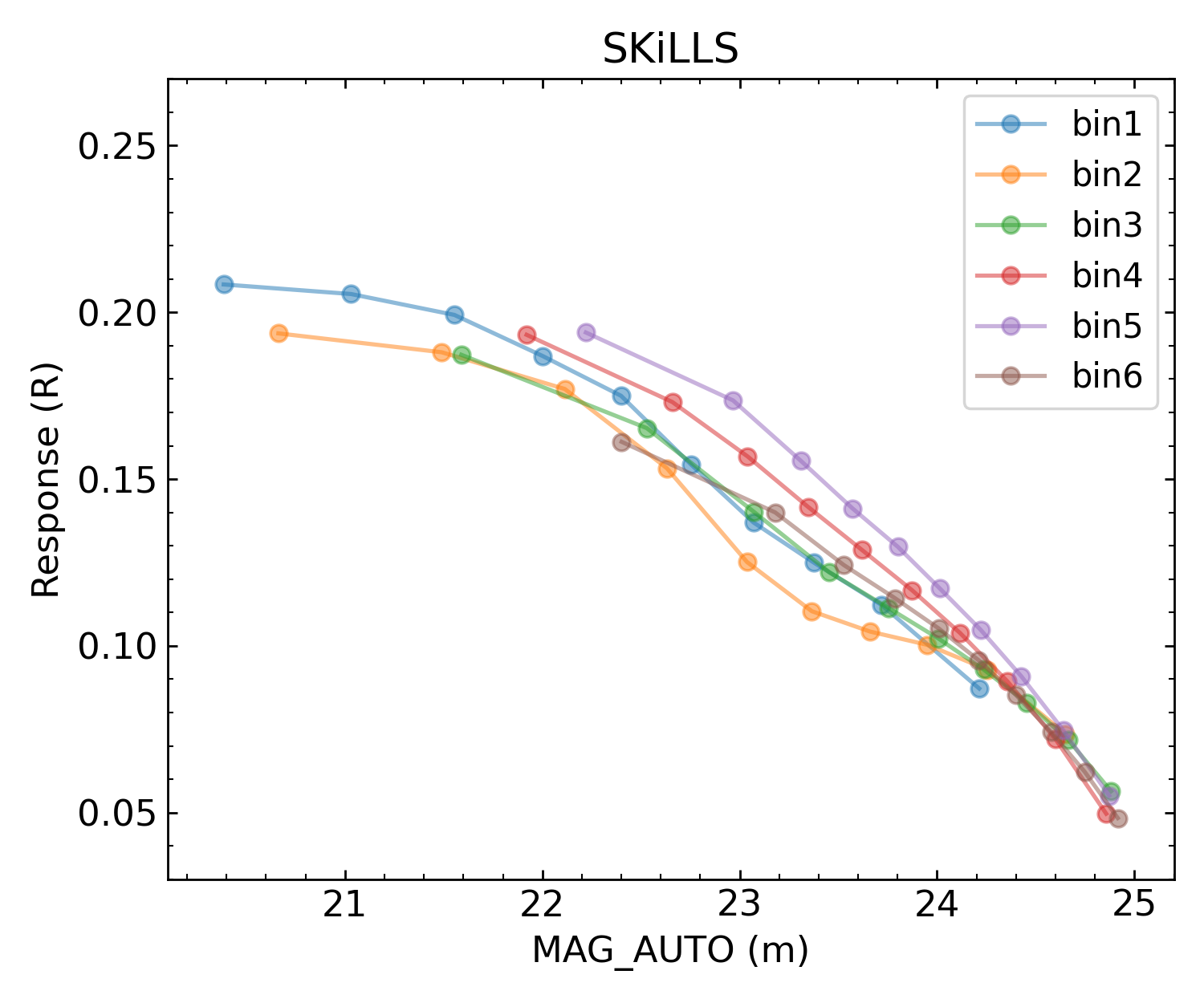}
\includegraphics[width=6cm]{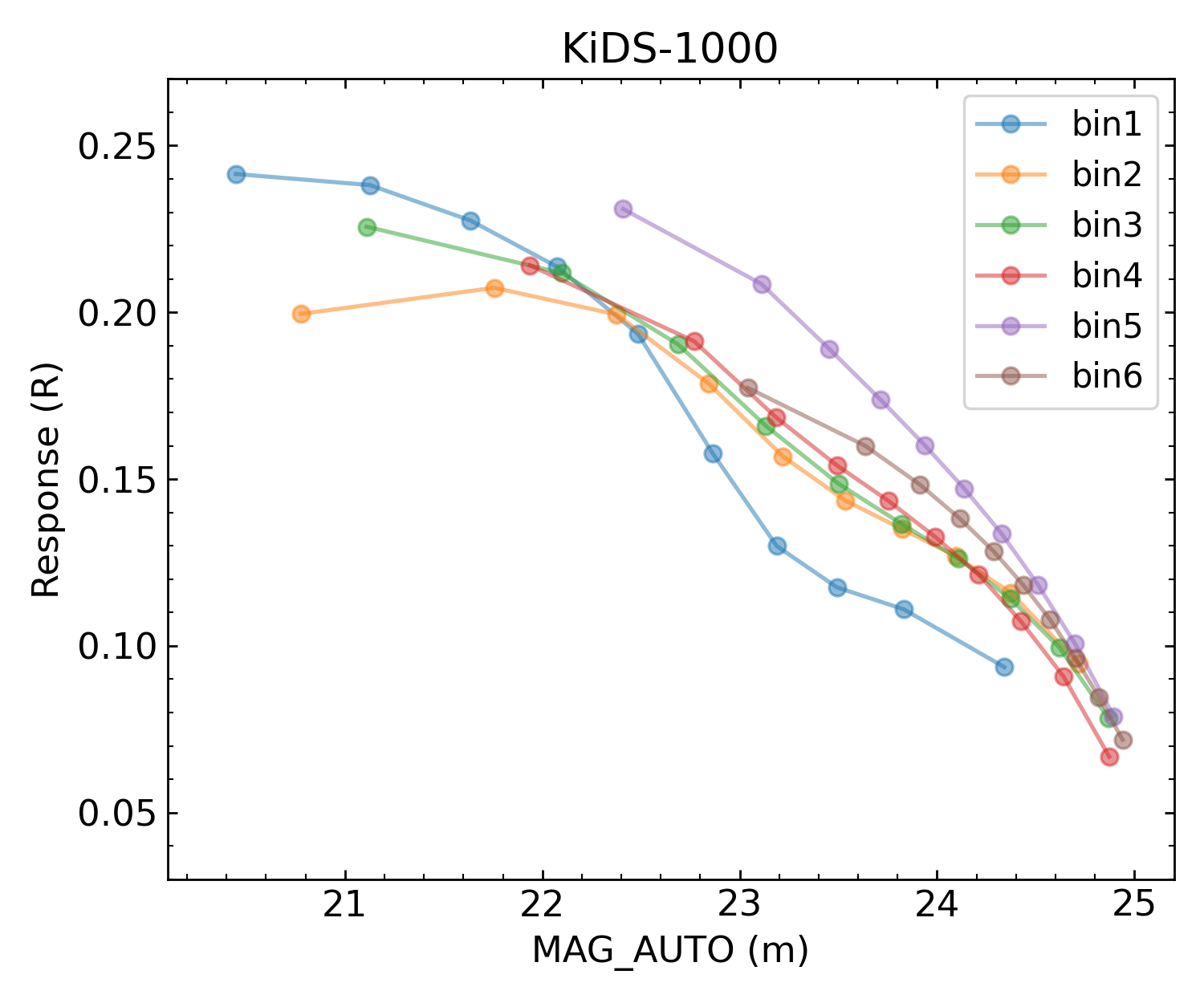}
\caption{Mean shear response, $R^\gamma$, as a function of
apparent magnitude for the six tomographic bins of the simulated SKiLLS galaxies (left) and real KiDS-1000 galaxies (right). We used ten equi-populated subsamples for each tomographic bin to compute the averages.} 
\label{fig:response}
\end{figure*}

For each postage stamp, we generated the corresponding PSF model image using the location of the postage stamp. Here, 
we used the same PSF model that was used in previous KiDS-1000 analyses based on \lensfit. 
We then stacked the galaxy postage stamps, as well as the corresponding PSF images. In \cref{app:psf_residual_check}, we discuss testing the performance of the PSF stacking procedure on the KiDS data.

In our pipeline, we accounted for PSF anisotropy by convolving the galaxy images with a kernel that circularises the (stacked) PSF.
Following \cite{Kuijken2006}, we used a two-Gaussian model fitted to the PSF images. We then estimated the fitted model-based PSF size and determined a convolution kernel that yielded a round PSF. We find that this two-Gaussian model performed better than a kernel based on a single Gaussian. To demonstrate the performance of the circularisation step, we show the results for a representative pointing in \cref{fig:psf_circ_one_pointing}.
The original PSF anisotropy varies spatially, although the amplitude is low. After the PSF circularisation kernel was applied, the resulting shapes were consistent with zero.

We applied the PSF circularisation kernels to the corresponding galaxy images and used the result for the main \metacal step.
For this, we used the functions that are available as part of the {\tt ngmix} package \citep{Sheldon2017, Huff2017}.
To determine the shear response, ${\mathsf R}^\gamma_{ij}$, we adopted a shear of 0.01, so that $\Delta\gamma_i=0.02$
(see Eq. \ref{eq:response}) and use the {\tt MetacalGaussPSF}
option for the slightly dilated reconvolution kernel. Both the circularisation and the shearing introduce anisotropic correlations in the noise. In principle, this can be accounted for, but we chose not to (i.e. we use {\tt fix\_noise=False}), because it leads to noisier measurements.
Importantly, the resulting bias is captured by the image simulations, and thus is naturally accounted for during the calibration using SKiLLS. 

For the actual shape measurement of the \metacal outputs, we used the {\tt hsm} module \citep{Hirata2003,Mandelbaum2005}
in {\tt GALSIM} \citep{Rowe15}. The choice of the weight function used to measure the galaxy shapes also matters. It can be matched to the source in size, shape or both, to maximise the S/N of the estimate. However, such optimisation introduces additional selection bias, because the weighting depends on the source ellipticity. As shown by \cite{Sheldon2017}, the total shear response consists of two terms: $\bm{\mathsf{R}}=\bm{\mathsf{R}}^\gamma+\bm{\mathsf{R}}^{\rm S}$, where the second term captures how the selection, or weighting scheme, depends on the shear. If the weight function is fixed, this sensitivity is minimised. 

As shown in \cite{Hoekstra2021a}, $\bm{\mathsf{R}}^{\rm S}$ accurately captures the selection bias for isolated sources, thus reducing the multiplicative bias after \metacal. 
However, blended sources bias $\bm{\mathsf{R}}^{\rm S}$ \citep{Hoekstra2021a}, so that image simulations are nonetheless needed to capture the total selection bias.\footnote{The detection bias can be captured using \metadetect \citep{Sheldon2020}, although the impact of redshift-dependent blending still leaves a small residual bias that must be accounted for.} This selection bias is captured well by the image simulations, because it is less sensitive to the input morphologies of the galaxies compared to the actual shape estimation. As the latter dominates the accuracy of the final results, we do not compute $\bm{\mathsf{R}}^{\rm S}$, because it simplifies the pipeline. Consequently, the results presented here essentially test the impact of model bias \citep[e.g.][]{Kacprzak2014} for the baseline KiDS results, which are based on \lensfit.

These considerations also motivate our choice of a fixed circular Gaussian with a dispersion of 2.5 pixels. Although this is generally not the optimal choice, the distributions of \texttt{FLUX\_RADIUS} in \cref{fig:com_skills_k1000} show that it matches well the typical size of the galaxies in all tomographic redshift bins. 
This is sufficient, because we are interested in robust shear estimates rather than precise shapes for individual galaxies. For larger, brighter galaxies the uncertainty in the shear estimate is determined by the intrinsic galaxy shape, irrespective of the weight function.
As the tests in \cref{app:test_circular_width} show, the width of the weight function does not affect the S/N of the cosmic shear signal, but it can help reduce the sensitivity to blended sources, in addition to the masking we already applied.

\Cref{fig:response} shows the shear response, $R^\gamma$, as a function of apparent magnitude for the six tomographic bins for SKiLLS (left) and KiDS-1000 (right). The shear response is a strong function of magnitude, but the results are quite similar between tomographic bins, especially for SKiLLS. Although the overall behaviour is similar between simulated and real data, we observe generally higher values for KiDS-1000. We attribute this to the fact that the morphologies in SKiLLS do not match the data perfectly, as alluded to in \cref{fig:com_skills_k1000}. In particular, for bright, resolved galaxies the use of a S{\`e}rsic profile may not be sufficient. We explore the implications of this in \cref{sec:sensitivity}.

\subsection{Weighting scheme for the ensemble of galaxy shapes} 
\label{sec:e_weighting}

In the previous section, we defined a simple estimate for the galaxy shapes, using \cref{eqn:e_definition} and a fixed circular weight function in \cref{eqn:q_definition}. 
In the absence of noise and detection bias, an unbiased estimate for the shear $\gamma_i$ is given by $\hat{e}_i/R_{ii}^\gamma$, where $\hat{e}_i$ is measured from the \metacal `no\_shear' image (this image is not sheared, but is convolved by the slightly dilated PSF used in the \metacal step). Unfortunately, the estimate for the shear response, $R_{ii}^\gamma$, is noisy for individual galaxies, and the ensemble average of the individual values for $\gamma_i$ will therefore be biased. 

Instead, following \cite{Sheldon2017}, we can consider the ensemble average of \cref{eqn:metacal}. Ideally, this should give us an unbiased estimate of the shear, but in practice blending and detection biases also contribute to multiplicative bias \citep[e.g.][]{Hoekstra2021a}. To explicitly account for this, we consider the following unbiased shear estimator:
\begin{equation}
\hat{\gamma}_i = \frac{\hat{e}_i}{\left\langle (1+m_i) R_{ii}^\gamma\right\rangle}
\approx \frac{(1 + m_i)\, {{R}}_{ii}^\gamma}
{\left\langle (1+m_i) R_{ii}^\gamma\right\rangle}
\, \gamma_i \, +\, \frac{e_i^{\rm noise}}{\left\langle (1+m_i) R_{ii}^\gamma\right\rangle},
\label{eq:raw_estimator}
\end{equation}
where $e^{\rm noise}_i$ quantifies the uncertainty in the shear estimate. The intrinsic shape of the galaxy is the dominant contributor, but for fainter galaxies, the noise in the images can also become important. 

The advantage of using \cref{eq:raw_estimator} is that it yields 
an unbiased estimate, $\hat{\gamma}_i$, while avoiding taking ratios of noisy quantities for individual galaxies. However, the shears are now weighted by the product of $(1+m_i)$ and $R^\gamma_{ii}$. This needs to be accounted for in the modelling of the lensing signal, as it modifies the effective source redshift distribution. In general, for narrow tomographic bins, we expect the impact to be negligible, because the shear bias tends to be small, while \cref{tab:mbias} shows that $\langle R^\gamma\rangle$ does not differ much between tomographic bins (also see \cref{sec:m_bias_static}).

As shown by \cref{fig:response}, $R^\gamma$ conveniently down-weights fainter galaxies, for which the shape measurements are noisier, but the precision of the shear signal can be improved further by using the inverse variance of the shear as an additional weight, $w_j$, for the $j$-th galaxy. Using \cref{eq:raw_estimator} we find that the
variance of the shear estimate is given by 
\begin{equation}
\label{eqn:shear_dispersion}
\sigma^{2}_{\hat\gamma_i} \approx \frac{\left\langle \hat{e}_i^2\right\rangle}{\left\langle (1+m_i)R_{ii}^\gamma \right\rangle^2} \,,
\end{equation}
where we assumed that the observed variance of $\hat e_i$ can be used as a proxy for $e_i^{\rm noise}$. This ignores sub-percent broadening of the observed distribution due to cosmic shear. 

Estimating the weight for individual galaxies is noisy, and tends to make the weight depend on ellipticity. Instead, we consider the variance for subsamples of galaxies. Specifically, we expect the measurement uncertainty to be mostly a function of the flux (or magnitude) and we therefore determine for each tomographic bin, $\mathrm{t}_z$, the weight for ensembles binned by $m_{\rm AUTO}$, the apparent magnitude determined by \se. We thus computed the weight as
\begin{equation}
w_j \left( m_{\rm AUTO}, \mathrm{t}_z \right) = \sigma_{\hat\gamma}^{-2} \left( m_{\rm AUTO}, \mathrm{t}_z \right) = \left[ \sigma_{\hat{e}}^{-2} \left\langle {R}^\gamma \right\rangle^{2} \right] \left( m_{\rm AUTO}, \mathrm{t}_z \right) ,
\label{eqn:weight_inv_var}
\end{equation}
where
\begin{equation}
\sigma_{\hat{e}}^2 (m_{\rm AUTO}, \mathrm{t}_z) = \frac{1}{2} \left[\frac 
{\sum_j {(\hat{e}_{1,\, j})^2+(\hat{e}_{2,\, j} )^2}}
{N_{\rm gal}} \right]\,. 
\label{eqn:variance_ellip}
\end{equation}
We ignored the multiplicative bias in the estimate of the weight, because it is small and does not vary much with magnitude
(see \cref{sec:m_bias_static}). This weight is assigned to all the $N_{\rm gal}$ galaxies in each magnitude and tomographic bin. Our approach is similar to \cite{Gatti2021}, who bin in S/N, but also resolution. We decided to ignore the resolution, because it tends to be noisier and does not improve the weighting much. 

Our final estimator for the shear of an ensemble of sources is then
\begin{equation}
    \left\langle \hat\gamma_i \right\rangle = \frac{ \left\langle w\, \hat{e}_i \right\rangle }{ \left( 1 + \left\langle m_i \right\rangle \right) \left\langle w\, {R}^\gamma_{ii} \right\rangle } \, ,
    \label{eq:shear_estimator}
\end{equation}
where we used that the value of $m_i$ is determined for the ensemble by applying a linear regression of the input shear to $\langle w  e_i \rangle / \langle w  R_{ii}^\gamma \rangle$ (see \cref{sec:m_bias_static} for details). 
In practice, we estimated $\langle m\rangle$ by taking the arithmetic mean of $\langle  m_1 \rangle$ and $\langle  m_2 \rangle$, which are consistent with one another for each tomo-bin. This is expected based on symmetry arguments, but is also supported by image simulations \citep{Hoekstra2021a}. 

\Cref{eq:raw_estimator}, or alternatively \cref{eq:shear_estimator},  can be used to compare the various approaches for shear estimation. In the end, each algorithm is defined by a shape estimate, a correction that is derived from the data, and a correction derived from image simulations. These appear in \cref{eq:raw_estimator} as $\hat{e}_i$, $R^\gamma_{ii}$, and $m_i$, respectively.\footnote{This interpretation changes the meaning of $R^\gamma$ somewhat.} 
The strength of \metacal is that $R^\gamma_{ii}$ is determined directly from the observed images for any estimator, which results in small values for $m$
\citep{Sheldon2017, Hoekstra2021a}. However, detection bias and blending ultimately limit the accuracy, and the residual bias needs to be determined using image simulations. However, these are relatively small corrections, and the sensitivity to the simulation input is reduced \citep{Hoekstra2021a}. 

A drawback of our simple shape estimator, \cref{eqn:e_definition}, is the dependence of the shear response on size and magnitude, as this leads to an additional weighting of the shear.
If possible, having an estimator with $R_i^\gamma\approx 1$ (or constant) for all galaxies would be preferable.
However, in doing so, one would need to ensure that the multiplicative bias remains small. Exploring this is beyond the scope of this paper.

Finally, it is worth reflecting on machine-learning approaches. These aim to learn the mapping between the shape estimate and the shear \citep{Tewes2019}. In the most extreme case, no additional information about the object is used, so that $R^\gamma_i=1$, and the mapping is completely determined from the image simulations. Given the challenge to match the simulations and the data \citep{Li2023a}, this is undesirable, because it maximises the sensitivity to the limitations of the simulated data. In contrast, \metacal minimises the sensitivity to the image simulations. Only \metadetect \citep{Sheldon2020} can reduce this further, at the expense of a more complex data flow.

\section{Metacalibration shear bias from SKiLLS}
\label{sec:comparison}

Here, we discuss applying our \metacal pipeline to the SKiLLS data to determine the (residual) multiplicative and additive shear biases. Following, \cite{Li2023a}, we did this in two steps. First, we used simulations with a constant shear applied to all the sources. Motivated by the findings of \cite{Hoekstra2021b}, we calibrated the multiplicative and additive biases separately. The results for the 
multiplicative bias are discussed in \cref{sec:m_bias_static}. In \cref{sec:psf_leakage} we quantify the additive bias, with a particular focus on the bias introduced by PSF anisotropy. The constant-shear simulations account for most of the biases, but as the shear applied does not depend on redshift, it cannot correctly capture the blending of sources at different redshifts \citep{MacCrann2022,Li2023a}. As described in \cref{sec:m_bias_variable}, this second order correction to the multiplicative bias is determined using a second set of SKiLLS with variable, redshift-dependent input shears. 

\begin{table*}
    \centering
    \caption{\metacal summary statistics for the six tomographic bins. }
    \label{tab:mbias}
    {\renewcommand{\arraystretch}{1.2}
    \begin{tabular}{cccccccccccc}
        \hline\hline
        Bin & Photo-$z$ range & 
       $\langle z\,\sigma_\gamma^{-2}\rangle$ &  $\langle z\,(1+m) R^\gamma\sigma_\gamma^{-2}\rangle$ & 
        $\langle R^\gamma\rangle$&  $n_{\rm eff}~[{\rm arcmin}^{-2}]$ & $\sigma_{\gamma,i}$ & $m_{\rm fix}$& $m_{\rm final}$ & $\sigma_{m}$  \\
        \hline
        1   & $0.1 < z_{\rm B} \leq 0.3$ & 0.338 & 0.292 & $0.21$& $1.01$ & $0.330$  & $ -0.0086$  & $-0.0082$ & $0.0044$  \\
        2   & $0.3 < z_{\rm B} \leq 0.5$& 0.414 & 0.410 & $0.19$& $1.86$ & $0.331$  & $-0.0227$ & $-0.0219$ & $0.0027$\\
        3   & $0.5 < z_{\rm B} \leq 0.7$& 0.580 & 0.577 & $0.20$ & $2.77$ & $0.351$  & $-0.0072 $ & $-0.0037$ & $0.0034$ \\
        4   & $0.7 < z_{\rm B} \leq 0.9$& 0.792 & 0.788 & $0.19$ & $1.80$ & $0.342$ & $ \hphantom{-}0.0051 $ &  $\hphantom{-}0.0034$ & $0.0038$ \\
        5   & $0.9 < z_{\rm B} \leq 1.2$& 1.046 & 1.039 & $0.20$ & $1.84$ & $0.364$& $ -0.0071 $ & $-0.0024$ & $0.0051$ \\
        6   & $1.2 < z_{\rm B} \leq 2.0$ &1.447 & 1.411 & $0.16$ & $1.43$ & $0.451$ & $-0.0106$& $-0.0121$ & $0.0057$\\
        \hline
    \end{tabular}
    }
    
    \renewcommand{\arraystretch}{1.0}
    \tablefoot{Galaxies are assigned to tomographic bins based on their photometric redshift $z_{\rm B}$. For each bin, we report the mean true redshift of the sources in SKiLLS, weighted by the inverse shear variance, $\sigma_\gamma^{-2}$, as well as the results when the multiplicative bias and shear response are included in the weight. Other columns list the mean shear response, $\langle R^\gamma\rangle$, the effective source density, $n_{\rm eff}$, and the uncertainty in the shear estimates of individual galaxies (each component), $\sigma_{\gamma,i}$. Finally, we report the multiplicative bias using a fixed shear, $m_{\rm fix}$, which needs adjustment to account for the variable shear, as detailed in \cref{sec:m_bias_variable}, and obtain the final shear bias ($m_{\rm final}$) and corresponding uncertainty ($\sigma_m$), both of which are used in our cosmological analysis. } 
\end{table*}

\subsection{Multiplicative bias}
\label{sec:m_bias_static}

For a given number of simulated sources, the uncertainty in the multiplicative bias, $m$, is determined by the shear applied; the larger the shear, the better $m$ is determined. However, if the shear is too large, non-linearities in the shear measurement algorithm can bias the results \citep{Kitching2022, Jansen2024, Li2024}.

Following previous KiDS studies \citep{Kannawadi2019, Li2023a}, we applied a constant shear $\gamma_i=\pm 0.028$ (for a total shear $|\gamma|=0.04$), resulting in four shear realisations. These values are higher than the average shear we expect in the data, but the results from \cref{app:test_high_amp_shear} show that it yields accurate estimates for $m$, while reducing the required simulation volume by more than an order of magnitude compared to the case with more realistic, lower shear amplitudes. To improve the precision further, we created pairs of images, where the sources were rotated by $90^\circ$, which dramatically reduces the shape noise contribution \citep{STEP2, FenechConti2017}. We used the weighting scheme outlined in \cref{sec:e_weighting} to compute the mean shear for these pairs of images. Linear regression using \cref{Eqn:shear_calibration} then yields estimates for $m$ and $c_i$. 

\begin{figure}
\centering
\includegraphics[width=0.49\textwidth]{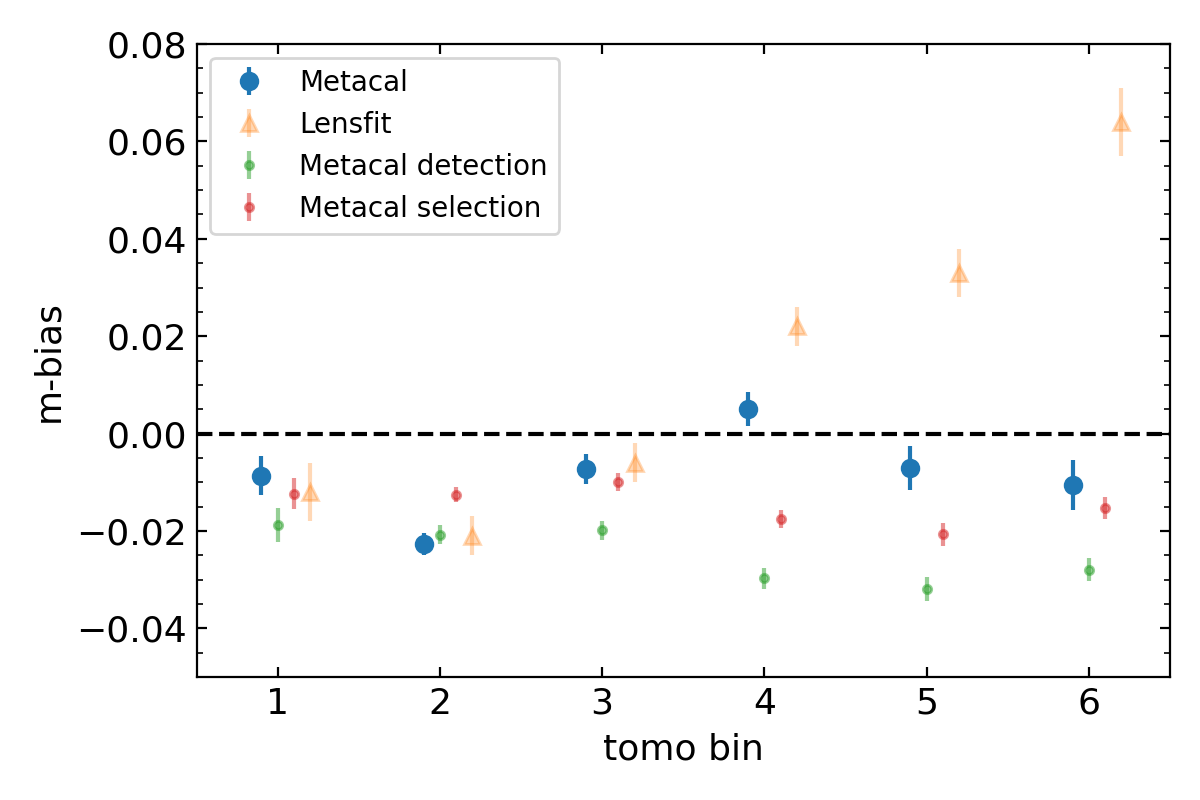}
\caption{Multiplicative bias as a function of tomographic bin estimated from fixed-shear simulations. We find that the biases for \metacal (blue points) are small ($|m|<0.023$) and agree well with the level of bias expected for a perfect shape measurement method after accounting for selection bias (red points). The green points indicate the corresponding detection bias from \se. For the first three bins, the multiplicative bias for \lensfit (orange triangles) is similar but increases for more distant galaxies.}
\label{fig:m_bias}
\end{figure}

\Cref{tab:mbias} lists the multiplicative bias values for the fixed shear, $m_{\rm fix}$, for the six tomographic bins. We find that the average bias in each tomographic bin is small, with the largest value (bin~2) only $m=-0.0227\pm0.0023$, mainly due to contamination by stars (discussed in \cref{app:star_contamination}). \Cref{fig:m_bias} compares the \metacal results (blue points) to the biases determined for \lensfit (orange triangles). For the first three bins, the multiplicative biases are comparable, but  for the highest three redshift bins the values are considerably smaller for \metacal, highlighting the robustness of the approach.

The bias values reported here are the combination of detection bias and the bias subsequently introduced by the shape measurement algorithm. Following \cite{FenechConti2017} we estimate the detection bias using the input (sheared) ellipticities, assigning equal weights to all galaxies. The resulting detection bias is indicated by the green in points in \cref{fig:m_bias}. 
As the actual shear estimates use a weighting scheme, a more appropriate comparison includes the weights as well (the red points in \cref{fig:m_bias}).
The fair agreement between the red and the blue points demonstrates that \metacal behaves as a nearly perfect shape measurement algorithm. Some deviation is expected because of blending \citep{Hoekstra2021a}, but the impact is small.

\begin{figure}[t]
\includegraphics[width = 0.48\textwidth]{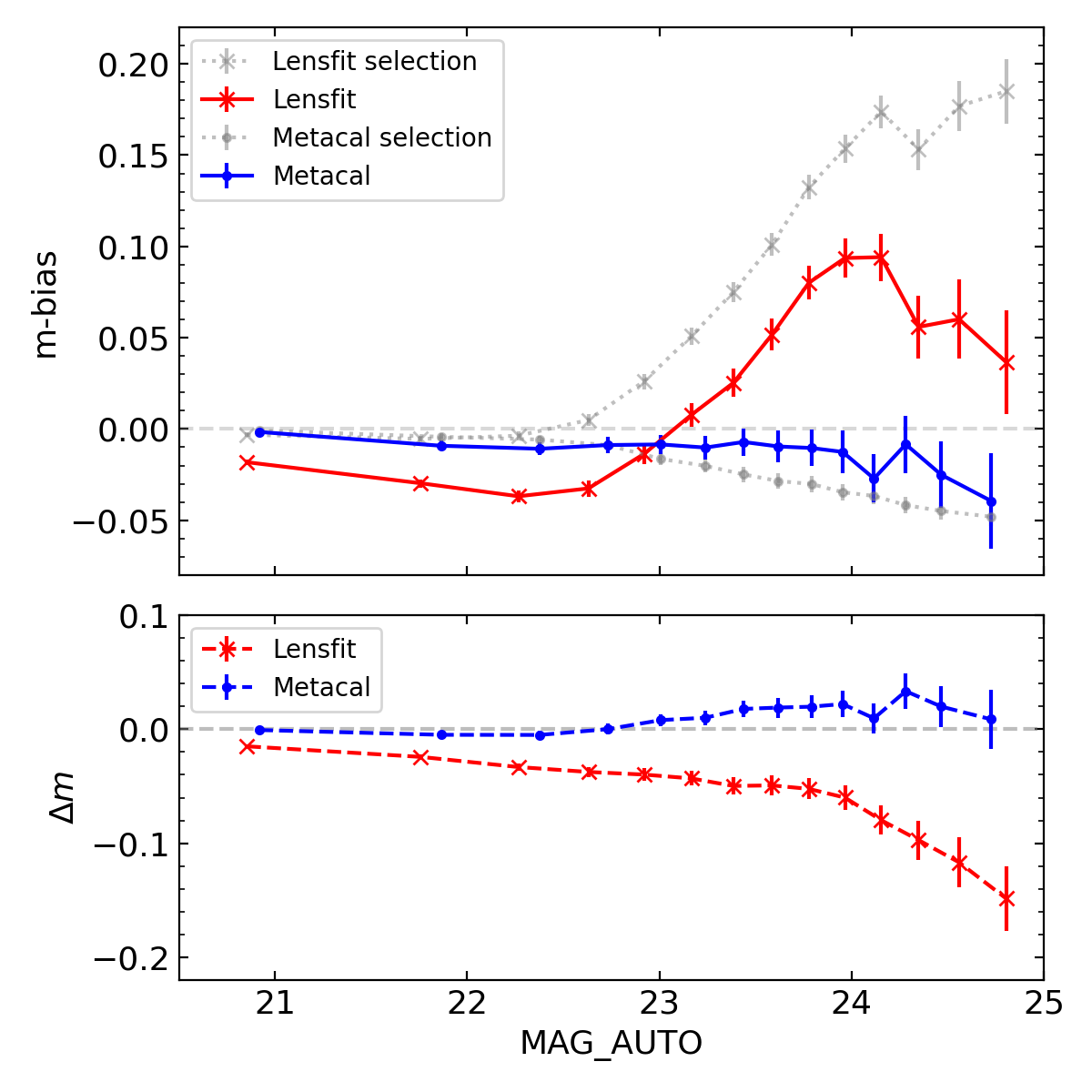}
\caption{{\it Top panel}: Multiplicative shear bias as a function of apparent magnitude \texttt{MAG\_AUTO} for \metacal (blue) and \lensfit (red). For reference, the corresponding selection bias is also indicated (grey).
{\it Bottom panel}: Difference, $\Delta m=m_{\rm fix}-m_{\rm sel}$, highlighting the excellent performance of \metacal. }
\label{fig:m_bias_vs_mag_auto}
\end{figure}

\begin{figure*}
\sidecaption
\includegraphics[width=12cm]{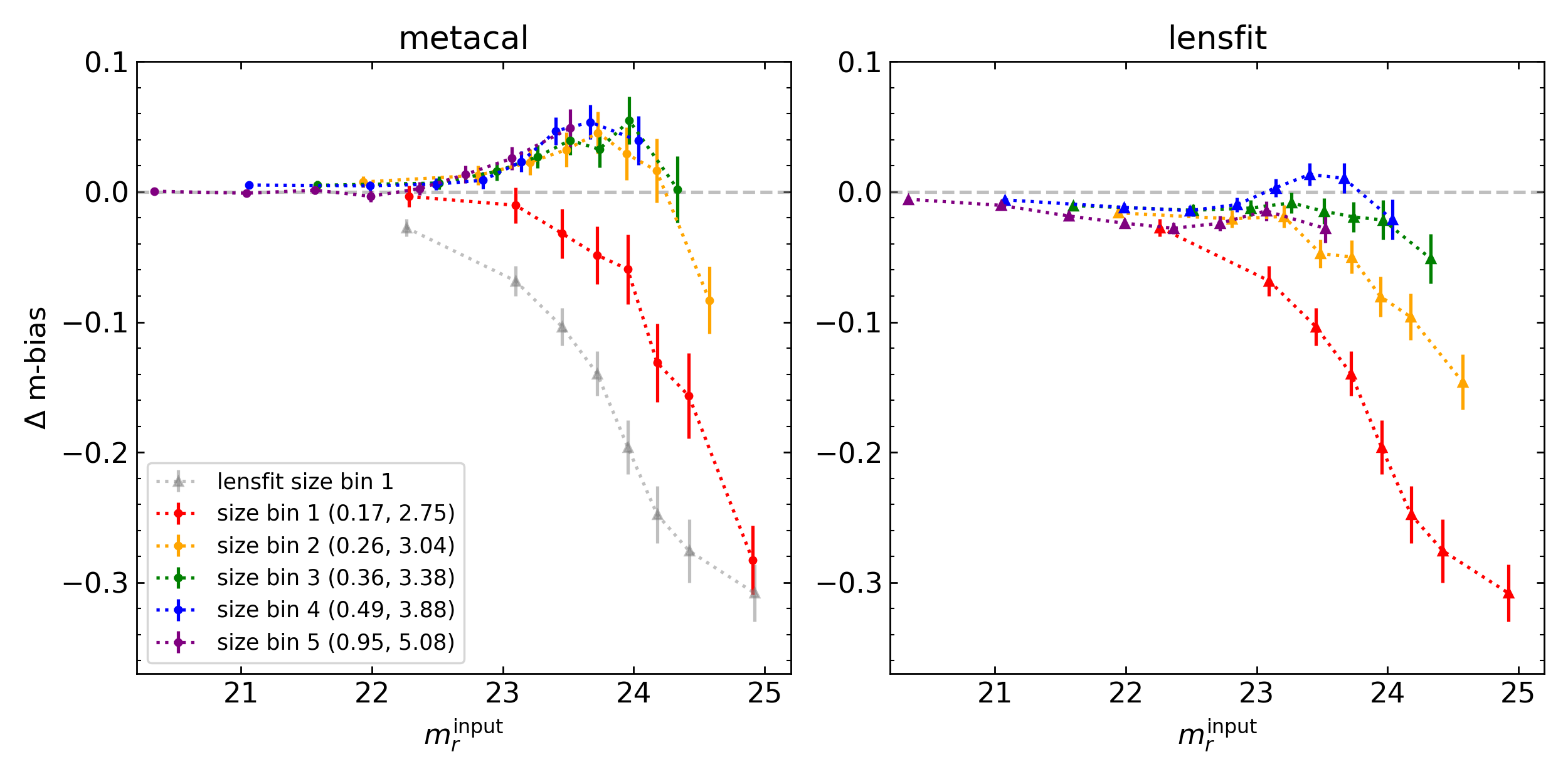}
\caption{Difference between $\Delta m=m_{\rm fix}-m_{\rm sel}$ as a function of $m_r^{\rm input}$ , i.e. the input apparent magnitude when the sources are split into quantiles by input size. For reference, we also indicate the mean input effective radius in arcseconds, and the corresponding mean \texttt{FLUX\_RADIUS} in pixels for each size bin. Left: Results for \metacal. Right: Results for \lensfit. These two sets of results use the sources in common between both catalogues. For reference, the grey points in the left panel correspond to the \lensfit results for the smallest input galaxies and should be compared to the red points for \metacal. 
The bump between $23<m_r^{\rm input}<24$ is caused by blending \citep{Hoekstra2021a}.}
\label{fig:m_bias_vs_mag_size}
\end{figure*}

The values for $m_{\rm fix}$ give a fair estimate of the shear bias in KiDS-1000, but they require some further revision for the cosmological analysis. First of all, following \cite{Li2023a}, we must account for the blending of sources at different redshifts, which we investigate in \cref{sec:m_bias_variable}. Moreover, additional selections are made. For instance, the scheme to correct for additive bias eliminates some sources, as does the redshift calibration. We discuss these below, and report $m_{\rm final}$, the final value of the multiplicative bias, in 
\cref{tab:mbias_final}.

It is desirable if the average shear bias is as small as possible, as it appears in the denominator for our estimator, \cref{eq:shear_estimator}. However, this does not necessarily mean the shear bias is small for any subset of sources. For instance, \cite{FenechConti2017} examined the dependence of $m$ with galaxy size and S/N, which showed that $m$ is still significant for faint, small galaxies. To explore this further, we show in the top panel of  \cref{fig:m_bias_vs_mag_auto} the average multiplicative bias as a function of \texttt{MAG\_AUTO} for \metacal (blue) and \lensfit (red). For reference, we also indicate the corresponding selection bias as a function of apparent magnitude (grey). The difference, $\Delta m=m_{\rm fix}-m_{\rm sel}$, is shown in the bottom panel. It shows that \metacal returns near-perfect shear estimates in the absence selection bias, whereas \lensfit is increasingly biased for faint galaxies. As a consequence, the calibration of \metacal is more robust against mismatches in the magnitude distribution between data and simulations (see \cref{fig:com_skills_k1000}). 
We return to this in \cref{sec:sensitivity}. 

Generally, we expect the shear bias to also depend on galaxy size.
In the case of \metacal, the observed size is strongly correlated with $R^\gamma$, which complicates such a comparison. However, to test the dependence on galaxy size, we can use the input properties instead. In \cref{fig:m_bias_vs_mag_size} we show the difference $\Delta m=m_{\rm fix}-m_{\rm sel}$ as a function of the input $r$-band magnitude when we split the sample into quintiles by input size. The left panel of \cref{fig:m_bias_vs_mag_size} shows the results for \metacal.  The bias is generally small for all subsets, but only for the smallest, faintest galaxies do we observe a significant bias. For the other size bins, the value of $\Delta m$ are consistent with one another, and thus do not depend on input size. The bump between $23<m_r^{\rm input}<24$ is caused by blended sources \citep{Hoekstra2021a}.
The right panel in \cref{fig:m_bias_vs_mag_size} shows the same, but for \lensfit. We observe more spread between size bins,
which indicates a slight dependence on input size. The biases for both methods are generally small, except for the smallest galaxies, where \metacal performs somewhat better (compare the grey and red results in the left panel of \cref{fig:m_bias_vs_mag_size}).
In practice, however, these galaxies carry little weight in the analysis.

\begin{figure}
\includegraphics[width = 0.47\textwidth]{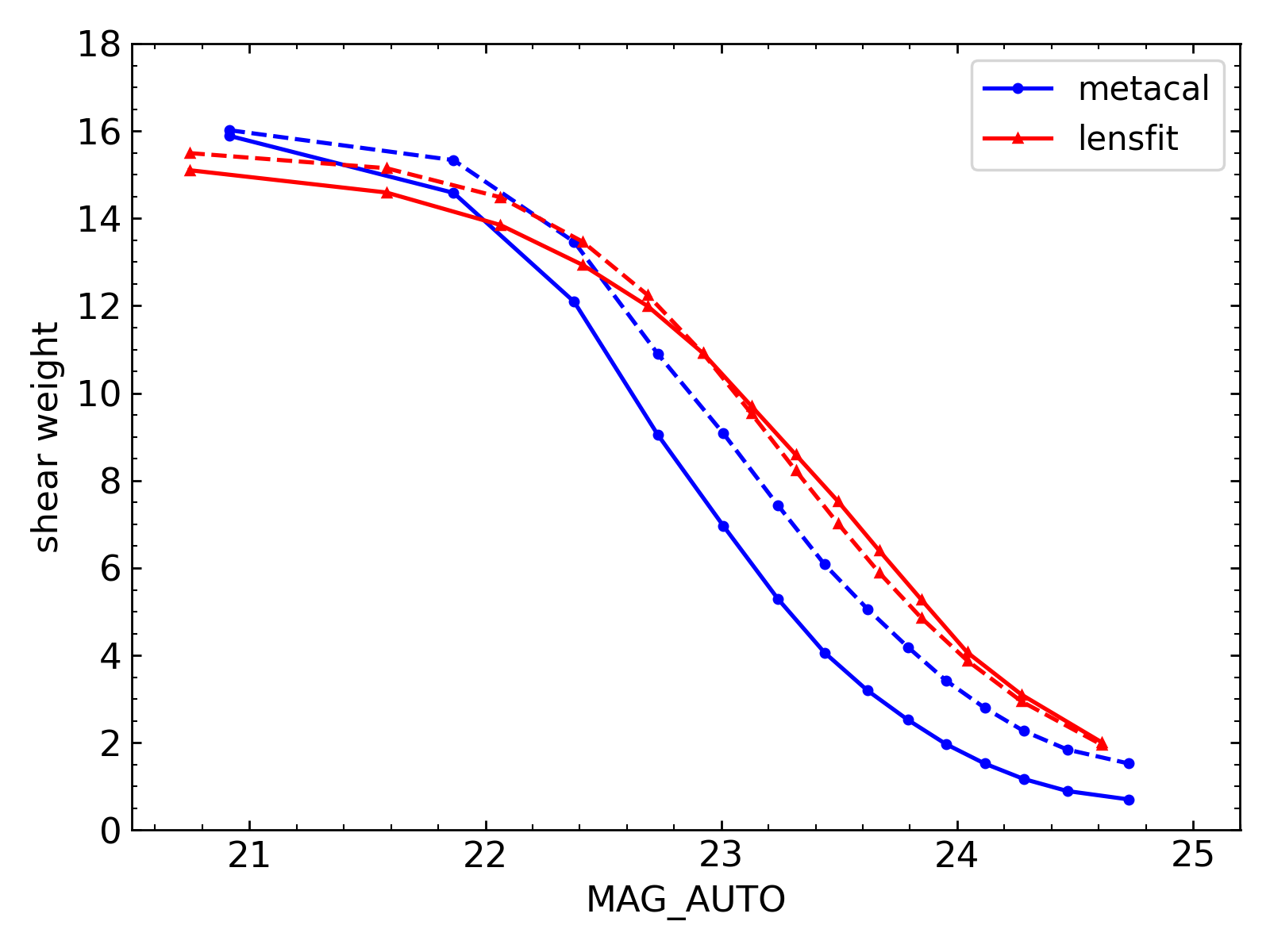}
 \caption{Effective shear weight, $(1+m)R^\gamma\sigma^{-2}_\gamma$, as a function of magnitude (solid blue line) compared to \lensfit (solid red lines). For comparison, the dashed lines denote the corresponding inverse variance weights, $1/\sigma^2_\gamma$, which are used to compute the effective number density of sources. For \lensfit the two weights are similar; however, \metacal suppresses the contribution from faint sources, because of the extra factor, $R^\gamma$. As discussed in the text, this slightly changes the effective source redshift distribution.}
\label{fig:weight}
\end{figure}

As discussed in \cref{sec:e_weighting}, we weighted the sources based on the measurement uncertainty using an inverse variance weight that depends on apparent magnitude for a given tomographic bin. 
The dashed lines in \cref{fig:weight} show this for all bins combined for \metacal (blue) and \lensfit (red). The behaviour
for both methods is similar. We used these results to compare the statistical  constraining power of our \metacal catalogue. To this end, we determined the uncertainty in the shear estimate from a single galaxy, $\sigma_{\hat\gamma,i}$ and the effective number density, $n_{\mathrm{eff}}$ using the definition in \cite{Heymans2012}. The results for each tomographic bin are reported in \cref{tab:mbias}.

\begin{figure}[t]
\includegraphics[width = 0.47\textwidth]{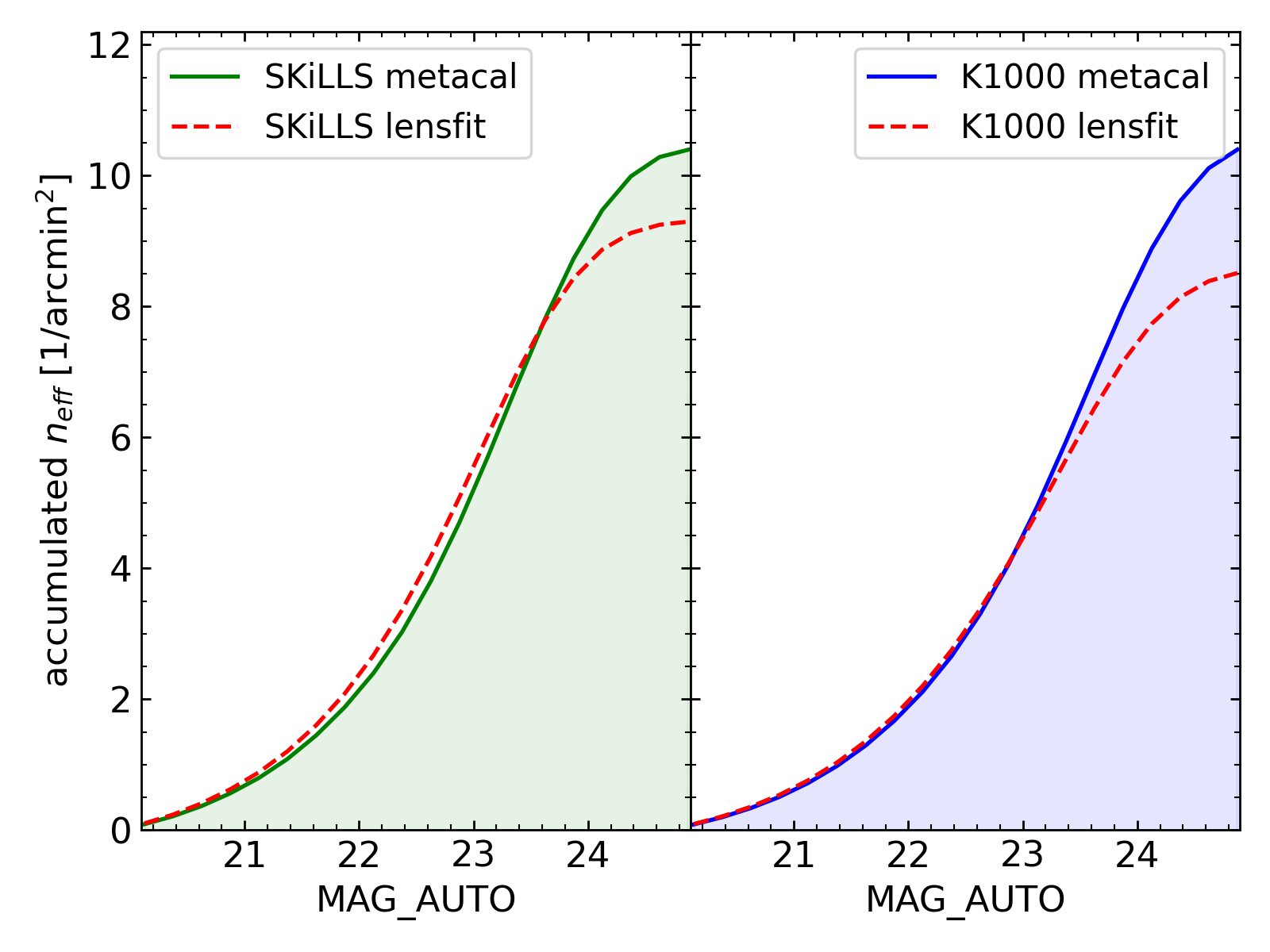}
\caption{Accumulated effective number density, $n_{\rm eff}$, as a function of apparent magnitude for \metacal (solid line) and \lensfit (dashed line). \metacal yields higher effective source densities for both SKiLLS (left) and KiDS-1000 (right) for faint galaxies.}
\label{fig:neff_comp}
\end{figure}

\cref{fig:neff_comp} shows the cumulative effective number density as a function of apparent magnitude, for \metacal and \lensfit for both SKiLLS (left) and the KiDS-1000 data (right). 
For both SKiLLS and KiDS, \metacal yields  $n_{\mathrm{eff}} \simeq$ 10.4 arcmin$^{-2}$, while \lensfit yields $n_{\mathrm{eff}} \simeq$ 9.3 and 8.5 arcmin$^{-2}$, respectively. The 22\% higher
$n_{\rm eff}$ for \metacal when applied to KiDS-1000 is attributed to the higher fraction of sources for which shapes could be measured, which in turn is a consequence of our simple shape measurement and our estimate for the inverse variance.

The solid lines in \cref{fig:weight} show how the effective shear weight\footnote{We call this the effective shear weight, which should be used to weight the redshift distributions used for the cosmological inference. This distinguishes it from the shear weight used to weight galaxies in the summary statistics.}, $(1+m)R^\gamma\sigma^{-2}_\gamma$, depends on apparent magnitude. 
For \lensfit the weighting of the shears is similar to that of the shape estimates, so that $R^\gamma\equiv 1$, but our implementation of \metacal suppresses the contribution from faint sources, because of the extra factor $R^\gamma$.
This changes the effective redshift distribution used to compute the predicted signal for a given sample. To stay close to the analysis of \cite{Li2023b} we ignore this here, and we used the inverse variance weights to compute the effective source redshift distribution. In \cref{tab:mbias} we report the corresponding mean weighted redshifts for SKiLLS for the tomographic bins. 
 
For completeness, in \cref{tab:mbias} we also report the weighted mean redshifts when the additional weighting of the shears by $(1+m)R^\gamma$ is included.
For bins 2--5 the shifts are less than 0.01, comparable to the overall uncertainties in the mean redshifts. We observe an unexpected large change for the first bin. When examining the mean shear response as a function of redshift in SKiLLS we find that it varies rapidly within the bin, which is not physical.\footnote{The shear response is mostly a function of galaxy size. We expect the mean galaxy size to be a smooth function of redshift.}
Rather, we conclude this is a feature of SKiLLS, perhaps owing to the relatively small volume probed by the training set. This requires further study for stage IV projects, for which the observed changes in mean redshift exceed requirements \citep{Mellier2024}. For our purpose, we conclude that using the full shear weights or just the inverse variance weighting has no significant impact.

\subsection{Correction for PSF anisotropy} 
\label{sec:psf_leakage}

We can expand the term describing the additive bias in \cref{Eqn:shear_calibration} to distinguish between the various sources of additive bias. 
Specifically, we must quantify the bias that arises from an incomplete correction for PSF anisotropy, also referred to as `PSF leakage' and errors in the PSF model itself. To this end, we modified \cref{Eqn:shear_calibration} to 
\begin{equation}
\left\langle\epsilon_i^{\rm obs}\right\rangle\equiv
\frac{\left\langle \hat{e}_i \right\rangle}{\left\langle R^\gamma\right\rangle}
= (1 + m_i)\, \gamma^{\mathrm{true}} + \alpha_i\, e_i^{\mathrm{PSF}} + \beta_i \delta e_i^{\rm PSF} +\tilde{c}_i \, ,
\label{eqn:alpha}
\end{equation}
to express the result as observed shear and ellipticity, where $\alpha_i$ quantifies the leakage of the PSF anisotropy, $e_i^{\rm PSF}$, into the shear estimate, while $\beta_i$ captures errors in the PSF model. In the remainder of the paper we drop the subscript because we found that both ellipticity components respond the same. After our empirical correction for residual PSF leakage, errors in the PSF model may become dominant. However, its impact was 
 studied in \cite{Giblin2021} and \cite{Li2023a}, who found it to be negligible for KiDS.
Finally, any intrinsic additive bias and detector-related effects are captured by $\tilde{c}_i$.

\begin{figure}[t]
\centering
\includegraphics[width=0.45\textwidth]{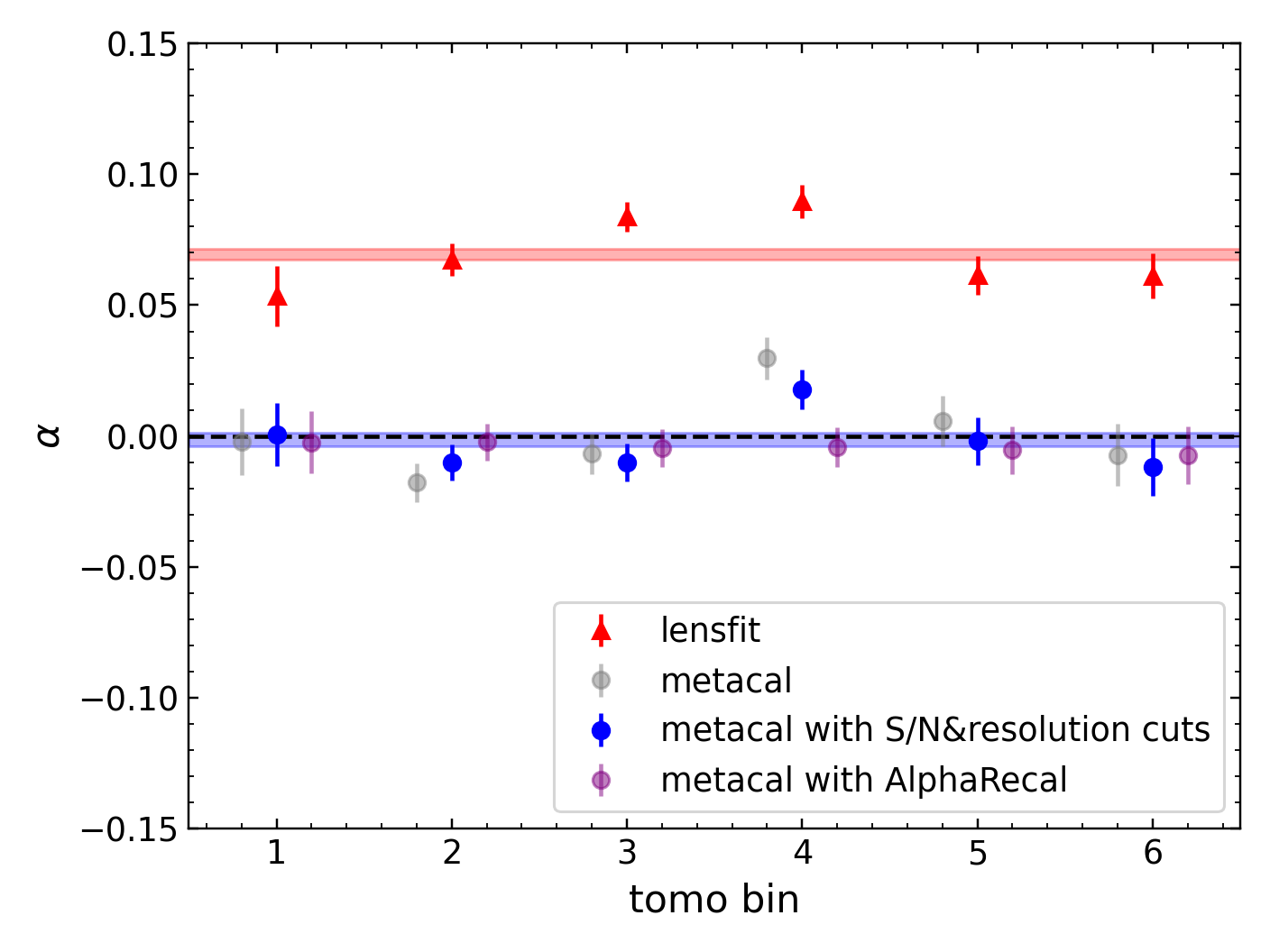}
\caption{$\alpha$ values for the different tomographic bins. The raw values from \metacal and \lensfit are indicated by the grey and red points, respectively.
The raw \metacal values are close to zero, and even more so after applying the cuts to the signal-to-noise ratio and resolution (blue points; see text for details). After AlphaRecal, $\alpha$ is consistent with zero for all bins (purple points). To enable a fair comparison, we scaled the PSF ellipticities measured by the \metacal pipeline to match the \lensfit values.}
\label{fig:raw_alpha}
\end{figure}

In our setup of \metacal, we circularised the PSF before measurement. The results presented in \cref{fig:psf_circ_one_pointing} suggest that this works well, resulting in PSFs with negligible PSF anisotropy. In \cref{fig:raw_alpha} we show the value for $\alpha$ for the  tomographic bins. As anticipated, we find very small values. For reference, we also show the corresponding values for \lensfit for the five bins used for KiDS-1000 \citep{Li2023b}. To enable a fair comparison, we scaled the PSF ellipticities measured by the \metacal pipeline to match the \lensfit values. This is needed, because of the circular weighting applied to the \lensfit PSF measurements \citep{Li2023a}. 

Although the recovered values are small, we observe a non-vanishing signal for the fourth tomographic bin. Moreover, the performance on actual data may differ. 
Therefore, to account for any residual PSF leakage, we followed \cite{Li2023a} and applied an empirical correction. To this end, we began by exploring how the PSF leakage varies with galaxy properties in SKiLLS. We focused in particular on signal-to-noise ratio and resolution, which are the dominant quantities for PSF-related shear bias.  
To quantify S/N, $\nu_{\rm SN}$, in our \metacal pipeline we used the following definition:
\begin{equation}
   \nu_{\rm SN} \equiv \frac{\int \mathrm{d}^2\bm{x}\, W(\bm{x})\,I_{\rm obs}(\bm{x})/\sigma^2_{\rm noise}(\bm{x})}
    {\sqrt{\int \mathrm{d}^2\bm{x} \,W^2(\bm{x})/\sigma_{\rm noise}^2(\bm{x})}},
\label{eq:snr}
\end{equation}
where, $I_{\rm obs}(\bm{x})$ is the observed image surface brightness, and $\sigma_{\rm noise}(\mathbf{x})$ is the corresponding noise in the image. The optimal choice for the spatial weight function, $W(\mathbf{x})$, is the noiseless image, but here we use the same Gaussian weight function that was used in
\cref{eqn:q_definition} to measure the quadrupole moments. 

We expect PSF leakage to be smaller for well-resolved objects. 
To capture this aspect, we defined the resolution\footnote{Our definition differs from that used in \cite{Li2023a}, because they based it on the parameters returned by \lensfit. However, we expect that the rank-order of objects is largely similar in practice.} as
\begin{equation}
    {\cal R} \equiv \frac{\sigma^2_{\rm PSF}}{(\texttt{FLUX\_RADIUS})^2}\ ,
    \label{eq:resolution}
\end{equation}
where the size of the galaxies is given by the half-light radius (\texttt{FLUX\_RADIUS}) given by \se.
The PSF size is quantified by $\sigma_{\rm PSF}$, which is computed using 
the stacked PSF model for individual sources, thus taking into account only the exposures that are used for \metacal.
To determine $\sigma_{\rm PSF}$, we fitted an elliptical Gaussian to the PSF image using the \texttt{galsim.hsm.FindAdaptiveMom} routine in GalSim \citep{Rowe15}, which returns the quadrupole moments. These were used to compute $\sigma_{\rm PSF}\equiv |Q_{11}Q_{22} - Q_{12}Q_{21}|^{1/4}$.

\begin{figure}
\centering
\includegraphics[width=0.45\textwidth]{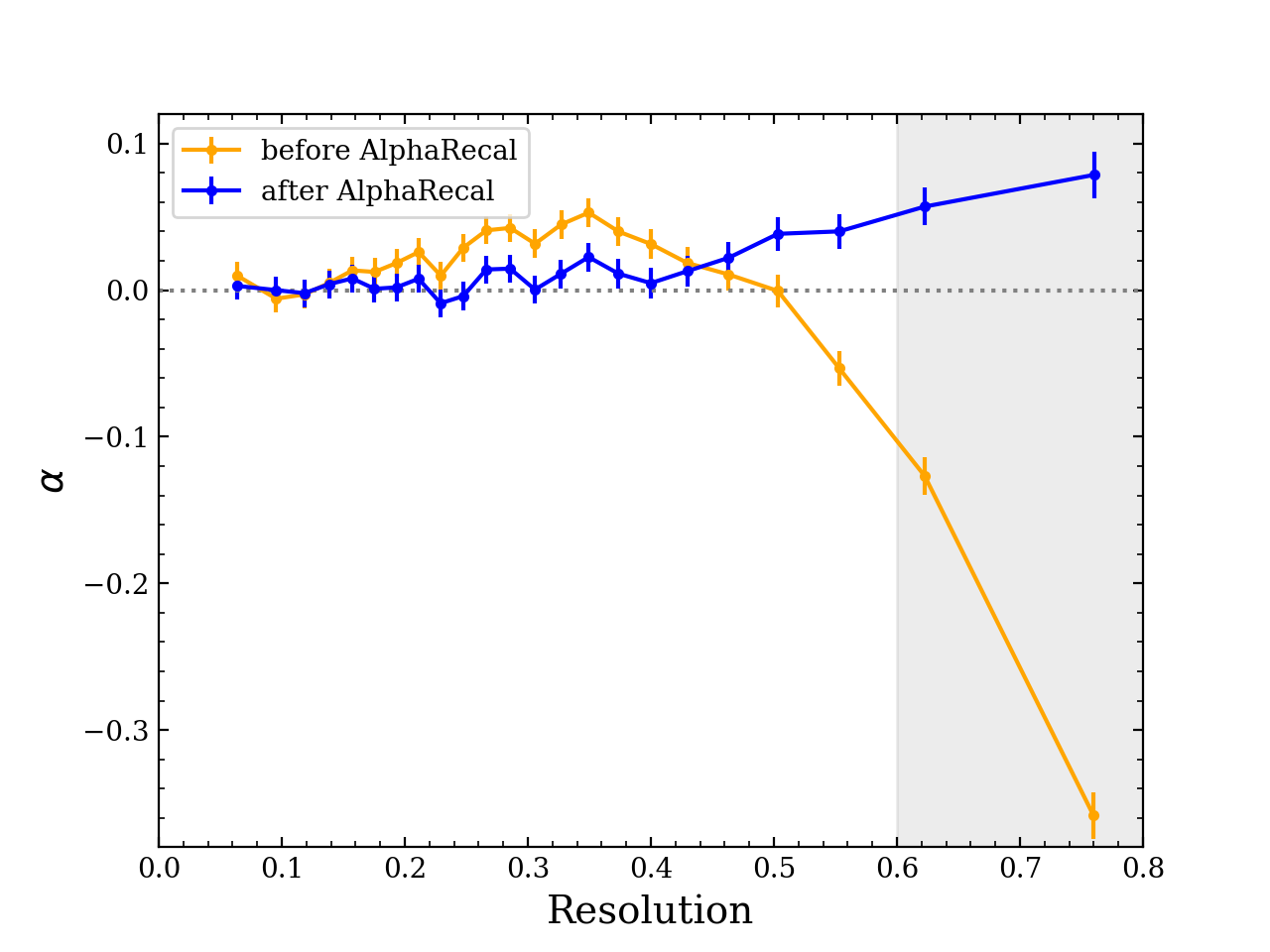}
\includegraphics[width=0.45\textwidth]{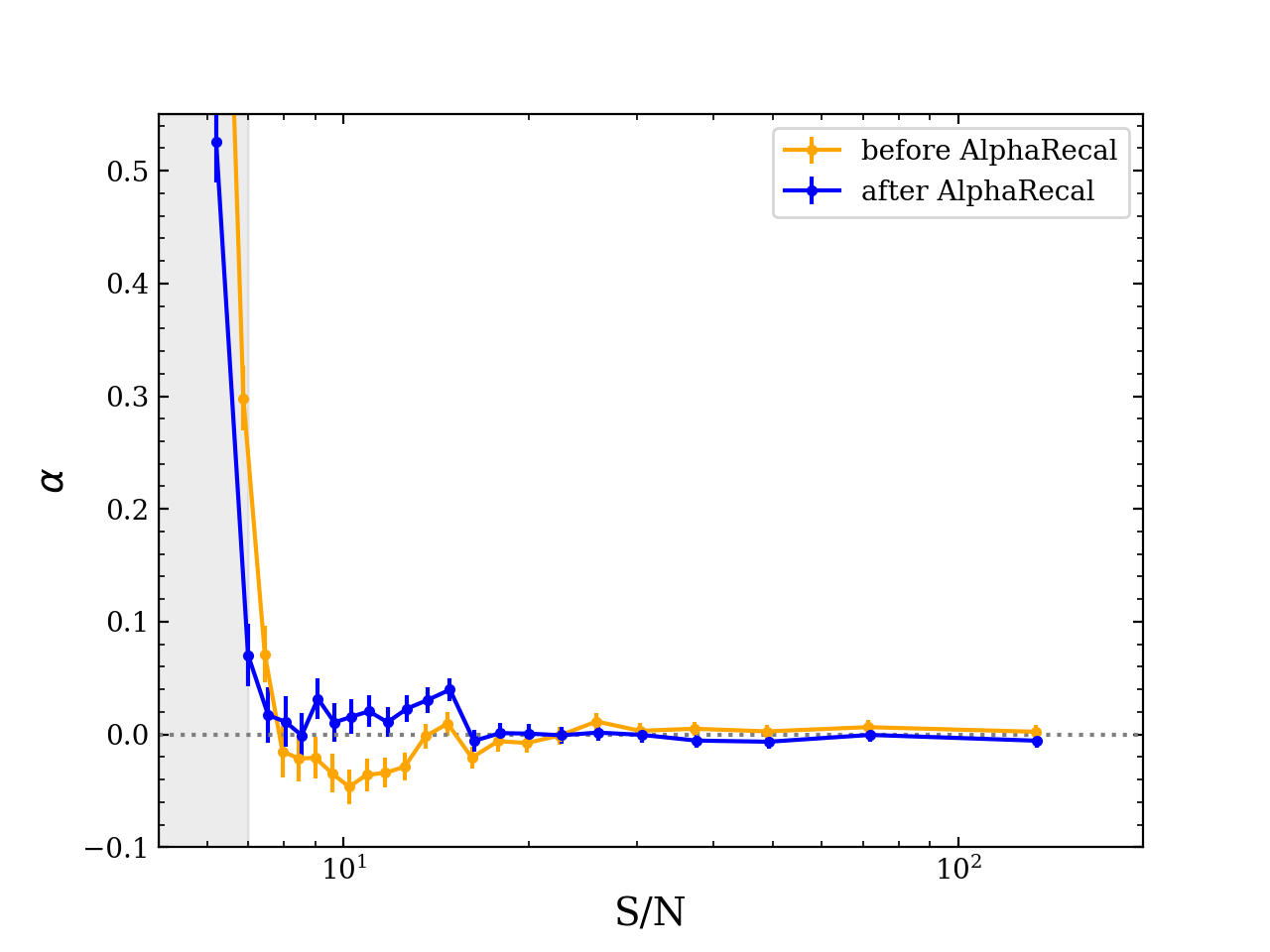}
\caption{1D projection of the alpha term measurement as a function of resolution and S/N, respectively. After applying the AlphaRecal process, alpha terms are consistent with zero in most bins (blue). We applied cuts based on remaining alpha terms for the low-$\nu_{\mathrm{SN}}$ and low-resolution sources ($\nu_{\mathrm{SN}} \leq 7$ and $R \geq 0.6$, grey shade).}
\label{fig:alpha_AlphaRecal}
\end{figure}

\Cref{fig:alpha_AlphaRecal} shows 1D projections of $\alpha$ as a function of resolution (top) and S/N (bottom). The `raw' values are indicated by the orange points, which show that the amplitude of $\alpha$ quickly increases for small and noisy galaxies. To capture the PSF leakage dependence as a function of $\nu_{\rm SN}$ and $\cal{R}$, we measured $\alpha$ on a 20\texttimes20 grid, and fitted the following function \citep{Li2023a}
\begin{equation}
   \alpha \left( \nu_{\mathrm{SN}}, \mathcal{R} \right) = a_0 + a_1\, \nu_{\mathrm{SN}}^{-2} + a_2\, \nu_{\mathrm{SN}}^{-3} + b_1\, \mathcal{R} + c_1\, \mathcal{R}\, \nu_{\mathrm{SN}}^{-2} \, .
    \label{eq:alpharecal}
\end{equation}

The observed large values for $\alpha$ for small, faint galaxies, highlight the challenge of measuring their shapes robustly even after correcting the observed values for the dependence of $\alpha$ on ${\cal R}$ and $\nu_{\rm SN}$ (referred to as `AlphaRecal'; see blue points in \cref{fig:alpha_AlphaRecal}). We therefore cut our source sample, requiring $\nu_{\rm SN} > 7$ and ${\cal R} < 0.6$.
The applied cuts result in a 16\% reduction in the number of sources. However, as \cref{fig:weight} shows, those are down-weighted in practice. Hence, the impact on the effective number density of sources is negligible, while improving the robustness of the measurements. 
This cut reduces the mean $\alpha$ for the tomographic bins (blue points in \cref{fig:raw_alpha}) for the second and fourth tomographic bins. After accounting for AlphaRecal, the values are reduced further (see blue points in \cref{fig:alpha_AlphaRecal}) and consistent with zero for all tomographic bins, as indicated by the purple points in \cref{fig:raw_alpha}.

\subsection{Shear calibration parameters from SKiLLS with variable shear}
\label{sec:m_bias_variable} 

Up to this point, we have discussed using image simulations with the same shear applied to all galaxies in the scene. In reality, the shear varies spatially and as a function of redshift. In principle, this is fine for \metacal, as its response to the applied shear is linear (see \cref{app:test_high_amp_shear}), but only for isolated galaxies. 
However, the shear response is expected to be changed for blended sources at different redshifts \citep{MacCrann2022}, modifying the multiplicative bias when the redshift dependence of the input shear is included. To quantify this, we followed \cite{Li2023a} and re-simulated only the sources identified as blended in the input catalogue, applying a realistic redshift-dependent shear. 

We used the same simulations as \cite{Li2023a} to calibrate the  \lensfit results but adopted a different approach to calculate the change in the multiplicative bias. We take advantage of the robustness of our shape estimator, which allows us to measure the relevant quantities for individual sources. This approach is viable because the source detection from the fixed-shear simulation almost overlaps with the realistic-shear simulation. We verified that only less than $0.7\%$ sources are not matched between fixed-shear and realistic-shear simulations, while the multiplicative biases are small ($|m|<0.05$). As this only applies to the 35\% of blended sources, the missing objects would bias our estimates by less than $1.2\times 10^{-4}$, well below the level of correction we obtain. We note that a more careful comparison suggests a much lower bias. Therefore, we can use the fact that the source detections are almost identical to determine how the multiplicative bias changes for the blended sources using more realistic shears, $\gamma_{\rm var}$, and compare the ellipticity difference for the $i^{th}$ source, $\Delta e_i=e_i^{\rm var}-e_i^{\rm fix}$. Using \cref{eq:raw_estimator}, this difference can be expressed in terms of the changes in $m$ and $R^\gamma$ for the ensemble of sources:
\begin{equation}
\Delta e_i =(1+m+\Delta m)(R_{i}^\gamma+\Delta R^\gamma_i)(\gamma+\Delta\gamma_i) - (1+m)R_{i}^\gamma\gamma,
\end{equation}
where we dropped the subscript `fix' for the quantities measured from the simulations with fixed shear. The change in multiplicative bias $\Delta m=m_{\rm var}-m_{\rm fix}$ is then given by
\begin{equation}
    \Delta m  =  \frac{\langle \Delta \hat{e}\rangle}{\gamma\,\langle R^\gamma\rangle}-(1+m +\Delta m)\left[ \frac{\langle \Delta\gamma \rangle}{\gamma} \left(1 + \frac{\langle\Delta R^\gamma\rangle}{\langle R^\gamma\rangle} \right) +\frac{\langle\Delta R^\gamma\rangle}{\langle R^\gamma\rangle}\right].
\end{equation}
We assume that $|m|\ll 1$, while the changes in $m$ and $R^\gamma$ are small so that products of differences can be ignored. Hence, we can write
\begin{equation}
   \Delta m\,\gamma \approx \frac{\langle \Delta \hat{e}\rangle}{\langle R^\gamma\rangle}- \left[\Delta\gamma +\frac{\langle\Delta R^\gamma\rangle}{\langle R^\gamma\rangle}\gamma\right]\,,\label{eqn:blending_shear}
\end{equation}
so that $\Delta m$ can be determined through linear regression of the right hand side against the input fixed shear, analogous to \cref{Eqn:shear_calibration}.  

\begin{figure}
\centering
\includegraphics[width=0.46\textwidth]{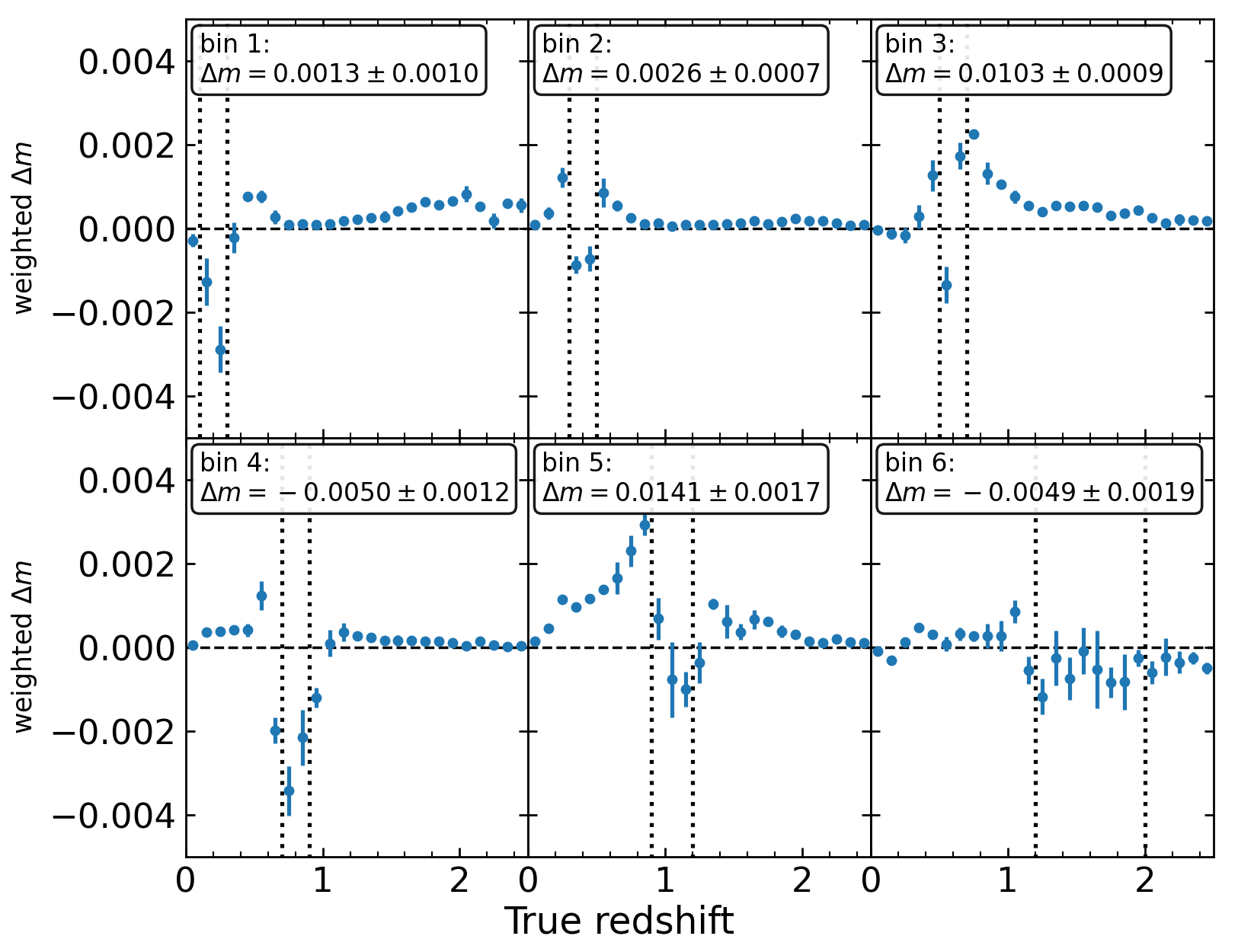}
\caption{Change in multiplicative bias, weighted by the normalised source redshift distribution, for blended-only sources with a realistic shear applied, as a function of redshift for each tomographic bin. The bin boundaries are indicated by the vertical dashed lines. $\Delta m=m_{\rm var}-m_{\rm fix}$ was computed 
using \cref{eqn:blending_shear}. The sum of all the data points in each tomographic bin yields the overall shift $\Delta m$ for blended sources, which is reported in each panel.} 
\label{fig:delta_m_tomo}
\end{figure}

\begin{figure}
\centering
\includegraphics[width=0.45\textwidth]{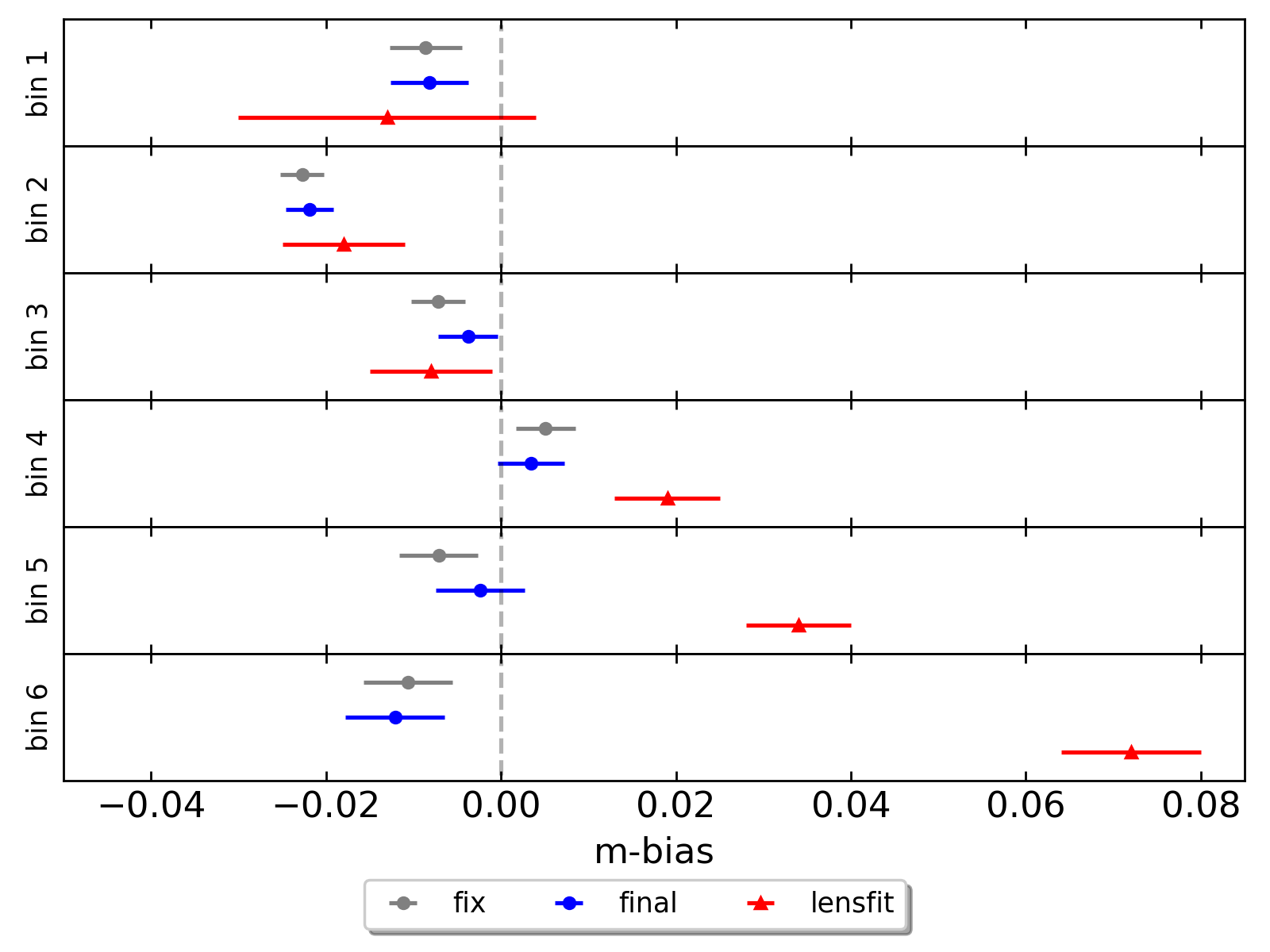}
\caption{Final estimates of the multiplicative bias, $m_{\rm final}$, for each tomographic bin. The blue points account for redshift-dependent blending, which slightly shifts the values compared to the fixed-shear case (grey). For reference, we show the $m_{\rm final}$ for \lensfit (red).} 
\label{fig:m_bias_final}
\end{figure}

\Cref{fig:delta_m_tomo} shows the values of $\Delta m$, multiplied by the normalised redshift distribution of the simulated blended sources, as a function of true redshift, to highlight the sources that introduce biases. The sum of the points for each tomographic bin yields the total shift in multiplicative bias for the blended sources.
We find that the largest deviations occur for bins~3 and~5, but the impact on the final bias estimates, $m_{\rm final}$ is small, as shown in \cref{fig:m_bias_final}, because the fraction of galaxies with a neighbour within $4''$ is only about one third. Compared to the fixed shear simulations (grey points), accounting for the redshift-dependent blending shifts the values by at most $1\sigma$ (blue points).

For reference, \cref{fig:m_bias_final} also shows $m_{\rm final}$ for \lensfit from \cite{Li2023a}, who determined $m_{\rm var}$ for the blended sources. As those objects have small shears, the resulting uncertainty in $m$ is large for a fixed number of sources. The advantage of the direct estimation of $\Delta m$ is that the results are independent of the variable shear amplitude, resulting in  much smaller uncertainties, especially for lower redshift bins. Thanks to the improvement in precision, we required far fewer simulated sources: we reduced the numbers by a factor of eight compared to \cite{Li2023a}.

\subsection{Sensitivity tests}
\label{sec:sensitivity}

As shown in \cref{fig:com_skills_k1000}, SKiLLS is able to reproduce the observed distributions of key galaxy properties well, but not perfectly. \Cref{eqn:m_inaccuracy} implies that this generally results in a change in the value of $m$.
It is therefore useful to quantify the sensitivity of the multiplicative bias to such deviations. \Cref{fig:m_bias_vs_mag_auto} suggests that small mismatches in the distribution of apparent magnitudes will not have much impact, but \cref{fig:com_skills_k1000} also indicates differences in the morphologies of the galaxies, such as their sizes and shapes. These also change the recovered biases \citep{Hoekstra2017}. We therefore follow \cite{Li2023b}, and quantify the sensitivity of \metacal to rather extreme changes in the half-light radii, axis ratios and S{\`e}rsic indices of the input galaxies. We used the same input distributions as those used to test the sensitivity of \lensfit in Appendix~C of \cite{Li2023b}. 

\begin{figure}[h]
\centering
\includegraphics[width=0.49\textwidth]{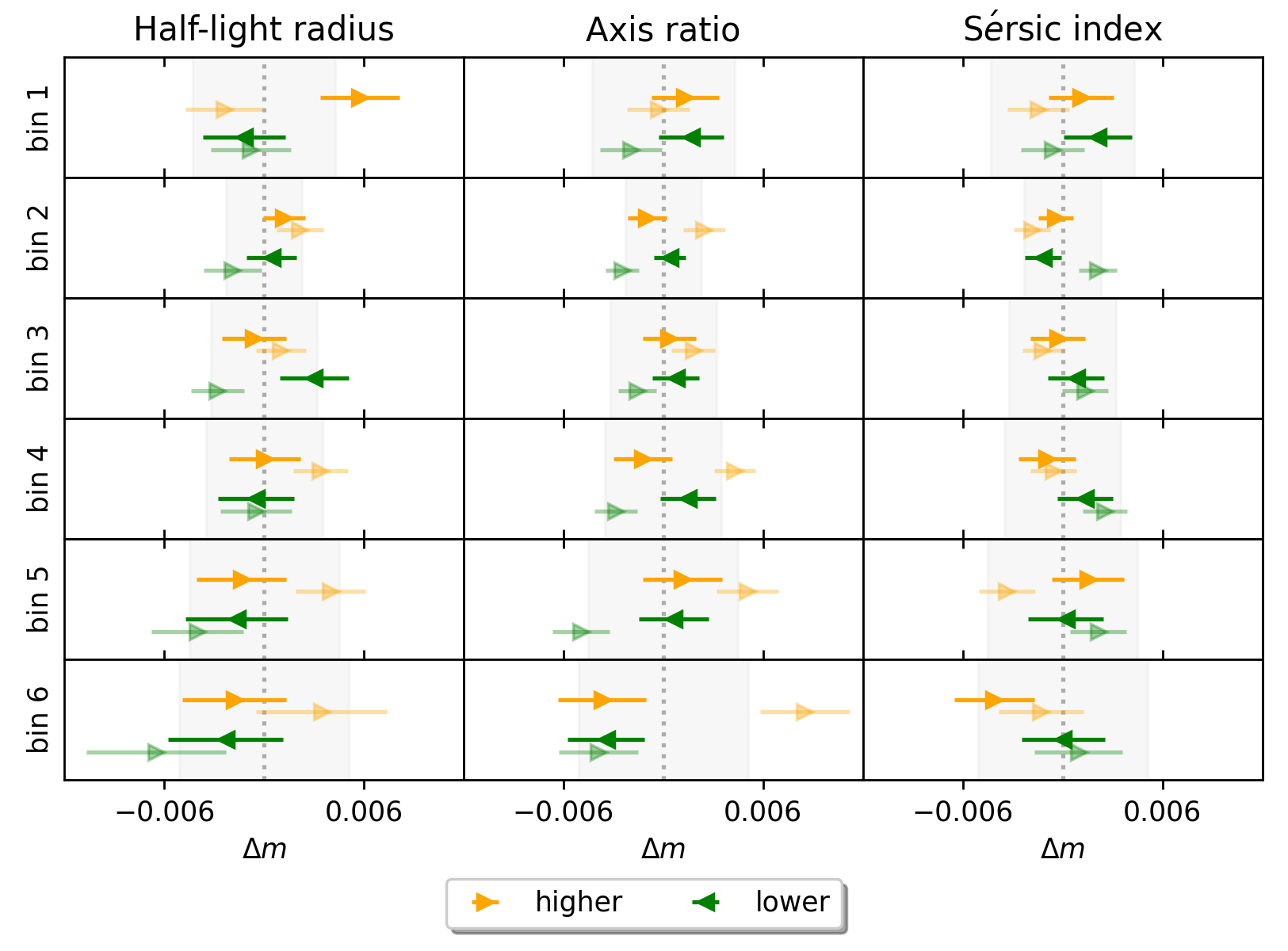}
\caption{Change in multiplicative bias as a result of the extreme changes in the input morphologies. The setup is the same as that used by \cite{Li2023b}. Their results for \lensfit are shown as the fainter points with error bars. From left to right, we shifted the half-light radii, axis ratios or S{\'e}rsic indices to higher (orange) or lower (green) values. The sensitivity of both \metacal and \lensfit to these extreme changes is well within the estimated errors (grey).}
\label{fig:extreme_sample_sensitivity}
\end{figure}

In \cref{fig:extreme_sample_sensitivity}, we show the resulting shifts in multiplicative bias, $\Delta m$, for the different extreme setups. From left to right, we shift the half-light radii, axis ratios or S{\'e}rsic indices to higher (orange) or lower values (green) values. The corresponding \lensfit results from \cite{Li2023b} are shown by the light-coloured points. The sensitivity of both \metacal and \lensfit to these extreme changes is well within requirements. 
Nonetheless, in the case of the source axis ratios, we find that the \metacal results are more robust, with \lensfit showing an increased sensitivity for the higher redshift bins.

These tests suggest that our results are robust, 
but \cite{Csizi24} recently showed that considering more complex morphologies may still impact the results. The changes are, however, sufficiently small, so that we can conclude that the galaxy morphologies in SKiLLS are sufficiently realistic for KiDS cosmic shear measurements. 

\begin{figure}\centering
\includegraphics[width=0.475\textwidth]{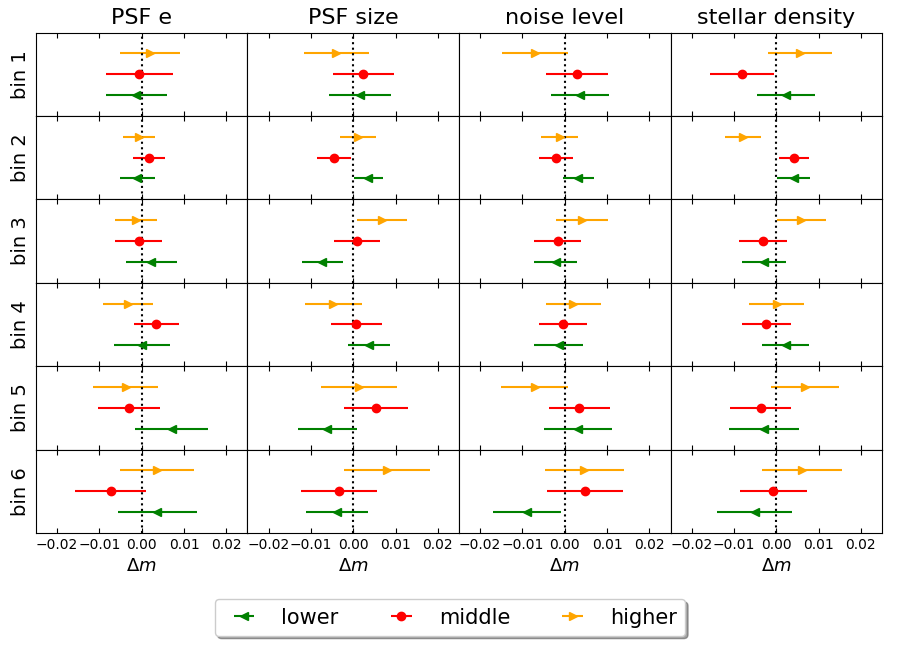}
\caption{Change in multiplicative bias when the data are split by, from left to right, PSF ellipticity, PSF size, background noise level, and stellar density. In each case, the simulations were split into three equally populated samples. The results for the lower (green), middle (red), and higher (orange) values are consistent with the average bias for the full sample. Only for the second tomographic bin do we observe a hint of a sensitivity to the star density.}
\label{fig:subsample_sensitivity}
\end{figure}

Changes in the observing conditions or the star density can also influence the shear bias \citep{Hoekstra2015, Hoekstra2017}. In principle, SKiLLS samples these variations, so that the final shear bias is representative for the survey as a whole. Provided the spatial variation in multiplicative bias is small, this is sufficient for cosmological parameter inference \citep{Kitching2019}. Nonetheless, it is useful to explore the sensitivity of the results to such variation.

We followed \cite{Li2023a} and quantified the sensitivity to four key parameters: PSF ellipticity, PSF size, background noise level, and stellar density. To this end, we split the sources into three subsamples based on the parameter of interest and compute the change in multiplicative bias with respect to the full sample for each tomographic bin. We ensured that the variation between the three subgroups is large enough to capture the diversity in the data.
For example, the star densities in the mid- and high-groups are two and four times that of the low-group (3 stars/arcmin$^2$), respectively. We do not observe significant changes in $m$, as shown by the results in \cref{fig:subsample_sensitivity}.
The largest deviation is seen for the
high star density case in the second tomographic bin. We explore this further in \cref{app:star_contamination} but conclude that our \metacal shear estimates are robust against the variations we expect across the survey. 

\section{KiDS data preparation for cosmological analysis}
\label{sec:datavector}

The main aim of this paper is to verify the methodology of \metacal and to investigate its impact on the cosmological inference. Therefore, we limit our cosmological analysis as close as possible to the KiDS-1000 cosmic shear analysis with \lensfit, using the five tomographic bins.
Although \metacal provides accurate shear estimates for the sixth bin, including these measurements would have required
changing the redshift calibration methodology for the \metacal catalogue, which we leave for future work.

Nonetheless, the source selection differs because \metacal provides shape estimates for objects that are excluded by \lensfit. We discuss the impact of these in more detail in \cref{app:source_selection}, reporting how each step affects the source sample.
We applied the S/N and resolution cuts as described in \cref{sec:psf_leakage}. Another important selection in the source number relies on the
reliability of the photometric redshift calibration. As discussed in \cref{sec:nz_estimation}, this step filters out the sources that lack sufficient counterparts with spectroscopic information. 

\begin{table*}[h]
    \centering
    \caption{\metacal summary statistics of five tomographic bins for the cosmological analysis. }
    {\renewcommand{\arraystretch}{1.2}
    \label{tab:mbias_final}
    \begin{tabular}{cccccccccc}
        \hline\hline
        Bin & Photo-$z$ range & $\langle R^{\gamma} \rangle$ & $n_{\rm eff}$ (metacal) & $n_{\rm eff}$ (\lensfit) & $\sigma_{\gamma,i}$ & $m_{\rm SOM, gold}$&$m_{\rm final}$  & $\sigma_{\rm m, final}$ & $\delta z = z_{\rm est} - z_{\rm true}$\\
        \hline
        1   & $0.1 < z_{\rm B} \leq 0.3$ & $0.21$& $0.74$ & 0.68 & $0.308$  & $-0.0061$ & $-0.0056$ & 0.0056 & $\hphantom{-}0.000 \pm 0.0096$ \\
        2   & $0.3 < z_{\rm B} \leq 0.5$& $0.19$& $1.44$ & 1.30 & $0.289$  & $-0.0178$ &$-0.0169$&  0.0027 & $\hphantom{-}0.002 \pm 0.0114$ \\
        3   & $0.5 < z_{\rm B} \leq 0.7$&$0.20$ & $2.28$ & 1.97 & $0.313$  & $-0.0093$ & $-0.0054$&  0.0039 & $\hphantom{-}0.013 \pm 0.0116$\\
        4   & $0.7 < z_{\rm B} \leq 0.9$& $0.19$ & $1.57$ & 1.39 & $0.293$ & $-0.0050$ &$-0.0069$&  0.0042 & $\hphantom{-}0.011 \pm 0.0084$\\
        5   & $0.9 < z_{\rm B} \leq 1.2$&$0.20$ & $1.74$ & 1.35 & $0.330$  & $-0.0091$ &$-0.0036$&  0.0057 &  $-0.006 \pm 0.0097$ \\
        \hline
        1-5 & $0.1 < z_{\rm B} \leq 1.2$& $0.19$ & $7.77$ & 6.69 & $0.306$ & -0.0104 & -0.0089& 0.0018 & N/A   \\
        \hline
        
    \end{tabular}
    }
    \tablefoot{This is the summary statistics of the KiDS-1000 metacal catalogue with the selections made for the cosmic shear cosmological analysis (comparable to the KiDS-1000-v2 catalogue \citep{Li2023b}). The m bias (gold) was recalculated based on the selections made during SOM photo-z calibration and AlphaRecal (the residual PSF correction). Through the extra selections, $\sim25\%$ of effective number density is reduced (see Table~\ref{tab:mbias}). The total source number density ($n_{\rm eff}$) is 7.77~${\rm arcmin}^{-2}$ for the effective area 771.90 ${\rm deg.}^2$ The effective source numbers were calculated based on the shear dispersion ($\sigma_\gamma$), ~\cref{eqn:shear_dispersion}. Compared to the \lensfit catalogue, total $n_{\rm eff}$ increased by $\sim16\%$.}     
\end{table*}

Once the final sample is selected, we apply the AlphaRecal step to remove the residual PSF leakage (described in \cref{sec:psf_leakage}). There was no need to filter out objects with extreme shears after this correction. The same procedures were applied to the SKiLLS data to derive the final
values for $m$ for each tomographic bin. We report these as $m_{\rm final}$ in \cref{tab:mbias_final}.

\subsection{Source redshift distributions}
\label{sec:nz_estimation}

To assign galaxies to a tomographic bin, we used the photometric redshift, $z_{\rm B}$, determined by BPZ \citep{Benitez2000}. 
However, \metacal yields a different shape catalogue with modified shear weights. As a result, the corresponding source redshift distributions are different from those used by \citep{Li2023b}. To account for this, we used the same redshift calibration methodology as described in \cite{Hildebrandt2021, vdBusch2022}. We used the \metacal weighting defined by \cref{eqn:weight_inv_var} but ignored the re-weighting of the shears caused by the redshift dependence of $R^\gamma$ and $m$, the factor (1+$m_i$)$R_{ii}^\gamma$ described in \cref{eq:raw_estimator}. 
Fully accounting for the latter would require changing the redshift calibration pipeline, while the impact is negligible for the first five tomographic bins.\footnote{For reference, we computed the impact using SKiLLS and report the resulting mean redshifts in \cref{tab:mbias}. This shows that we can safely ignore the additional redshift weighting of the shear for KiDS-1000 when using the first five bins. As discussed in \cref{sec:shear_calibration} we attribute the change in the first bin to the limitations of the input catalogue of SKiLLS.}

\begin{figure}
\centering
\includegraphics[width = 0.49\textwidth]{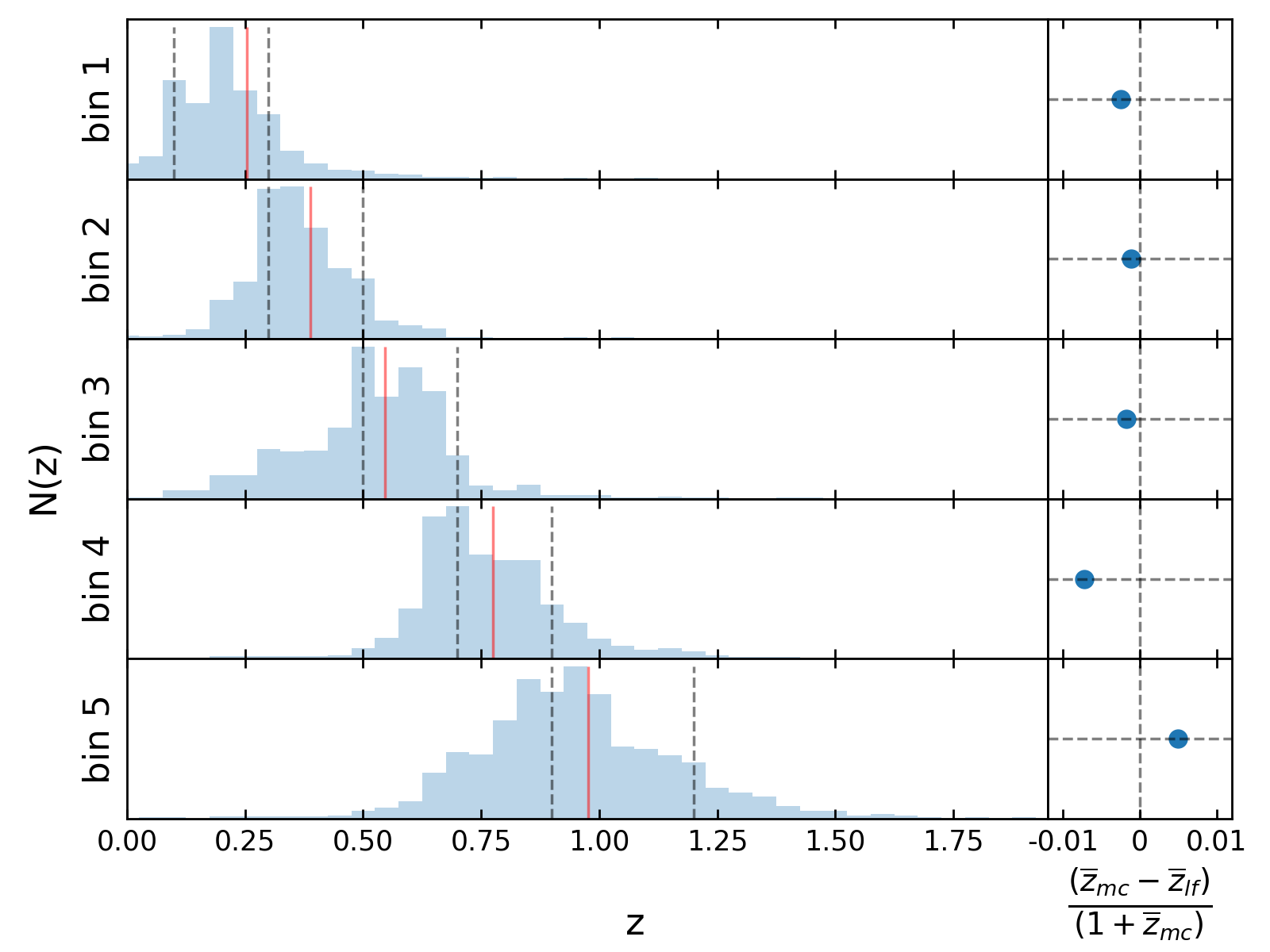}
\caption{Redshift distributions of KiDS-1000 \metacal (left panels) for each tomographic bin. Sources were selected based on their $z_{\rm B}$, with the bin limits indicated by the dashed vertical lines. The vertical red line corresponds to the mean redshift of the bin. The change with respect to the \lensfit catalogue is shown on the right. The mean redshifts are similar.}
\label{fig:nz_kids}
\end{figure}

The calibration of the bin-wise redshift distributions follows the method developed in \cite{Wright2020}. A SOM was trained on a compilation of spectroscopic datasets using the $r$-band magnitude and colours derived from the nine KiDS photometric bands, resulting into 37 input entries for the SOM. For the purpose of consistency with previous work, we adopted the SOM trained by \cite{vdBusch2022} on an extended compilation of spectroscopic data (see appendix~B therein). We then parsed the KiDS source sample into the SOM and identify regions of the colour space that are not sufficiently covered and represented by the calibration data \citep[see][Sect.~3.2]{vdBusch2022}. We removed all galaxies in these parts of the colour space, per tomographic bin, sacrificing a small fraction of sources for a more robust redshift calibration of the remaining `gold-sample' selected based on `gold-flag'. 

After applying the gold selection, we obtained the final redshift estimate for each tomographic bin by computing the weighted redshift distribution of the spectroscopic calibration data. These so-called SOM weights are constructed by weighting the calibration data in each SOM cell such that their number density matches the \metacal. As shown in \cref{fig:nz_kids}, mean redshifts of the five tomographic bins are very similar between \metacal and \lensfit. The deviation of mean redshifts for each tomographic bin from the \lensfit values are -0.003, -0.002, -0.003, -0.013, 0.010, respectively. Among them, the fourth tomographic bin deviates the most, with  $dz/(1+\bar{z})= -0.007$. An important aspect of the redshift distribution estimation and calibration procedure within KiDS is the incorporation of informative priors on the bias of each tomographic bin's $\langle{\rm z}\rangle$. In \cite{Hildebrandt2021}, these informative priors are derived for each tomographic bin using MICE2 simulations \citep{Fosalba2015, Fosalba2015b}, and are sensitive to the make-up of the wide-field shear sample and the sample of calibrating sources used to estimate the $N(z)$. 

Since we used a different wide-field sample (the galaxies are selected and weighted using \metacal, rather than \lensfit), the biases estimated in \cite{Hildebrandt2021} are not directly applicable to the source galaxies here. We did not recompute the $N(z)$ bias parameters using the simulations for this work. Instead, we adopted the bias estimates and prior ranges from \cite{Hildebrandt2021}, listed in the last column of \cref{tab:mbias_final}. We confirmed that the impact of the priors on the cosmological analysis is insignificant by checking the posteriors of the photo-$z$ shift parameters, which turned out not to correlate with the $S_8$ constraint. In the future analysis of KiDS-Legacy with \metacal, this will be improved through the recomputation of bespoke bias parameters for the galaxy sample. 

The effective number density $n_{\rm eff}$, average shear response $\langle R^\gamma\rangle$, and shear variance $\sigma_{\gamma,i}$, are reported in Table~\ref{tab:mbias_final} for each tomographic bin.
We also present results for the joint \metacal sample, which yields $n_{\rm eff}=7.77\text{arcmin}^{-2}$, 
$\sim$16\% larger than \lensfit. Compared to the results reported in \cref{tab:mbias}, the improvement in number density is somewhat reduced, because the SOM (\texttt{Gold}) selection preferentially removes faint sources. As shown in \cite{Wright2025a}, this can be alleviated by using a weight rather than a binary selection for the faint sources.

\subsection{Cosmic shear two-point statistics}

The cosmic shear signal can be quantified using the two-point shear correlation functions,  $\xi^{ij}_\pm(\theta)$, for samples of galaxies in tomographic bins $i$ and $j$, separated on the sky by an angle $\theta$. This can be estimated from the shape catalogue using 
\begin{equation}
\label{eq:2ptcorr}
    \hat{\xi}^{ij}_{\pm}(\theta)=\frac{\sum_{ab}w_aw_b\left[e_t^i(\bm{x}_a)e_t^j(\bm{y}_b)\pm e_{\times}^i(\bm{x}_a)e_{\times}^j(\bm{y}_b)\right]}{\sum_{ab}w_aw_b R^\gamma_a R^\gamma_b}~,
\end{equation}
where the weight $w$ is the shear weight given by
\cref{eqn:weight_inv_var}. The tangential and cross ellipticities $\epsilon_{t, \times}$ are computed with respect to the vector $\bm{x}_a - \bm{y}_b$ that connects the galaxy pair $(a,b)$. To compute $\hat{\xi}^{ij}_{\pm}(\theta)$ for our analysis we used \texttt{TreeCorr}\footnote{https://rmjarvis.github.io/TreeCorr/\_build/html/correlation2.html} \citep{Jarvis2004}.

For the cosmological inference, we used the complete orthogonal sets of $E/B$-integrals \citep[COSEBIs;][]{Schneider2010}. These provide a convenient way to compress the information contained in $\xi_\pm(\theta)$, and cleanly decompose the signal into  $E$- and $B$-modes, which correspond to curl-free gradient distortions and curl distortions, respectively. The COSEBIs are given by
\begin{align}
E_n &= \int_{\theta_{\min}}^{\theta_{\max}} \diff \theta\, \theta \left[ T_{+n}(\theta)\, \xi_{+}(\theta) + T_{-n}(\theta)\, \xi_{-}(\theta) \right] , \label{eqn:Emode}\\
B_n &= \int_{\theta_{\min}}^{\theta_{\max}} \diff \theta\, \theta \left[ T_{+n}(\theta)\, \xi_{+}(\theta) - T_{-n}(\theta)\, \xi_{-}(\theta) \right] .
\end{align}
Here, $T_{\pm n}(\theta)$ are filters defined over a chosen angular range ($\theta_{\mathrm{min}}$, $\theta_{\mathrm{max}}$). We used the range $2-300$ arcmin, and the maximum COSEBI mode was chosen to be $n_{\rm max}=5$, following the setup of \cite{Li2023b} for consistency.
This choice of scale range reduces the impact of the uncertainty in modelling the baryonic feedback on scales smaller than 2 arcmin.

The measured COSEBIs using our \metacal catalogue are presented in \cref{fig:2pt_measurement_metacal} for the various tomographic bin combinations. The blue lines in the panels showing $E_n$ indicate the best-fit model to the measurements, with a reduced $\chi^2\simeq1.2$ (see below for more details). The uncertainties in the measurements are estimated with an analytical covariance, whose recipe is described in more detail in \citet{Asgari2020,Joachimi2021}. Although the analysis of KiDS DR5 \citep{Wright2024} employs a new covariance code by \citet{Reischke2024}\footnote{\href{https://github.com/rreischke/OneCovariance}{https://github.com/rreischke/OneCovariance}}, we continued to use the previous code as it is implemented in the KiDS-1000 pipeline. The two codes have been verified to be consistent with both the new implementation and simulation results. In the implementation used in this work, the main difference is the idealised treatment of the survey geometry in the mixed term and the model for non-linearities. These effects, however, were found to not affect the cosmological results. We therefore can safely use the old covariance code for the analysis here. For completeness, we also carried out a single run with the updated covariance matrix, as discussed in \Cref{sec:cosmology} (see \cref{fig:internal_consistency}).

The covariance consists of four distinct components: the super-sample covariance, which arises from modes outside the survey volume; the non-Gaussian covariance, which is due to the non-linearity of structure formation; the multiplicative shear bias uncertainty; and the disconnected (Gaussian) part. The first two components are calculated using the halo model \citep[see][for a review]{Asgari2023}. For the Gaussian covariance, we specifically account for the exact impact of the mask and weights (e.g. shape measurement weights) in the shape noise term by directly using the weighted pair counts. All other terms are projected from the harmonic space covariance, $C(\ell)$, to the COSEBIs through \cref{eq:cosebis_e_modes}. For brevity, we direct readers to \citet{Reischke2024} for more detailed information on the implementation and formalism.

\subsection{COSEBIs $B$-mode null test}

\begin{figure*}[t]
\centering
\includegraphics[width = 0.8\textwidth]{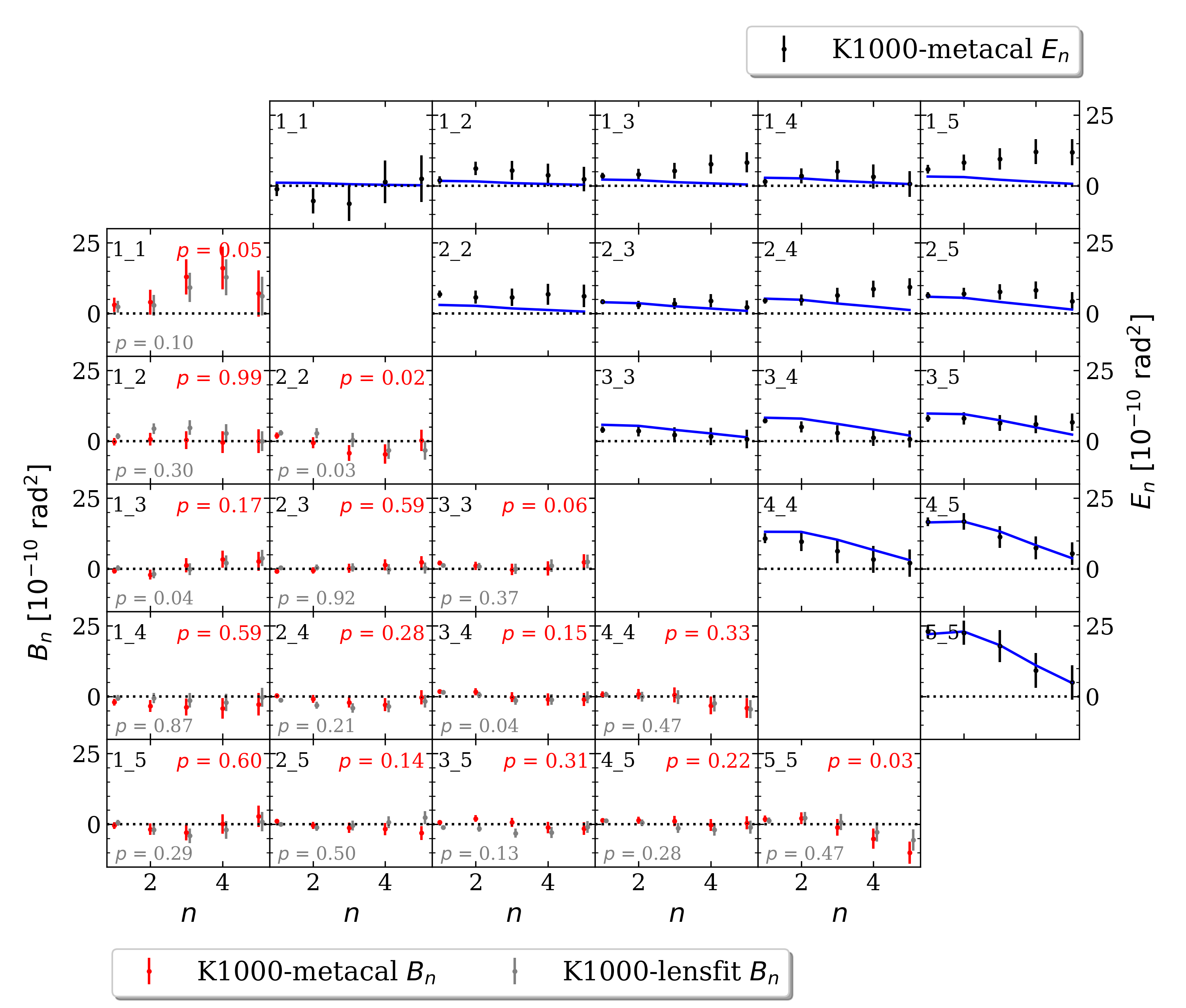}
\caption{Measurements for the first five COSEBI $E_n$ (black) with the best-fit theory lines (blue) and $B_n$ (red) measurements with their $p$-values (red) from the KiDS-1000 \metacal catalogue, converted from the real scale signals, $\xi_{\pm}(\theta)$ with the scale range, [2.0, 300.0] arcmin. As a comparison, the \lensfit $B_n$ measurements were denoted in grey with each corresponding $p$-value (grey). The theory with the best-fit values from the fiducial cosmology analysis is fitted with the $\chi^2 \simeq 87.5$.}
\label{fig:2pt_measurement_metacal}
\end{figure*}

To first order, weak lensing introduces correlations in the galaxy shapes that are curl-free. Some real-life effects, such as source redshift clustering, complicate this simple picture \citep[e.g.][]{Schneider2002}. But we can safely ignore these for our analysis \citep{Linke2025}. Hence, we expect $B_n$ to be consistent with zero, thus providing a non-trivial test for residual systematics.

The measurements for $B_n$ are shown as the red points in \cref{fig:2pt_measurement_metacal}, as well as the corresponding $p$-values of the null-test. We used the same range in angular scales ($2-300$ arcmin) and maximum COSEBI mode ($n_{\rm max}=5$) as used to compute $E_n$ (black points). This differs from the setup used by \cite{Li2023b} for this test.
Therefore, we also recalculated the $B_n$ for \lensfit for the same scale range and the number of COSEBI modes. The results are shown by the grey points. For our \metacal measurements, the smallest, $p=0.02$ is found for the auto-correlation of the second tomographic bin (we consider the value, 0.02, acceptable), where the contamination from stars is largest (check \cref{app:star_contamination} for the details of star contamination). Although the other auto-correlations also tend to have small $p$-values, the other bin combinations show acceptable values. Moreover, the 
$p$-values are overall comparable to the \lensfit results,
while the estimated uncertainties are in fact smaller for \metacal. Hence, the robustness of the \metacal signal is at least comparable to that of \lensfit. Following previous papers, we also checked the B-modes for $n_{max} =20$.
In this case, we typically find larger $p$-values, with the exception of 
the correlation of bin 2 and 5 ($p=0.01$).  We consider this acceptable.
 
\subsection{Accuracy requirement and validation of the first-order systematics model}
\label{sec:xisys}

\begin{figure}
\centering
\includegraphics[width = 0.5\textwidth]{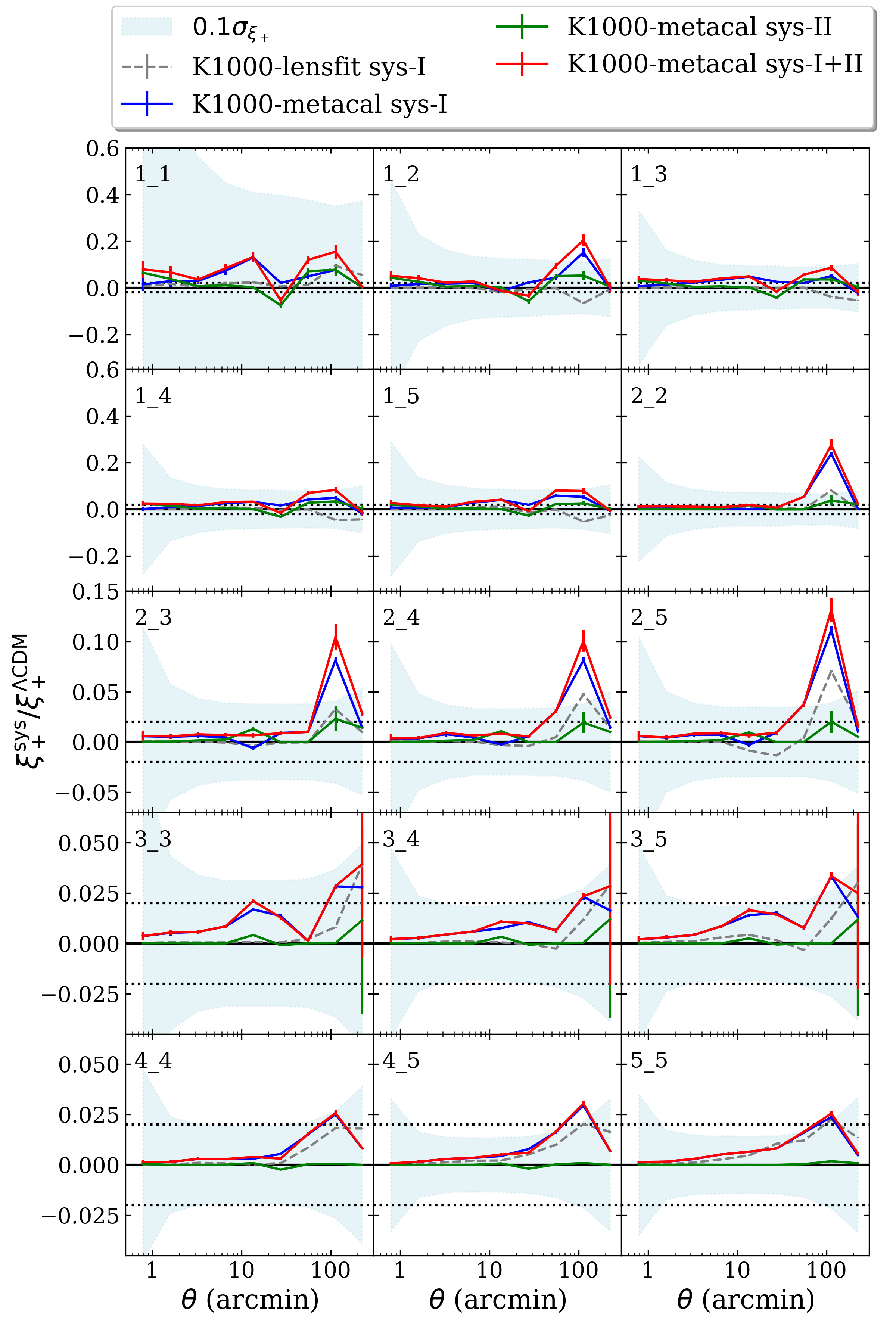}
\caption{Ratio of the PSF systematic contribution to the amplitude of the cosmic shear signal with the fiducial cosmology. The red line shows $\xi_+^{\rm sys}$, as defined by \cref{eq:xisyst}. For comparison to previous work
\citep{Giblin2021}, we also show the first term of this equation (grey and blue for \lensfit and \metacal, respectively).
For completeness, the green line shows the second term of \cref{eq:xisyst} for the \metacal measurements. These results show that the systematic signals from \metacal are an order of magnitude smaller than $\sigma_{\xi_+}$, the uncertainty in the cosmic shear signal.}
\label{fig:Csys_metacal}
\end{figure}

The PSF anisotropy is the dominant source of additive bias. Although the measurements of $B_n$ provide a useful test, any curl-free contributions would not be captured, or underestimated \citep[see e.g.][]{Hoekstra2004}. Therefore, we also directly examined the contribution from PSF systematics to  $\xi_+$. We began from \cref{eqn:alpha} and computed the contribution to the two-point ellipticity correlation function arising from PSF systematics. This yields
\begin{equation}
\xi^{\rm sys}_+ =  \frac{\left\langle \epsilon^{\rm obs} \epsilon ^{\rm PSF} \right\rangle^2}{ \left\langle \epsilon^{\rm PSF} \epsilon^{\rm PSF}\right\rangle}+
\cfrac{\left[ 
\langle \epsilon^{\rm obs} \delta \epsilon^{\rm  PSF}\rangle -\cfrac{\langle  \epsilon^{\rm obs}\epsilon^{\rm  PSF}\rangle \langle  \epsilon^{\rm  PSF} \delta \epsilon^{\rm  PSF}\rangle}{\langle \epsilon^{\rm  PSF} \epsilon^{\rm  PSF}\rangle} \right]^2}{\left[{\langle \delta \epsilon^{\rm  PSF} \delta \epsilon^{\rm  PSF}\rangle - \cfrac{\langle  \epsilon^{\rm  PSF} \delta \epsilon^{\rm  PSF}\rangle^2}{\langle \epsilon^{\rm  PSF} \epsilon^{\rm  PSF}\rangle}} \right]}.
\label{eq:xisyst}
\end{equation}
The first term corresponds to the expression derived by \cite{Bacon2003} and used by \cite{Giblin2021} to examine the PSF systematics in KiDS. However, \cref{eq:xisyst} accounts for all contributions arising from PSF modelling errors. 

In \cref{fig:Csys_metacal} we compare $\xi^{\rm sys}$ to the predicted cosmic shear signal for the tomographic bin combinations we use for the cosmological inference. 
The red line shows the ratio of $\xi^{\rm sys}$ and the cosmic shear amplitude predicted for our fiducial cosmology.
We find that the amplitude of $\xi^{\rm sys}$ is well
within $10\%$ of the uncertainty in the observed $\xi_+$ (indicated by the light blue regions) for most scales and tomographic bin combinations. For comparison to previous KiDS studies, we also show the first term of \cref{eq:xisyst} for both our \metacal measurements (red line) and \lensfit (dashed black line), while the green line shows the second term for the \metacal catalogue.

As is the case for previous KiDS analyses, we do observe an increase around 100 arcminutes, which coincides with the range where the denominator of the first term in \cref{eq:xisyst} almost vanishes. The deviation is largest for the second tomographic bin, which may be impacted by our suboptimal selection of stars using the \lensfit star flag (see \cref{app:star_contamination}).\footnote{Here, we focus on additive bias. However, stars also lower the lensing signal, since they are not lensed. Moreover, stars are down-weighted, because \metacal should yield $R^\gamma=0$. In practice, their contribution may not vanish completely because of the noise in the measurements. However, as SKiLLS samples the distribution of stars across the survey, this residual contribution is accounted for in our estimates of $m$.}

To explore this further, we examined possible contributions to the numerator. If the fraction of stars is $f_{\rm star}$, we have
\begin{align}
\langle \epsilon^{\rm obs} \epsilon^{\rm PSF}\rangle  & =  (1-f_{\rm star})  \alpha_{\rm gal} \left\langle \epsilon^{\rm PSF} \epsilon^{\rm PSF}\right\rangle +
f_{\rm star} \alpha_{\rm star} \left\langle \epsilon^{\rm PSF} \epsilon^{\rm PSF}\right\rangle \notag\\
 & ~~~~ + \left\langle \delta \epsilon^{\rm PSF}\epsilon^{\rm PSF}\right\rangle\notag\\
& =  \alpha \left\langle \epsilon^{\rm PSF}\epsilon^{\rm PSF}\right\rangle + \left\langle \delta \epsilon^{\rm PSF}\epsilon^{\rm PSF}\right\rangle,
\label{eq:leakage}
\end{align}
where we explicitly distinguished between the PSF leakage for a galaxy and that of a star in the first step. This suggests that PSF leakage is naturally accounted for using our empirical `AlphaRecal' procedure, as we expect $\alpha\approx 0$. 

Interestingly, we find that the amplitude at $100'$ is higher for pointings with low star densities. For such pointings, the PSF model itself is noisier, thus supporting the notion that the bump in $\xi_+^{\rm sys}$ is caused by residual PSF modelling errors. However, given the negligible impact on the cosmological results, further exploration is beyond the scope of the paper.

\section{Cosmological results}
\label{sec:cosmology}

Our inference of cosmological parameters follows the previous KiDS-1000 analyses closely \citep[for details, see][]{Asgari2021}. Consequently, we only summarise the most salient aspects here.
The cosmological information is contained in the values for $E_n$, \cref{eqn:Emode}, presented in \cref{fig:2pt_measurement_metacal}.
These are related to the $E$-mode angular power spectrum, $C_{EE}$, through 
\begin{equation}
\label{eq:cosebis_e_modes}
E_n = \int_{0}^{\infty} \frac{ \diff \ell\, \ell }{2\pi} C_{EE}(\ell)\, W_n(\ell) ,
\end{equation}
where the weight functions $W_n(\ell)$ are Hankel transforms of $T_\pm (\theta)$ \citep{Asgari2012}.

$C^{ij}_{EE}(\ell)$ for the $i$th and $j$th tomographic bins is the sum of three terms that capture the contribution from gravitational lensing (G), intrinsic alignments of galaxies (I) and their cross-correlation:
\begin{equation}
C^{ij}_{EE}(\ell)=C^{ij}_{\rm GG}(\ell)+C^{ij}_{\rm GI}(\ell)+C_{\rm II}^{ij}(\ell).
\end{equation}
Using the Limber approximation \citep{Loverde2008, Kilbinger2017}, these angular power spectra can be computed using the matter power spectrum, $P_{\mathrm{m}}(\ell,\chi)$, through
\begin{equation}
C^{ij}_{\mathrm{XY}} (\ell) = \int_{0}^{\chi_{\mathrm{H}}} \diff \chi\, \frac{ q^i_{\rm X}(\chi) q^j_{\rm Y}(\chi) }{ f_K^2(\chi) } P_{\mathrm{m}} \left( \frac{\ell + 1/2}{f_K(\chi)}, \chi \right) \,.
\label{eqn:c_gg}
\end{equation}
where X and Y stand for G or I, $\chi$ is the radial comoving distance that is integrated from the observer ($\chi =0$) to the horizon ($\chi =\chi_{\mathrm{H}}$), and $f_K(\chi)$ is the comoving distance. The kernels, $q_{\rm X}^{i}$, depend on the source redshift distribution, $n_i(\chi)$. The lensing kernel is given by 
\begin{equation}
q_{\rm G}^{i}(\chi)=\frac{3H^2_0 \Omega_{\mathrm{m}}}{2c^2} \int_{\chi}^{\chi_H} \diff\chi' n^i(\chi')\frac{f_K(\chi' - \chi)}{f_K(\chi')}\,,
\end{equation}
and for the intrinsic alignment we chose a kernel that produces the NLA model \cite{Bridle2007}
\begin{equation}
q_{\rm I}^i(\chi)=    - A_{{\mathrm{IA}}} C_1 \rho_{{\mathrm{crit}}} \frac{\Omega_{\mathrm{m}}}{D_+(z)} \left( \frac{1+ z}{1+z_0}\right)^\eta n^i(\chi),
\label{eqn:nla_model}
\end{equation}
with $A_{{\mathrm{IA}}}$ the intrinsic alignment parameter. The other parameters are $C_1 = 5 \times 10^{-14}\;h^{-1}\,{\mathrm{Mpc}}^3$, $\rho_{{\mathrm{crit}}}$ is the critical density today, and $D_+(z)$ is the linear growth factor. 

As our fiducial model, we adopt $\eta = 0$ and a pivot redshift $z_0=0.3$. Moreover, to ensure a direct comparison with \cite{Li2023b}, we use the same prior choices for the twelve free parameters (cosmological and nuisance parameters), which are summarised in \cref{tab:cosmo_priors}. 

The uncertainties in $m$ for each tomographic bin were included in the covariance matrix, while the two-point statistics were corrected using the value of $m$. The photo-$z$ calibration parameters from the clustering redshift ($\delta_{zi}$) were used as priors for the Markov Chain Monte Carlo (MCMC) runs. We used 
\texttt{polychord} \citep{Handley2015a, Handley2015b} as the fiducial sampling method for the MCMC runs, while the consistency tests in \cref{sec:interntal_consistency_test} used \texttt{multinest} \citep{Feroz2009} to speed up the computations. 

The main differences between the original KiDS-1000 analysis by \cite{Asgari2021} and the revision by \cite{Li2023b} pertain to changes in the baryonic feedback and intrinsic alignment models. \cite{Asgari2021} used the halo-model parametrisation from \cite{mead2015}.
Hence, we followed \cite{Li2023b} and used 
an updated baryonic feedback model  \citep[HMCode-2020;][]{Mead2021} based on the temperature of the active galactic nucleus (AGN) feedback ($T_{\rm AGN}$) derived from a suite of large-volume hydrodynamic simulations, BAryons and HAloes of MAssive Systems \citep[BAHAMAS;][]{McCarthy2017}. 

\cite{Asgari2021} used a wide prior $A_{\rm IA} \in$ [-6.0, 6.0], while much of this prior range is excluded by models that aim to account for available observational constraints \citep[e.g.][]{Fortuna2021}. Instead, \cite{Li2023b} adopted a prior $A_{\rm IA} \in$ [-0.2, 1.1], which is considerably narrower. In general, it is preferable to reduce the prior ranges whenever possible, to avoid issues with sampling the high-dimensional parameter spaces. 
We investigate the impact of the different choices for these baryonic feedback models and intrinsic alignment priors in \cref{sec:interntal_consistency_test}.

\subsection{Cosmological constraints from the \metacal pipeline}
\label{sec:constraints}

\begin{table}
    \centering
    \caption{\metacal fiducial cosmology priors}
    \label{tab:cosmo_priors}
    {\renewcommand{\arraystretch}{1.2}
    \begin{tabular}{cc}
        \hline\hline
        Parameter & Prior range \\
        \hline
        $S_8$ & $\bb{0.1,\,1.3}$ \\
        $\omega_{\rm c}$ & $\bb{0.051,\,0.255}$ \\
        $\omega_{\rm b}$ & $\bb{0.019,\,0.026}$ \\
        $h$ & $\bb{0.64,\,0.82}$ \\ 
        $n_{\rm s}$ & $\bb{0.84,\,1.1}$ \\

        $\log_{10}(T_{\rm AGN}[{\rm K}])$ & $\bb{7.3,\,8.0}$ \\ 
        $A_{\rm IA}$ & $\bb{-0.2,\,1.1}$ \\
        $\delta z_{i}$ & ${\cal N}(\vec{\mu_{z_i}};{\mathbf{C}_{\delta z}})$ \\
        \hline

    \end{tabular}
    }
    \tablefoot{The redshift shift ($\delta z_{i}$) for each tomographic bin, ${\cal N}(\vec{\mu_{z_i}};\mathbf{C}_{\delta z})$, refers to correlated priors among tomographic bins with means $\vec{\mu_{z_i}}$ and covariance $\mathbf{C}_{\delta z}$ reported in \cref{tab:mbias_final} and \cite{Hildebrandt2021}.}
\end{table}

The blue lines in \cref{fig:2pt_measurement_metacal} show the best-fit model to the measurements to $E_n$. The model fit with the final best-fit parameters is fair with $\chi^2=87.5$. For reference, the best fits using the \lensfit measurements correspond to $\chi^2=62.7$ \citep{Li2023b} with the same degrees of freedom (DoF) $\gtrsim 75 - 4.5$. Due to the narrower $A_{IA}$ prior that affected its constraint, the DoF is estimated slightly bigger than the previous KiDS studies, \cite{Asgari2021, vdBusch2022} with $DoF = 75 - 4.5$ that yielded $\chi^2=82.2$ and $\chi^2=63.2$, respectively.

\begin{figure}[h]
\centering
\includegraphics[width = 0.48\textwidth]{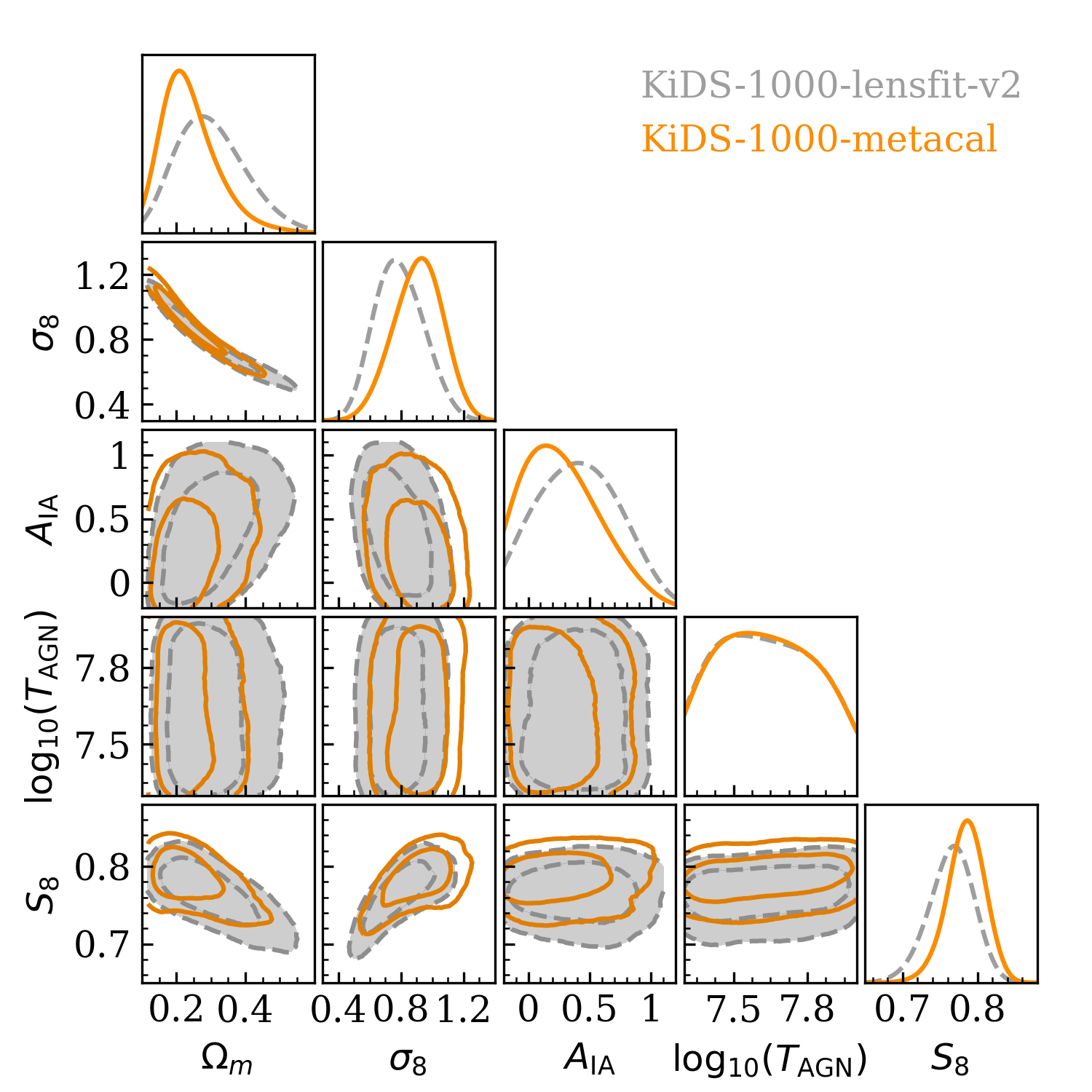}
\caption{\metacal cosmology compared to \lensfit, both from the fiducial setting using the \texttt{polychord} sampler and priors described in \cref{tab:cosmo_priors}. The parameter constraints are summarised in \cref{tab:cosmo_constraints} below. While the results from \lensfit and \metacal are consistent, the constraint on $S_8$ from \metacal is 28\% tighter than that from \lensfit and 0.5$\sigma$ closer to the Planck value.}
\label{fig:cosmology_comp_with_lensfit}
\end{figure}

We compared the posterior distributions for the parameters listed in \cref{tab:cosmo_constraints} for the \metacal and \lensfit analyses in \cref{fig:cosmology_comp_with_lensfit}. The results agree well, with \metacal yielding slightly tighter constraints. Compared to \lensfit, the \metacal results prefer a slightly higher $S_8\equiv \sigma_8 \sqrt{(\Omega_m/0.3)}$  with a lower value for $\Omega_m$, as the degeneracy between $\Omega_m$ and $\sigma_8$ steepens for lower $\Omega_m$.
Although both analyses are based on the same multi-band imaging data, and we attempted to stay close to the choices made by \cite{Li2023b}, we note that achieving this level of consistency is not trivial. For one, the \metacal catalogue includes more galaxies, especially for the high-redshift bins, while the shear calibration is quite different. Hence, the agreement is an important validation of the robustness of the calibration pipeline.

The inferred value for $A_{\rm IA}$ depends on the sample selection \citep{Fortuna2021}, but may also differ between shape measurement algorithms. For our \metacal measurements, we adopted a rather compact weight function, which minimises the sensitivity to the outer regions of galaxies, which may be more strongly aligned \citep{Georgiou2019}. Interestingly, we find a lower value for $A_{\rm IA}$ compared to \cite{Li2023b}. However, a more quantitative exploration of the choice of weight function and the constraints on $A_{\rm IA}$ is beyond the scope of this paper.

\begin{table}
    \centering
    \caption{\metacal fiducial parameter constraints.}
    \label{tab:cosmo_constraints}
    {\renewcommand{\arraystretch}{1.6}
    \begin{tabular}{cccc}
    \hline\hline    
    Parameter & MAP (fiducial) & Mean & Max\\
    \hline
    \multicolumn{4}{c}{Cosmological Parameter}\\
    \hline
    $S_8$ & $0.789_{-0.024}^{+0.020}$ &  $0.784_{-0.021}^{+0.024}$ & $0.787_{-0.025}^{+0.023}$ \\
    $\Omega_\mathrm{m}$ &$0.167_{-0.041}^{+0.097}$ & $0.240_{-0.090}^{+0.039}$ & $0.203_{-0.058}^{+0.086}$\\ 
    $\sigma_8$ & $1.057_{-0.220}^{+0.155}$& $0.908_{-0.136}^{+0.154}$ & $0.928_{-0.163}^{+0.142}$\\
    $\Sigma_8$ & $0.777_{-0.026}^{+0.018}$ & $0.778_{-0.020}^{+0.022}$ & $0.781_{-0.025}^{+0.020}$\\ 
    \hline
    \multicolumn{4}{c}{Astrophysical Parameter}\\
    \hline
    $A_{\rm IA}$ & $-0.105_{-0.095}^{+0.499}$ &$0.263_{-0.411}^{+0.188}$ &$0.092_{-0.228}^{+0.396}$ \\
    \hline 
    \end{tabular}
    }
    \tablefoot{The second column lists the multivariable maximum posterior (MAP) and its 68\% credible interval using its projected joint highest posterior density (PJ-HPD). The third and last columns list the mean and maximum values from the marginal posterior of each parameter, respectively. $\Sigma_8$ is defined as $\sigma_8(\Omega_\mathrm{m}/0.3)^{\alpha_{\rm best}}$, similar to the definition of $S_8$ using the power of 0.5, with the best-fit value of $\alpha_{\rm best} = 0.527$ in the plane of $\sigma_8$ and $\Omega_\mathrm{m}$ along the degeneracy direction of posterior contour.
    For the $A_{\rm IA}$ constraint, only the upper bound of is trustworthy, because the lower bound is limited by the prior limit. The baryonic feedback parameter, $\log_{10}(T_{\rm AGN})$, is not constrained, because of the narrow prior range.} 
\end{table}

\begin{figure}[t]
\centering
\includegraphics[width = 0.45\textwidth]{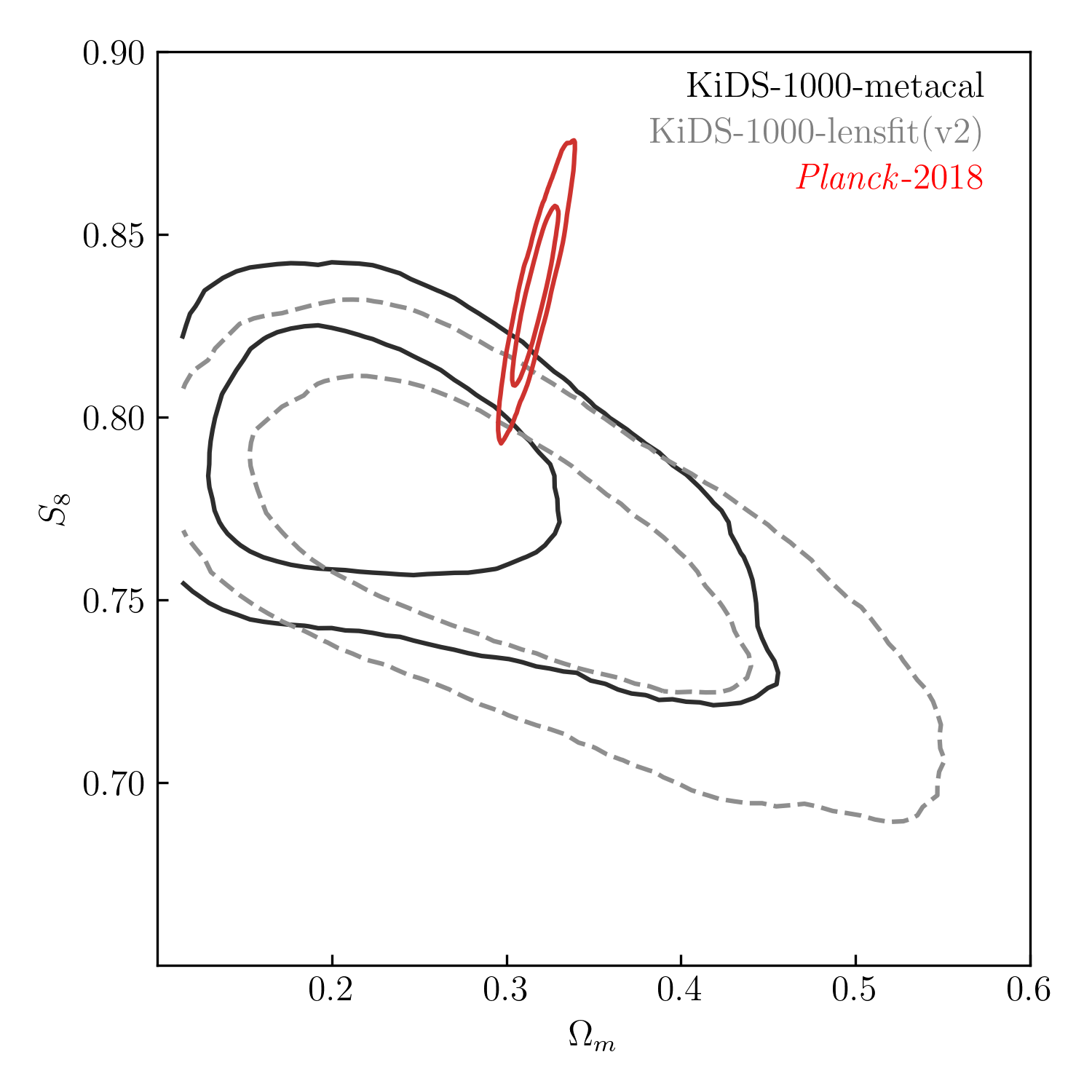}
\caption{Projected posteriors for $S_8 \equiv \sigma_8 \sqrt{\Omega_\mathrm{m}/0.3}$ and $\Omega_\mathrm{m}$) from the KiDS-1000 \metacal (black contours) and \lensfit \citep[dashed grey contours;][]{Li2023b} catalogues. The constraints agree well, but thanks to the higher source density, the \metacal measurements improve the constraints by 28\%.
The constraints prefer a lower $S_8$  than {\it Planck} (red contours),
although the difference in $S_8$ is reduced by 0.5$\sigma$ to 
1.8$\sigma$.}
\label{fig:cosmology_comp_Planck}
\end{figure}

Given the interest in the constraints on $S_8$, we present the corresponding projected 2D posterior distribution for the parameters $S_8$ and $\Omega_\mathrm{m}$ for the fiducial setup in \cref{fig:cosmology_comp_Planck}. We find $S_8=0.789_{-0.024}^{+0.020}$, using the multivariable maximum posterior (MAP) and its projected joint highest posterior density (PJ-HPD) \citep[see detailed discussion of the estimate in Sect. 6.4 of][]{Joachimi2021}. We also show the posteriors from the 
{\it Planck} CMB measurements \citep[Planck2018, TT,TE, EE + lowE + lensing;][]{PlanckParams2018}, which yield $S_8=0.832 \pm 0.013$. We also measured the value of $\Sigma_8\equiv \sigma_8(\Omega_\mathrm{m}/0.3)^{\alpha_{\rm best}}
=0.777_{-0.026}^{+0.018}$, where the best-fit value of $\alpha_{\rm best}=0.527$ was obtained from the orientation
of the $\sigma_8$ and $\Omega_\mathrm{m}$ posterior. The  corresponding {\it Planck} value is $\Sigma_8=0.835 \pm 0.016$. 
Similar to previous analysis of KiDS-1000 data \citep{Asgari2021, Li2023b} and other lensing studies \citep{Amon2022,Secco2022,XiangchongLi2023, Dalal2023}, we find a low value for $S_8$. However, compared to the findings of \cite{Li2023b}, the difference is reduced by 0.5$\sigma$ to 1.8$\sigma$, while the uncertainties are smaller thanks to the higher source number density in the \metacal catalogue. The difference remains at the 2.1$\sigma$ level when measured using $\Sigma_8$. The difference is quantified using the Hellinger tension measure \citep[see appendix G.1. of][]{Heymans2021}. 
Importantly, the good agreement between our results and those from \cite{Li2023b} suggests that the tension is not due to the shape measurements. Indeed, the most recent analysis of the KiDS-Legacy data \cite{Wright2025b} is based on \lensfit, but finds good agreement with {\it Planck} thanks to improvements in the redshift calibration and the addition of more sky area. 

\subsection{Internal consistency tests}
\label{sec:interntal_consistency_test}

\begin{figure}[th]
\centering
\includegraphics[width = 0.49\textwidth]{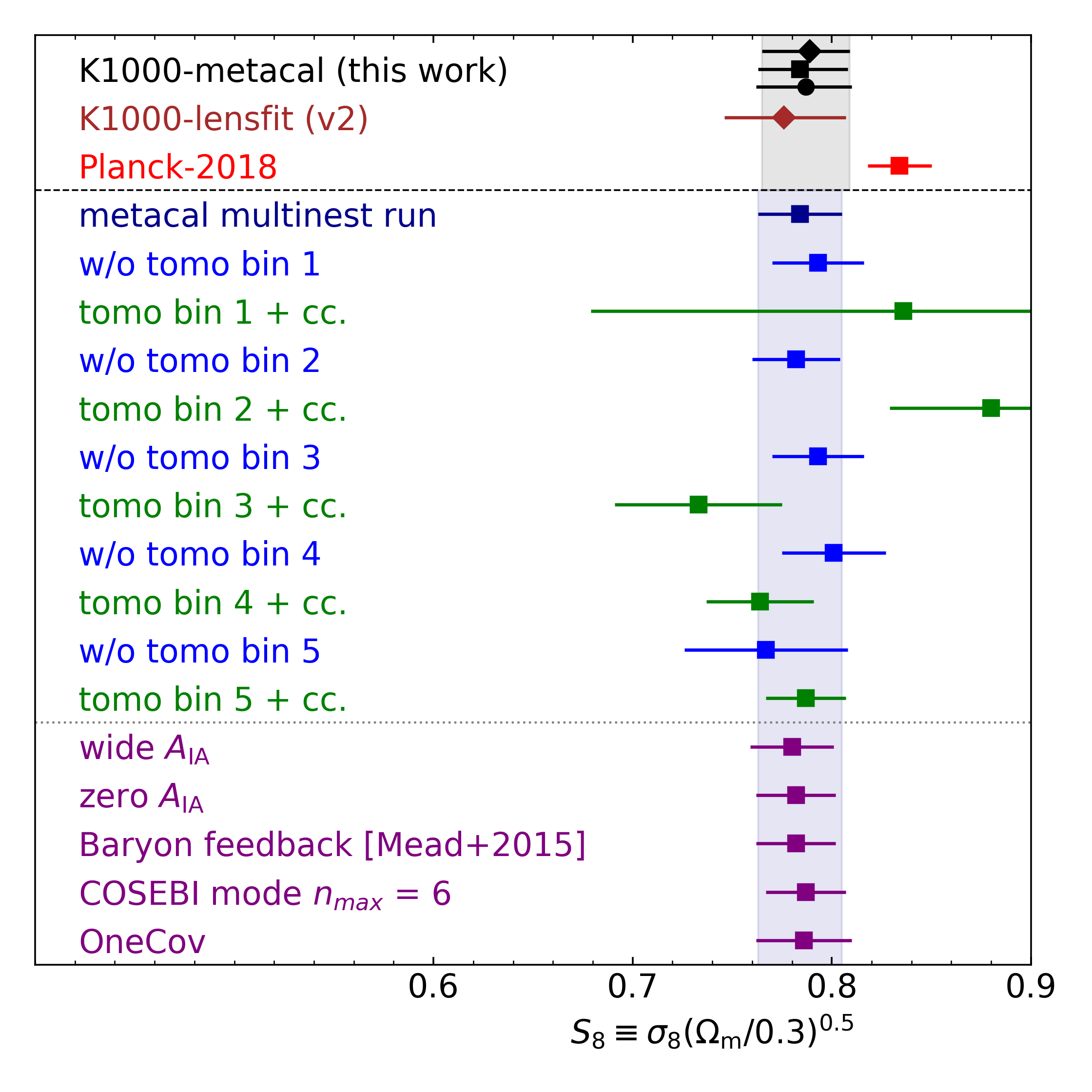}
\caption{Fiducial $S_8$ constraints from \metacal, \lensfit, and {\it Planck}, along with our \metacal internal consistency test results with different set-ups: tomo-bin splits, variation in intrinsic alignment priors, a baryonic feedback model from \cite{mead2015}, an extra COSEBI mode 6, and a new covariance estimation \citep{Reischke2024}. The consistency holds for all the varied settings, which validates the robustness of the cosmological analysis. Please note that for consistency tests, we used \texttt{multinest}, which yields slightly smaller errors than \texttt{polychord}. The diamond indicates the MAP and PJ-HPD (fiducial) estimate; the square denotes the mean of the marginal posterior; and the circle represents the maximum of the marginal posterior. }
\label{fig:internal_consistency}
\end{figure}

The fiducial setup used in the previous section reflects our current best knowledge of some of the nuisance parameters, and uses all the tomographic bins. However, it is useful to examine the robustness of the findings to some of these setup choices or specific subsets of the data. In this section we explore these, largely following what has been done in previous KiDS analyses \citep{Asgari2021, Li2023b, Stoelzner2025}. To speed up the calculations, we switched to 
\texttt{multinest} \citep{Feroz2009}, but note that it is known to underestimate the errors by 10\% compared to the fiducial run with \texttt{polychord} \citep{Handley2015a, Handley2015b}. However, this is not a concern as we are mostly concerned with overall shifts in posterior distributions. We describe each of the tests below. The findings are summarised in \cref{fig:internal_consistency},
which shows that the results are robust with respect to the fiducial setup.

\begin{figure}[th]
\centering
\includegraphics[width = 0.48\textwidth]{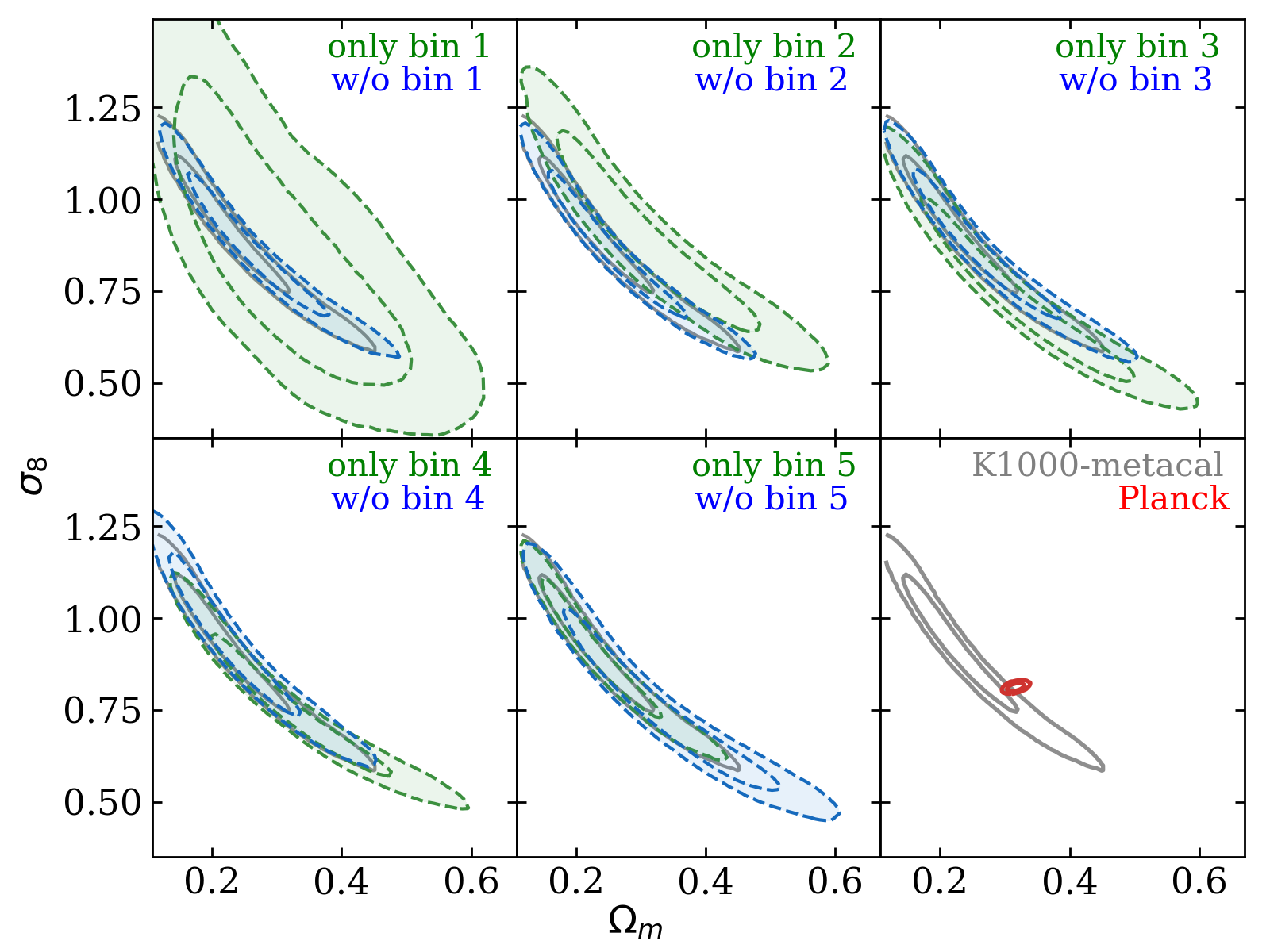}
\caption{Consistency between the different tomographic bin combinations. For each test run, the independent cosmological constraints (in addition to the baryonic feedback parameter and intrinsic alignment parameter) were obtained either by using a single bin and its cross-correlations
(green) or by excluding these (blue). The parameters shown 
($\Omega_\mathrm{m}$ and $\sigma_8$), as well as the values for $S_8$, are consistent among all the test sets. Although not significant, the second bin shows the largest deviation, because it prefers a higher intrinsic alignment contribution.}
\label{fig:tomo_consistency}
\end{figure}

We explored the constraining power of each of the tomographic bins. \cite{Asgari2021} find that the second tomographic bin yielded constraints that were somewhat inconsistent with the rest of the data (see their 
appendix~B.2 and figure~B.5), with that bin preferring 
a higher $S_8$ value. In contrast, the recent KiDS-Legacy cosmic shear analysis \citep[Sect.~4.1.1 and figure~1 in][]{Stoelzner2025} finds consistency among all tomographic bins. We therefore conducted the same test using our \metacal catalogue. To do so, we ran the cosmological inference pipeline for each tomographic bin and its cross-correlations with the other bins to ensure sufficient constraining power. These are reported as `tomo bin X + cc'. 
in \cref{fig:internal_consistency}, while the joint posteriors for $\sigma_8$ and $\Omega_{\rm m}$ are shown as green contours in \cref{fig:tomo_consistency}. We also show the results when a particular bin and its cross-correlations are excluded by the blue points and contours. These results show excellent agreement between all the bins.

\begin{figure}[t]
\centering
\includegraphics[width = 0.242\textwidth]{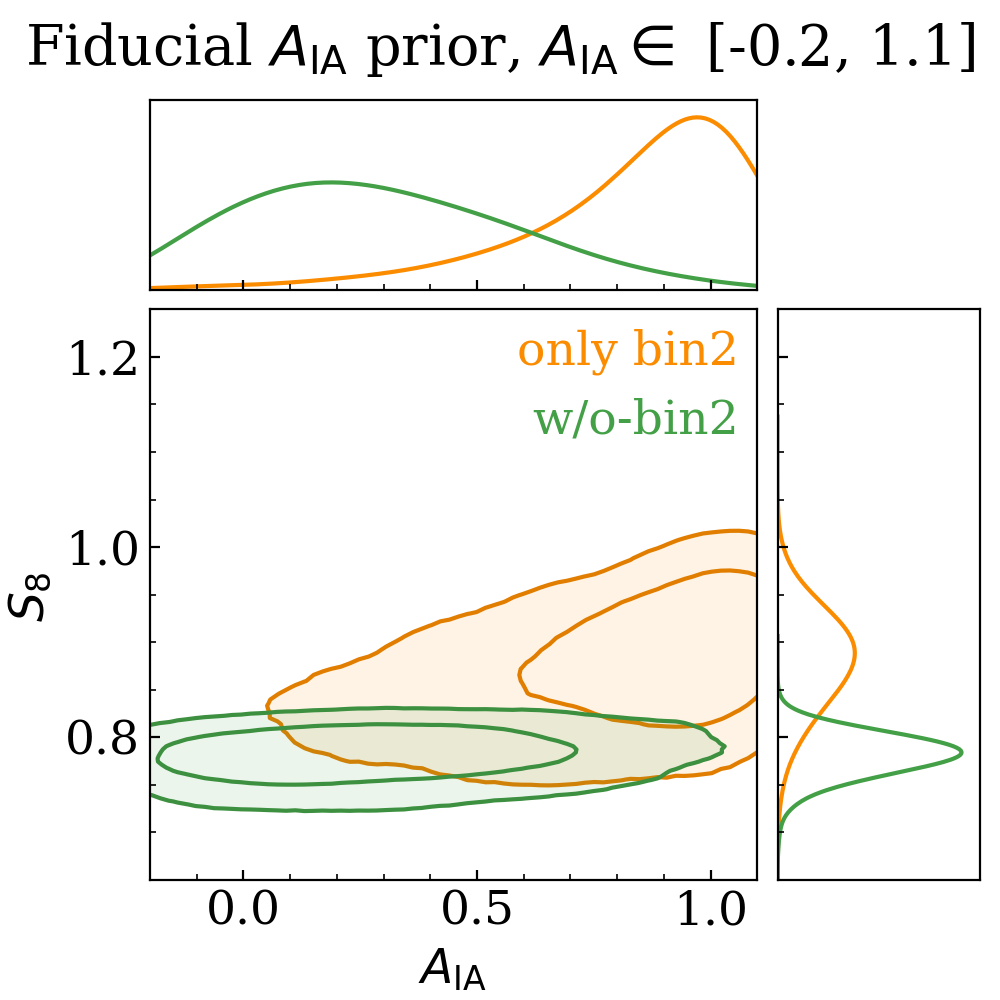}
\includegraphics[width = 0.242\textwidth]{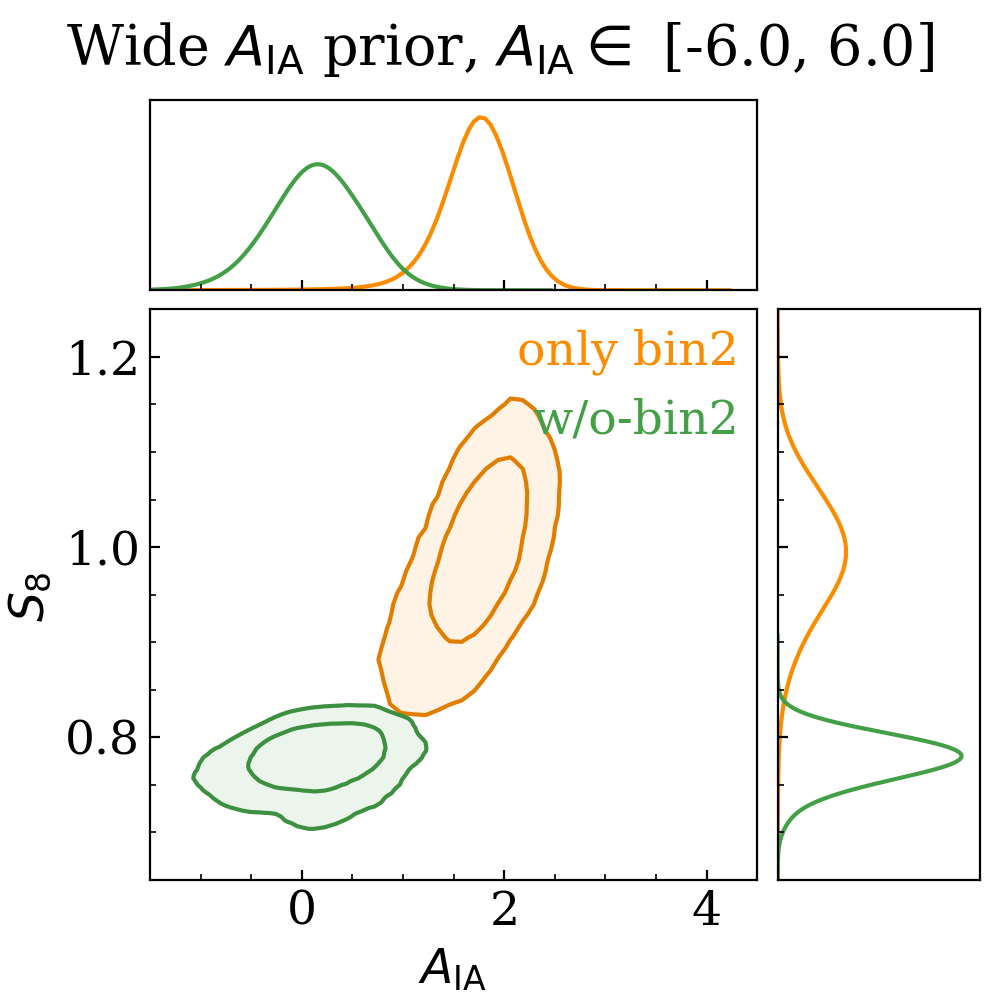}
\caption{Impact of the chosen IA prior for the internal consistency test. Left: Posteriors considering (orange) or excluding (green) bin 2 for our fiducial setup. Right: Same but using the wide IA prior from \cite{Asgari2021}.}
\label{fig:tomo_consistency_IA_prior}
\end{figure}

One change in our setup with respect to \cite{Asgari2021} is the prior for $A_{\rm IA}$. If we use the same wide prior
$A_{\rm IA} \in [-6.0, 6.0]$, we obtain a consistent $S_8$ value with a constraint on $A_{\rm IA} = 0.016 \pm 0.435$ (mean). Moreover, as shown in \cref{fig:tomo_consistency_IA_prior}, for this choice of prior we find that the `only bin 2' scenario prefers a higher value for $S_8$ that is in mild tension with the `without bin 2' results.
In addition to the wide prior, we also considered the impact of ignoring intrinsic alignments altogether, namely $A_{\rm IA}=0$. The results presented in \cref{fig:internal_consistency} show that the inferred values for $S_8$ is robust to the adopted $A_{\rm IA}$ prior, while the $S_8$ constraint is tighter by $\sim 5$ percent. This demonstrates the benefit of measuring the $A_{\rm IA}$ parameter externally to increase the constraining power on $S_8$, as was done by \cite{Wright2025b}

As shown in \cref{fig:cosmology_comp_with_lensfit}, the baryonic feedback-related parameter ($\log_{10}(T_{\rm AGN}$) is not constrained within our adopted narrow prior range, which in turn was based on the validity of the underlying model. To examine the impact, we tested a different (older) baryonic feedback model from \cite{mead2015}, which is based on the OverWhelmingly Large Simulations \citep[OWLS;][]{Schaye2010}. 
It has been used in a number of cosmological analyses \citep{Asgari2021, Yoon2021, Yoon2019}. However, the differences between the models impact mostly small scales. Thanks to our conservative scale-cut at 2 arcmin, the results are insensitive to the adopted baryonic feedback model.  

Finally, we examined the impact of settings that were adopted recently in the KiDS Legacy cosmic shear analysis \citep{Wright2025b}: including one more COSEBI mode $n_{\rm max} =6$, and a new covariance estimation, called `OneCov' \citep{Reischke2024}. Before running with the 6 COSEBI modes, we verified if this passes the $B$-mode test. All the tomographic bins passed, while the $p$-values increased for most bins. The $S_8$ constraint from the 6 COSEBI modes is consistent with the one from the fiducial run. While it decreases the errors on the $S_8$ constraint, the gain of constraining power is not significant. With the new covariance (OneCov), we also obtain a consistent $S_8$ constraint. 

\section{Conclusions}

We revisited the cosmic shear analysis of the KiDS-1000 data using an alternative shape measurement pipeline, based on \metacal, which is less sensitive to imperfections in the image simulations used to calibrate the remaining shear biases and, in particular, the multiplicative bias. Although \metacal provides a simple framework to infer the shear response from the actual data, the specific implementation of the \metacal pipeline---shape measurement method, weight function, and PSF treatment---still impacts the overall performance. 

We adopted a simple shape measurement, which has the benefit of increasing the number of galaxies for which shapes can be measured. Our setup enables an efficient determination of the bias for redshift-dependent blending, but it also down-weights fainter galaxies. Further optimisation may thus be possible. Nonetheless, we find that the effective number density is higher than that for the original KiDS-1000 analyses, based on \lensfit \citep{Asgari2021, Li2023b}, especially at high redshifts.
Moreover, the residual multiplicative shear biases are small for all tomographic bins, with the largest bias $m_{\rm final}=-0.0169$ observed for the second bin (see \cref{tab:mbias_final}). Compared to \lensfit, the multiplicative biases are considerably smaller for the highest redshift bins, while the dependence on apparent magnitude is much reduced.
Importantly, the calibration has negligible sensitivity to uncertainties in the properties of the galaxies used in the image simulations. 

To account for PSF anisotropy, we convolved the images with a kernel that circularises the PSF prior to the shape measurement. We find this works well, as the resulting additive bias is small. Afterwards, we applied an empirical correction to capture any residual PSF leakage and find that the additive bias is negligible. 

The \metacal measurements pass the tests that were also carried out in previous studies, and the constraint on 
$S_8=0.789_{-0.024}^{+0.020}$ is consistent with the results from \cite{Li2023b}. However, the constraints improve by 
about 28\% thanks to the increased effective source number density and the more robust shear calibration by \metacal. The difference with {\it Planck} is reduced by
0.5$\sigma$, but the difference seen in previous studies remains at a similar level. Hence, this is likely not caused by the shear measurement. Indeed, a recent
KiDS-Legacy analysis \citep{Wright2025b} agrees with 
{\it Planck}. The shift in these results is attributed to a combination of additional data and improved redshift calibration.

In this study we aimed to stay as close as possible to the previous KiDS-1000 analysis by \cite{Li2023b}. As a result, the star selection and redshift calibration were suboptimal. We will revisit these when we use our \metacal pipeline to analyse the KiDS-Legacy data. We expect to improve upon \cite{Wright2025b} thanks to the enhanced performance for the highest-redshift tomographic bins and the overall higher source number density.

\begin{acknowledgements}

MY, HHo, and SSL acknowledge support from the European Research Council (ERC) under the European Union’s Horizon 2020 research and innovation program with Grant agreement No. 101053992. 
SSL also acknowledges funding from the programme ``Netzwerke 2021,'' an initiative of the Ministry of Culture and Science of the State of Northrhine Westphalia.
LM acknowledges support from STFC grant ST/W000903/1.
MY, CH, and BS acknowledge support from the Max Planck Society and the Alexander von Humboldt Foundation in the framework of the Max Planck-Humboldt Research Award endowed by the Federal Ministry of Education and Research. 
CH also acknowledges support from the UK Science and Technology Facilities Council (STFC) under grant ST/V000594/1.
H. Hildebrandt and AHW are supported by a DFG Heisenberg grant (Hi 1495/5-1), the DFG Collaborative Research Center SFB1491, an ERC Consolidator Grant (No. 770935), and the DLR project 50QE2305.
BJ acknowledges support by the ERC-selected UKRI Frontier Research Grant EP/Y03015X/1 and by STFC Consolidated Grant ST/V000780/1.
MA acknowledges support by the UK Science and Technology Facilities Council (STFC) under grant number ST/Y002652/1 and the Royal Society under grant numbers RGSR2222268 and ICAR1231094.
RR is partially supported by an ERC Consolidator Grant (No. 770935).
     
\end{acknowledgements}

\bibliographystyle{aa} 
\bibliography{main}

\begin{appendix} 

\section{Source selection for the KiDS \metacal catalogue}
\label{app:source_selection}

\begin{table*}[h]
    \centering
    \caption{Metacal catalogue source selection.}
    \renewcommand{\arraystretch}{1.10}
    \label{tab:source_selection}
    \begin{tabular}{cccc}
        \hline\hline
    selection criteria & parameters & $\#$ of filtered sources & fraction [\%]\\
        \hline
        input KiDS-1000& -- &112,335,001 & 100\\

         masked area & ({\texttt{Mask}} \& 27676)>0 &32,644,554 & 29.06\\
         9 band GAAP flags & \texttt{FLAG\_GAAP\_anyband} != 0 &9,148,548 & 8.14\\
         stars & \texttt{fitclass} = 1 or 2 & 8,255,613 & 7.35\\
         asteroids & \texttt{MAG\_GAAP} (g-r) > 1.5 and \texttt{MAG\_GAAP} (i - r) > 1.5 &4,339,934 &3.86\\
       
         potential blends & \texttt{fitclass} = -10 or \texttt{contamination\_radius} $\leq$ 4.25 &29,487,117& 26.25\\        
         insufficient pixels on cutouts & & 1,666,036 & 1.48\\
         binary stars (for  $|e|\geq$ 0.3) &  \texttt{MAG\_AUTO} > - 1.728 $\times$ log(\texttt{FLUX\_RADIUS})+ 25.56  & 933,517 & 0.83 \\
         magnitude cuts & 20 < \texttt{MAG\_AUTO}  $\leq$ 25 & 19,113,615 & 17.01
         \\
       
        \hline
        all 6 selection criteria combined &  & 63,679,324  & 56.68 \\
        remaining sources      &  & 48,655,677& 43.31\\ 
        \hline
        tomographic binning (6 bins)  & $0.1 < z_{\rm B} \leq 2.0 $  & 47,080,020 & --\\ 
        \hline
        tomographic binning (5 bins)  & $0.1 < z_{\rm B} \leq 1.2 $  & 40,412,178& 100\\ 

        +S/N and resolution cuts & $\nu_{\mathrm{SN}}$ > 7 and $R$  < 0.6 & 33,906,706& 83.90\\
        +gold (photo-z SOM) selection& \texttt{Gold\_FLAG} = 1 & 27,523,376 & 68.11\\ 

        +AlphaRecal selection& calibrated e1, e2 < 1 &  27,523,376 &  68.11 \\

        \hline
        North patch & & 12,933,729 & 46.99\\
        South patch & & 14,589,647 & 53.01\\
         \hline
    \end{tabular}
    
    \renewcommand{\arraystretch}{1.0}
    \tablefoot{The initial KiDS-1000 catalogue, generated by the \se detection runs, contains sources from a total of 1000 tiles (see details in Appendix ~\ref{app:psf_residual_check}), which excludes those with high PSF residuals. The fraction of filtered sources for each selection is calculated relative to the total detected sources. Once all criteria are considered, 56.68\% of the original objects are filtered out. The final number of sources, 275,233,76, is 18\% more than that from the \lensfit v2 catalogue, 23,401,764.} 
\end{table*}

\begin{figure}[H]
\includegraphics[width=0.489\textwidth]{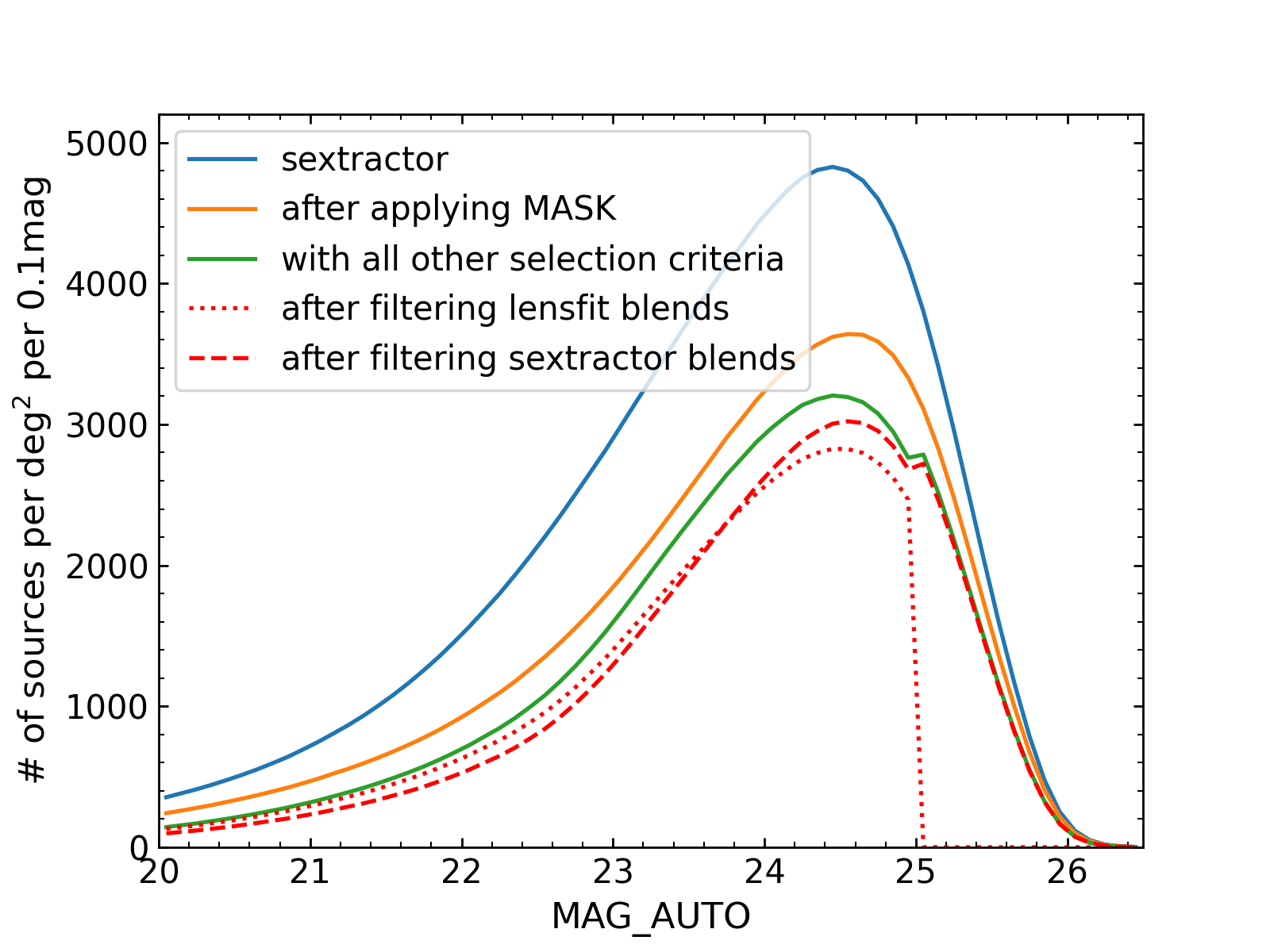}
\caption{Impact of various selection criteria on the number density of sources as a function of magnitude. The removal of objects in the masked regions and flagged objects have the biggest impact. The two different selections of blended sources are discussed in \cref{app:blending}.}
\label{fig:K1000_selection}
\end{figure}

Our analysis is based on the same imaging data, released as part of KiDS-DR4 \citep{Kuijken2019}, used in previous KiDS-1000 shear studies. We therefore start from the same catalogue of sources detected by \se. 
Moreover, to enable a more meaningful comparison of the performance of \lensfit and \metacal, where possible, we applied the same selections for various sources of contamination.
In particular, we took advantage of the fact that \lensfit classifies all objects in the input catalogue; the use of this \texttt{fitclass} allowed us to reproduce the star selection, without having to introduce new criteria. 

For our cosmological analysis, we used the flagging of blended objection by \lensfit, but we also explored an alternative approach based on the \se flags, which is detailed in \cref{app:blending}. As detailed in \cref{app:binary}, we did have to modify the removal of unresolved binary stars, because the impact depends on the shape measurement algorithm.

The sample of galaxies used for the lensing analysis is based on the criteria listed in \cref{tab:source_selection}, which also provides the number of sources affected by each selection. These results show that masking (to avoid bright stars, artefacts, etc.) and blended objects account for most of the reduction in sample size. As shown in \cref{fig:K1000_selection} those selections reduce the numbers over the full magnitude range.
Once all criteria are combined, about 43\% of the sources remain. This sample is reduced further after the photometric redshift calibration is considered, resulting in a total number of 30,401,986 sources. Although we closely follow the selection criteria used in previous KiDS-1000 analyses, the final \metacal sample is 30\% larger than the \lensfit v2 catalogue.

\begin{figure}
\includegraphics[width=0.45\textwidth]{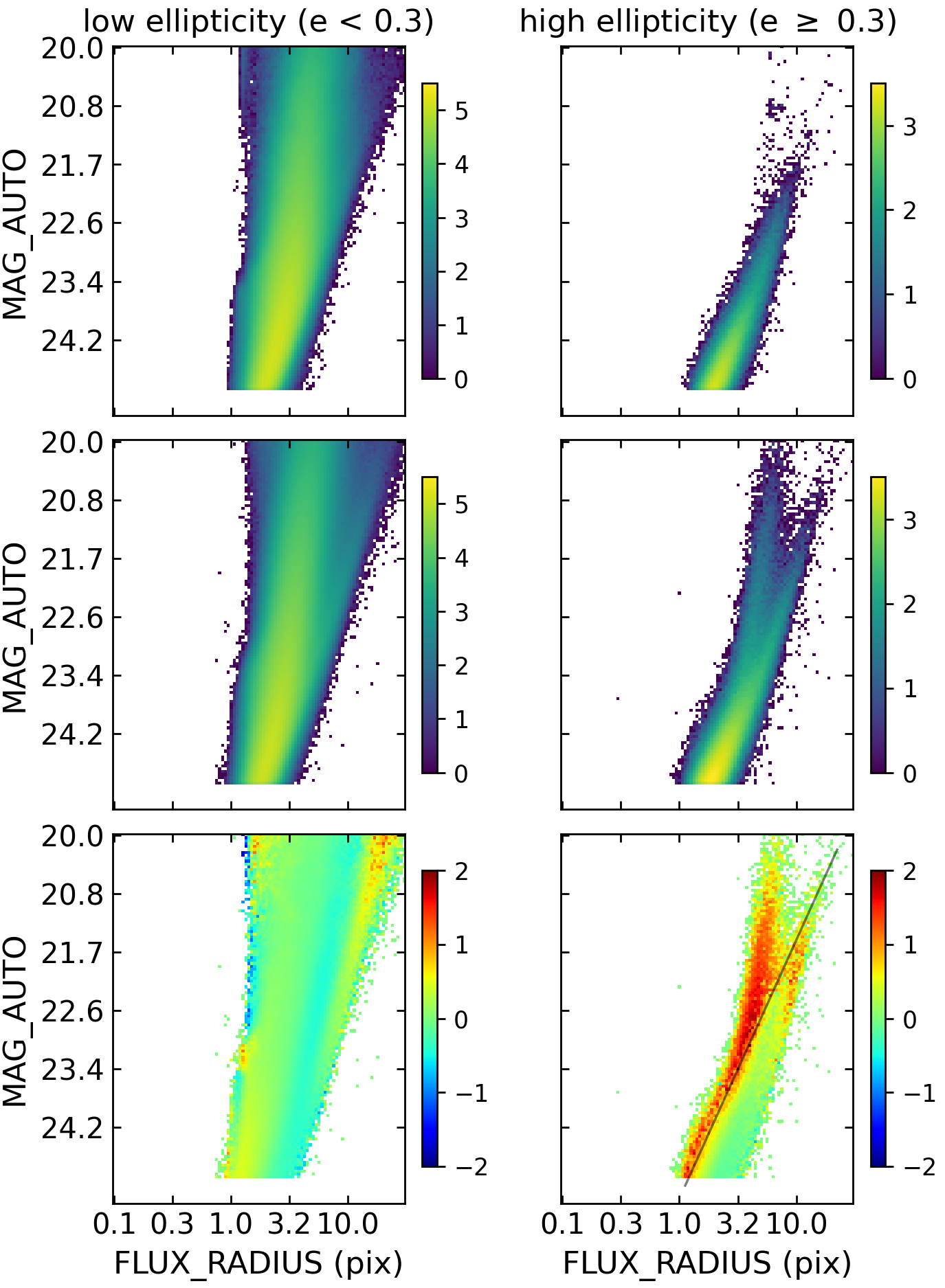}
\caption{Logarithm of the number density of sources as a function of size (\texttt{FLUX\_RADIUS}) and apparent magnitude (\texttt{MAG\_AUTO}), split by the observed third eccentricity, $|e|$, as defined by \cref{eqn:e_definition}.
Top and middle rows: Results for SKiLLS and KiDS, respectively.
Bottom row: Difference between the distributions (KiDS$-$SKiLLS), highlighting a population of bright, compact, but highly elliptical sources in the KiDS data that are lacking in SKiLLS. These are binary stars that contaminate the source catalogue. We removed objects lying above the line, drawn in the lower right panel, given by \cref{Eqn:binary}. The colour bars denote the logarithm of the number counts (base 10) in each size-magnitude bin, which are the same for all the subplots.}
\label{fig:K1000_binary_selection}
\end{figure}

\subsection{Unresolved binary stars}
\label{app:binary}

Unresolved binary stars show up as highly bright, compact, but highly elliptical sources. In previous analyses, the removal of these objects was based on \lensfit outputs \citep[see e.g. Fig. 30 in][]{Wright2024}. Although we expect many of these to be present in our catalogue, we decided to base our selection the shape estimates from \cref{eqn:e_definition}. 
To do so, we compared the number density of sources as a function of size and magnitude in both SKiLLS and KiDS, split by $|e|$. 
The results are presented in \cref{fig:K1000_binary_selection}. The bottom left panel shows that the distributions for objects with $|e|<0.3$ agree well between SKiLLS and KiDS-1000. In contrast, we observe an excess of bright, compact objects for elliptical sources ($|e|\geq 0.3$) in the actual KiDS data. Based on these results we defined the following empirical selection, and remove sources with $|e|\geq 0.3$ and (black line in \cref{fig:K1000_binary_selection})
\begin{equation}
        \mathrm{MAG\_AUTO} \leq - 1.728 \times \mathrm{ln(FLUX\_RADIUS}) + 25.56.  
\label{Eqn:binary}
\end{equation}

As reported in \cref{tab:source_selection}, this removes 0.83\% of the sources. This fraction is larger than the 0.14\% reported in \cite{Wright2024}, but this may well reflect the ability of \lensfit to reject some objects as bad fits at an earlier stage. In contrast, \metacal tends to return successful shape measurements for almost all bright sources.

\subsection{Selection of blended sources}
\label{app:blending}

Blended sources are a major source of shear bias, and the main reason for the difference between the multiplicative bias of \metacal and the detection bias from \se \citep{Hoekstra2021a}. A further complication is posed by blended sources at different redshift \citep{MacCrann2022, Li2023a}. Hence, removing blended sources may help reduce the overall bias and improve the robustness of the shear calibration. To this end, we use the segmentation map from \se to mask pixels from neighbouring detected sources (see \cref{sec:setup}), but this cannot fully eliminate the impact of blended sources. 

The image simulations provide a convenient test bed to explore the efficacy of blending selection criteria. In particular, the criteria could be optimised, and subsequently be applied to the data. However, it is important that the simulated data capture the clustering of galaxies adequately \citep{Martinet2019}. As shown in \cite{Li2023a}, SKiLLS reproduces the observed distribution of magnitudes and separations between neighbouring galaxies well.  

To explore the impact on the multiplicative bias, we considered two selection criteria to filter out the potential (recognised) blends:
\begin{itemize}
\item[1)] select sources with a \lensfit output flag \texttt{fitclass}!=-10 and a \lensfit contamination radius 
$\le 4.25$ pixels;
\item[2)] select the sources with \se output \texttt{FLAGS}=0.
\end{itemize}

The first is the selection used for the previous KiDS analyses, which are all based on catalogues produced with \lensfit \citep{Giblin2021}. However, as one might like the \metacal pipeline to be independent of \lensfit,  it is useful to compare the performance to the second criterium, which corresponds to the one used by the DES pipeline  \citep{Gatti2021}. As shown in \cref{fig:K1000_selection}, the selected sources have slightly different magnitude distributions, where the sharp drop for $m>25$ arises from an internal selection in \lensfit; most of them are assigned with \texttt{fitclass} = -1 (flag for objects with insufficient data) and also with \texttt{contamination\_radius} = 0.

\begin{figure}[h]
\includegraphics[width=0.489\textwidth]{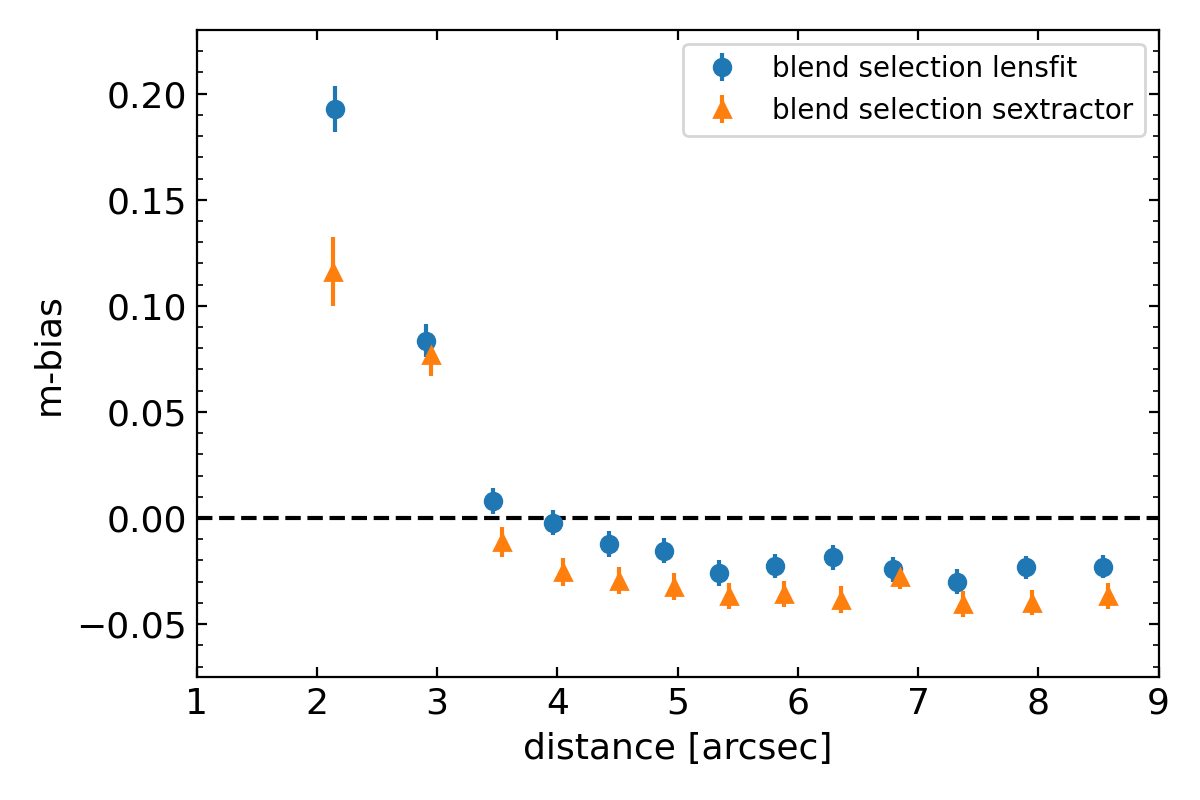}
\caption{Multiplicative shear bias as a function of distance to the nearest detected neighbour, after removing blended sources using the two criteria described in the text. The blue points correspond to the baseline KiDS-1000 analysis \citep{Giblin2021}, whereas the orange points reflect the DES pipeline \citep{Gatti2021}.}
\label{fig:blend_selection}
\end{figure}

\Cref{fig:blend_selection} shows the multiplicative shear bias as a function of distance to the nearest detected source, after one of the two criteria was applied. For apparently isolated sources the bias does not vanish, because many blends are not detectable \citep{Hoekstra2021a}. The smaller bias for \lensfit (blue points) for large separations, suggests that it recognises blends a bit better than our \se setup. Nonetheless, the difference between both criteria is small, also for small separations, where the biases increase rapidly. Hence, we do not expect the final cosmological results to depend on the selection.

However, we adopted the selection based on the \lensfit output, because it performs a bit better, but more importantly, because it enables a more direct comparison with previous work. As \cref{fig:blend_selection} shows,
it might still be beneficial for future KiDS analyses to consider eliminating all sources with neighbours detected within $3''$, as the bias rises quickly, although they may still be of interested for galaxy-galaxy lensing studies. 

\subsection{Star contamination}
\label{app:star_contamination}

Another important class of potential contaminants are stars that are not identified as such. At bright magnitudes, stars can generally be separated well based on their size, unless they are blended with another source. However, the size estimates are noisier for fainter stars, and overlap with the observed sizes of galaxies. As a result, the shear bias depends on the star density \citep{Hoekstra2015,Hoekstra2017}. In \cref{sec:sensitivity} we explored the sensitivity of the multiplicative bias to variations in star density, and found that the impact appears negligible. However, for the second tomographic bin,  \cref{fig:subsample_sensitivity} indicated a change in $m$ for the highest star density sample. Therefore, we explore the contamination by stars in more detail here.

\begin{figure}[h]
\includegraphics[width=0.489\textwidth]{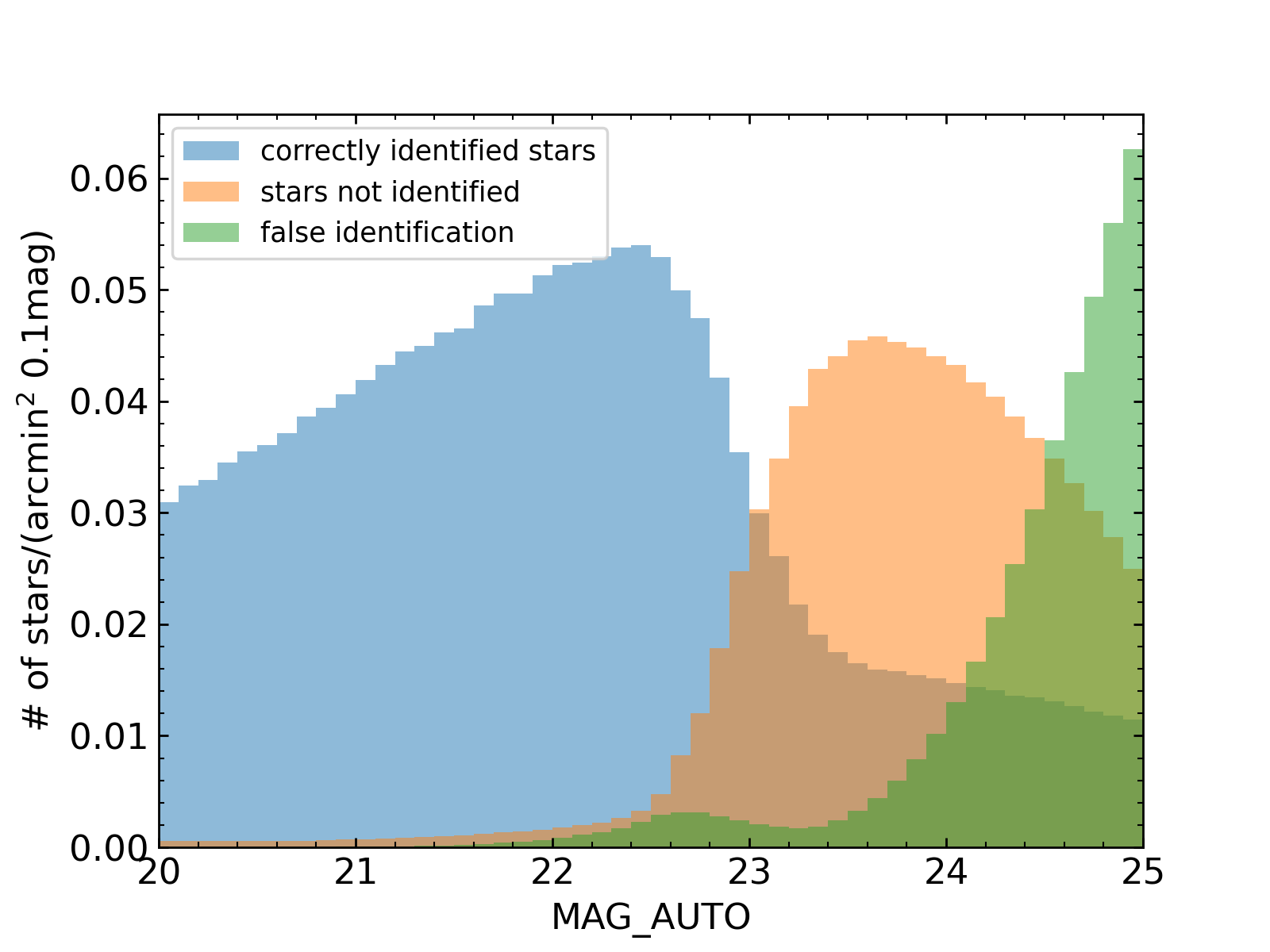}
\caption{Number density of the stars in the 
catalogue of sources detected by \texttt{sextractor} in SKiLLS. The sources were cross-matched with the input catalogues, resulting in a sample of correctly identified stars (\texttt{fitclass} = 1 or 2), missed stars, and incorrectly identified objects, denoted by the blue, orange, and green histograms, respectively. \lensfit flagging fails for stars fainter than \texttt{MAG\_AUTO}$> 22.5$.}
\label{fig:star_identification}
\end{figure}

As detailed in Sect.~2.2 of \cite{Li2023a},
SKiLLS includes a realistic population of stars, based on the population synthesis code \texttt{Trilegal} \citep{Girardi2005}. The star densities are sampled from the KiDS tiles to account for the variation with sky position.

To quantify the level of star contamination, we cross-matched the detected sources to the input catalogue of SKiLSS. This resulted in a sample of objects that were correctly flagged as stars by \lensfit (\texttt{lensfit\_class} = 1 or 2),
indicated by the blue histogram in \cref{fig:star_identification}, which works well for bright stars with \texttt{MAG\_AUTO}$< 22.5$.
At faint magnitudes, it is harder to distinguish stars from galaxies. The latter are more numerous and hence the rate of false positives increases, as indicated by the green histogram in
\cref{fig:star_identification}. As these objects are removed from the galaxy catalogues, they do affect the shear bias. 

The galaxy sample is contaminated by stars that are not identified as such, which is the case for most stars with \texttt{MAG\_AUTO}$>22.5$
(orange histogram in \cref{fig:star_identification}).
As a result, unidentified stars contaminate the tomographic bins by 2.9, 6.9, 4.3, 2.7, 0.3, and 0.3 percent, respectively.
In the case of \lensfit, the shear weights effectively reduce their impact to negligible levels, but as we opted for a weighting scheme that depends only on magnitude for \metacal, the 
contamination is reduced only slightly to 
2.7, 4.4. 3.1, 1.9, 0.3, and 0.3 percent, respectively.

Although this level of contamination is not ideal, we note that the star counts match those in the KiDS data, and hence is accounted for in our estimates for $m$. Moreover, the results presented in \cref{fig:subsample_sensitivity} show that the relatively high fraction of stars in the second tomographic bin has only a modest impact on the shear bias, as $\Delta m \simeq \pm0.006$. As an extreme test, we also computed the change in $m$ when all input stars are removed. In this case we find $\Delta m \simeq \pm0.0047$ for the second bin, consistent with our sensitivity test.

For a future analyses that does not use the \lensfit star selection, we plan to reduce the impact of stars by changing the weighting or selection schemes of stars. For instance, a simple cut based on the observed sizes (\texttt{FLUX\_RADIUS}) seems more effective (especially for the sources fainter than magnitude 22.5). We leave this for future work, as we prefer to stay as closely as possible to the \lensfit selection for a direct comparison. 

\section{Changing the size of weight function}
\label{app:test_circular_width}

\begin{figure}[h]
 \includegraphics[width = 0.45\textwidth]{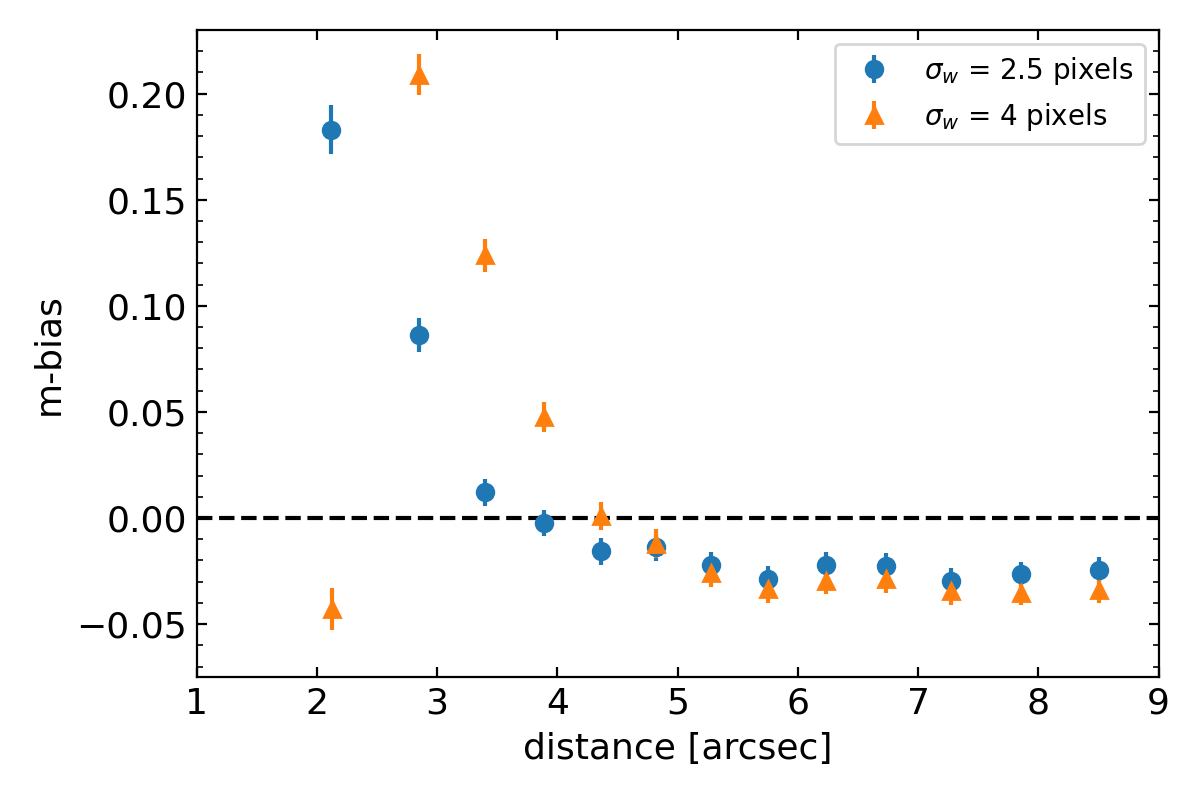}
    \caption{Multiplicative bias, $m$, measured as a function of distance to the nearest detected source (MAG\_AUTO < 24.5) using a circular weight function with $\sigma_w$ = 2.5 pixels (blue) and $\sigma_w=4$ pixels (orange). The bias converges to similar values for isolated sources, but the effect of blending appears for smaller separations in the case of the more compact weight function.}
\label{fig:m_bias_circular_width}
\end{figure} 

As described in Sect.~\ref{sec:setup}, we measured the quadrupole moments of the source surface brightness distributions using a fixed weight function. By construction, this removes the sensitivity of the weighting scheme to the shear. Based on the observed distribution of galaxy sizes
(see \cref{fig:com_skills_k1000}) we adopted a dispersion of
$\sigma_w=2.5$ pixels as baseline.

To quantify the impact of the width of the weight function, we repeated the \metacal shear calibration using a weight function with $\sigma_w=4$ pixels. 
We compared the uncertainty in the inferred multiplicative bias, which is a measure for the overall precision with which the lensing signal can be measured. We found that the uncertainty in the shear measurement increased by 16\% when the width was increased from $\sigma_{\rm w}=2.5$ to $\sigma_{\rm w}=4$ pixels. 

As the width of the weight function does not have a big impact on the precision of the shear estimate, other considerations can be taken into account to optimise the choice of weight function, such as the sensitivity to blending. A more compact weight function suppresses the light from close sources. To quantify the impact of blended sources, we determined the multiplicative bias as a function of distance to the nearest detected source for galaxies with MAG\_AUTO$< 24.5$. The results for the two weight functions are presented in \Cref{fig:m_bias_circular_width}. The multiplicative bias converges to similar values for isolated sources, but the effect of blending appears for smaller separations in the case of the more compact weight function (blue points). This further motivates our choice for the fiducial weight function with $\sigma_w=2.5$ pixels.

\section{PSF residual check}
\label{app:psf_residual_check}

\begin{figure*}
\centering
\includegraphics[width=0.98\textwidth]{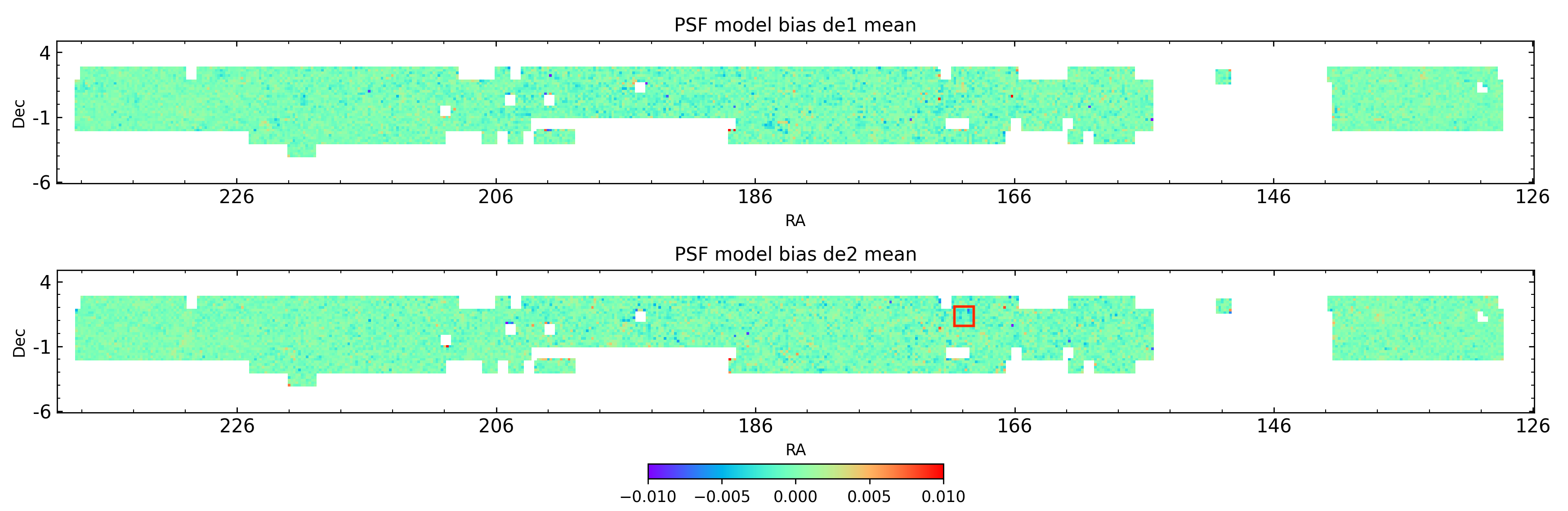}
\includegraphics[width=0.86\textwidth]{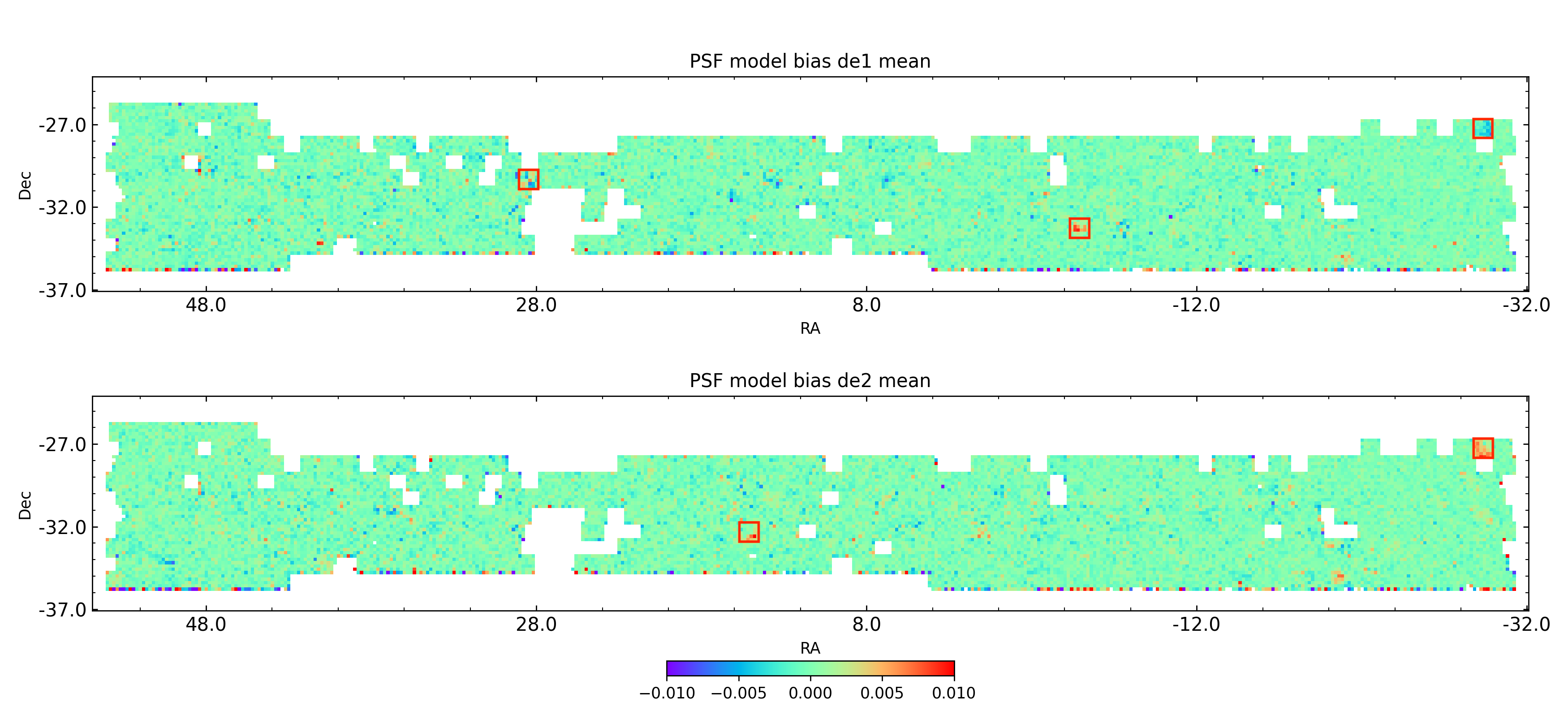}
\caption{PSF model residual over the KiDS survey coverage. The residual is averaged over stars, which is defined as the deduction of the measured ellipticity of the stacked star images from the ellipticity of the PSF modelled at the star location ($\rm{d}e$ = $\rm{e}_{PSF}$ - $\rm{e}_{star}$ ). Top: Northern fields, $\rm{d}e_1$, $\rm{d}e_2$. Bottom: Southern fields, $\rm{d} e_1$, $\rm{d} e_2$. This residual figure demonstrates the accuracy of astrometry as well as the accuracy of PSF models.}
\label{fig:PSF_model_residual_field}
\end{figure*}

Imperfect PSF modelling can introduce significant bias in shear measurement \citep{Hoekstra2004}. 
We ignored this when quantifying the shear bias using SKiLLS, because we assumed the PSF model is known perfectly. \cite{Li2023a} explored the sensitivity of \lensfit to PSF modelling errors using SKiLLS, and found that this can shift the multiplicative bias by about $\Delta m=0.005$.
Here, we do not revisit the actual PSF modelling, also  because previous work has validated the fidelity of the PSF model \citep[for details, see Sect. 3.3. in][]{Giblin2021}.
Instead we examine a different aspect that is relevant for our analysis of KiDS data.

The PSF model was developed for the \lensfit pipeline \cite[see][for details]{Miller2013, Kuijken2015} and derived from stars in the individual exposures. This is used by \lensfit to simultaneously fit a model of the galaxy surface brightness to the individual exposures \citep{Miller2013}. We use the same model, but our \metacal pipeline uses the stack of the postage stamp images. Therefore, we created a stacked PSF model for each galaxy. 

During the step of image stacking, we excluded the chip images with fewer than 30 stars to ensure the accuracy of PSF model. Such low numbers might indicate problems with the images (e.g. a faulty video board) or because bright stars render large areas useless. This matches a selection that is conducted internally by \lensfit, and excludes 756 individual chip images. Its necessity is demonstrated by a reduction in PSF residuals in the affected images. 

In principle, the PSF of the stacked images should match the average PSF model for that location. A direct comparison of the two images thus tests if the stacking is done correctly. However, astrometric errors can introduce a bias between the two PSFs, thus impacting the accuracy of the \metacal measurements. The SKiLLS setup assumes perfect astrometry, and thus cannot be used to test this specific aspect. We therefore validated this step using the KiDS data.

We measured the ellipticities of the model PSFs that were reconstructed from the parametrised equations using the coefficients determined during the PSF modelling. At the locations of the stars, we measured ellipticities using  \texttt{galsim.hsm.FindAdaptiveMom}, which uses adaptive elliptical weighting to estimate moment-based ellipticities. We repeated this measurement on the actual stacked KiDS $r$-band images, and show the difference 
(${\rm d}e = e^{\rm mod}_i - e^{\mathrm{obs}}_i$) averaged in bins of 12\texttimes12 arcmin in \cref{fig:PSF_model_residual_field}.

The top two panels show the results for KiDS-North, while the bottom two panels show the same for KiDS-South. The residuals are generally consistent with zero, but we did identify five pointings that showed significant deviations.\footnote{The following are the pointings: `170p0\_1p5', `15p3\_m32p1', `28p8\_m30p2', `330p8\_m27p2', and `355p2\_m33p1'.} These are marked by red squares in \cref{fig:PSF_model_residual_field}. Upon further inspection, we found problems with the astrometry between the exposures, and excluded them from the analysis. We also omitted pointing `356p6\_m29p2', because of a corrupted file. Hence, the final selection results in 1000 pointings and an effective area of 771.90 $\deg^2.$

\section{Linearity test of shear bias}
\label{app:test_high_amp_shear}

To determine the shear biases we apply a constant shear
of amplitude $|\gamma|=0.04$ to the simulated galaxies,
while we used more realistic, smaller values, to account for blended sources at different redshift in \cref{sec:m_bias_variable}. However, if the value of $m$ depends on the shear applied, our estimates may be biased \citep{Heymans2006}.
To validate the fidelity of the fixed input shear calibration, we quantify here the linearity of our \metacal implementation, using a wide range of shear inputs,
$|\gamma| \in [0, 0.2]$.

The setup of the image simulations is similar to the baseline, but we made some simplifications to reduce computing time. We only modify the $r-$band images, and do not repeat the determination of photometric redshifts from the multi-band images. Also, we fixed the PSF model. Otherwise, the analysis is unchanged, including the use of
multiple exposures. Further details, as well as the corresponding results from the same simulations for \lensfit can be found in Appendix~A of \cite{Li2024}.

\begin{figure}[h]
\includegraphics[width=0.46\textwidth]{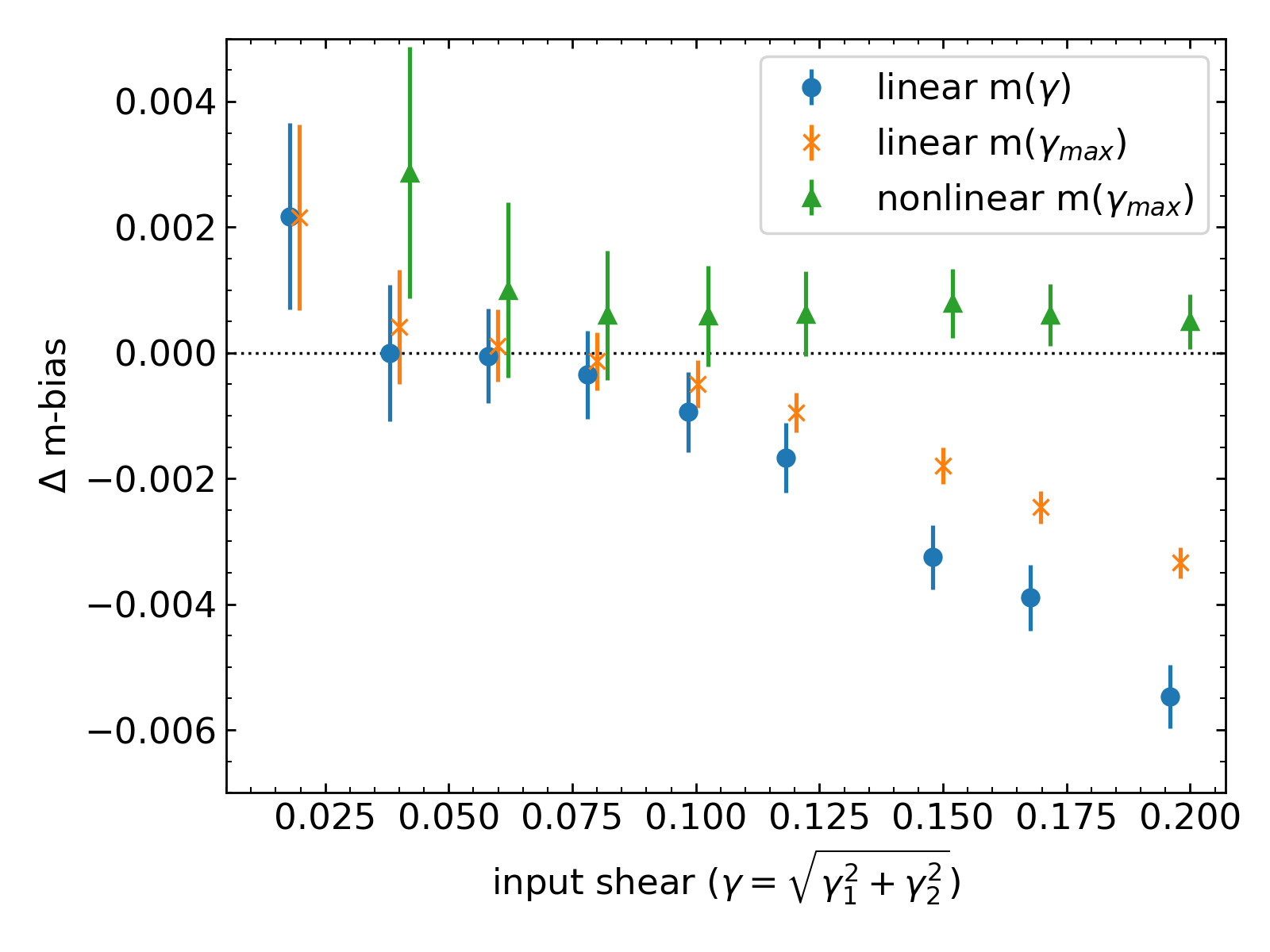}
\caption{Change in $m$ relative to the baseline input shear $|\gamma|=0.04$ as a function of input shear. The blue points correspond to our linear calibration procedure with a single constant shear.
The orange points indicate the estimated biases when fitted to all shears up to $|\gamma_{\rm max}|$, yielding similar but more precise results. Our \metacal pipeline is well within the linear regime for our setup. The green points show the residuals when considering a third-order polynomial.}
\label{fig:linearity_test}
\end{figure}

The change in $m$ with respect to the $|\gamma|=0.04$ case is shown in \cref{fig:linearity_test} as a function of the input shear. We assume that the change in $m$ is the same for both shear components. The blue points indicate the results if we simply repeat our calibration, assuming a linear relation, but with a larger shear. Alternatively, we can consider all the simulations up to $|\gamma_{\rm max}|$, indicated by the orange points. In both cases, we find that $m$ does not depend on the input shear up $|\gamma|=0.075$; for larger shears, the relation becomes non-linear. However, even for the largest shears that we considered, the changes are well within the requirements for KiDS-1000. Comparing to the findings by \cite{Li2024} using \lensfit, the level of non-linearity is comparable. Although \cite{Li2024} observed a small quadratic contribution, the dominant non-linearity depends on the cube of the input shear, which is expected based on symmetry arguments. Therefore, we also fit a third-order polynomial, and indicate the residuals by the green points in \cref{fig:linearity_test}. 

These results show that our baseline setup provides an accurate estimate for $m$ for the shears relevant for our cosmic shear analysis. In fact, given the linearity and the final calibration procedure, we could have used a somewhat larger shear for the fixed shear calibration, thus reducing the number of simulated images.

\end{appendix}

\end{document}